\documentclass[]{JHEP3}  
\usepackage{epsfig,amsbsy,calc,graphics,latexsym,graphicx,psfrag}

\def\aka{{\it a.k.a.}\ }
\def\ie{{\it i.e.}\ }
\def\eg{{\it e.g.}\ }
\def\etc{{\it etc.}\ }
\def\etal{{\it et al.}}
\def\cf{{\it cf.}\ }
\def\viz{{\it viz.}\ }
\def\be{\begin{equation}}
\def\ee{\end{equation}}
\def\bea{\begin{eqnarray}}
\def\eea{\end{eqnarray}}

\def\hS{\hat{S}}

\def\ker#1{\!\cdot\!#1\!\cdot\!}
\def\dDelta{\dot{\Delta}}
\def\sqr#1#2{{\vcenter{\vbox{\hrule height .#2pt
        \hbox{\vrule width .#2pt height#1pt \kern#1pt
              \vrule width.#2pt}
          \hrule height .#2pt}}}}
\def\Box{\,\sqr{6}{6}}
\def\ph#1{\phantom{#1}}
\def\ra#1{{\scriptsize \overrightarrow#1}}
\def\la#1{{\scriptsize \overleftarrow#1}}
\def\Ra#1{{\overrightarrow#1}}
\def\La#1{{\overleftarrow#1}}

\def\gt{\tilde{g}}
\def\bt{\tilde{\beta}}

\def\sig{\sigma}
\def\Lam{\Lambda}

\def\A{{\cal A}}
\def\S{{\cal S}}

\def\C{{\cal C}}
\def\D{{\cal D}}
\def\F{{\cal F}}

\def\J{{\cal J}}

\def\gap{\hspace{0.05in}}

\def\c{c^{-1}}
\def\ctil{\tilde{c}^{-1}}
\def\ct{\tilde{c}}
\def\rtil{\tilde{r}}
\def\rt{\rtil}

\def\tr{\mathrm{tr}}
\def\str{\mathrm{str}}

\def\one{\hbox{1\kern-.8mm l}}

\def\ds{\displaystyle}
\def\ldl{\Lambda \partial_{\Lambda}}
\def\eq#1{(\ref{#1})}
\def\eqs#1#2{(\ref{#1},\ref{#2})}
\def\sec#1{sec.\ \ref{#1}}
\def\fig#1{fig.\ \ref{#1}}
\def\ins#1#2#3{\hskip #1cm \hbox{#3}\hskip #2cm}

\title{A proposal for a manifestly gauge invariant and universal calculus
in Yang-Mills theory}

\author{Stefano Arnone, Antonio Gatti
and Tim R. Morris\\ 
Department of Physics and Astronomy,
University of Southampton\\
\hspace{.045em} Highfield, Southampton SO17 1BJ, U.K.\\

E-mails: \email{sa@hep.phys.soton.ac.uk, gatti@hep.phys.soton.ac.uk, 
T.R.Morris@soton.ac.uk}}  

\maketitle

\preprint{SHEP 02-22}
\abstract{
We uncover a method of calculation that proceeds
at every step without fixing the gauge or specifying details of
the regularisation scheme. Results are obtained 
by iterated use of integration by parts and gauge invariance
identities. The initial stages can even be computed diagrammatically.
The method is formulated within the framework of an exact
renormalization group for $SU(N)$ Yang-Mills gauge theory,
incorporating an effective cutoff through a manifest spontaneously
broken $SU(N|N)$ gauge invariance. 
We demonstrate the technique with a compact calculation of the one-loop
beta function, achieving a manifestly universal result, and without
gauge fixing, for the first time at finite $N$.
}

\begin{document}
\section{Introduction} 
\label{Introduction}

It need hardly be stressed that there is a clear need for a better
non-perturbative understanding of quantum field theory. Examples can be
given from all domains where quantum field theory is applicable: from
phase transitions in early universe cosmology, quantum gravity, QCD,
through to high temperature superconductivity, to mention just a few. 

The exact Renormalization Group (RG) \cite{Wil, weg}, the continuum version
of a Wilsonian RG, provides a powerful framework for considering
non-perturbative analytic approximations to quantum field theories
\cite{etc,morig,rev,jose,trunc}. This follows from the fact that solutions 
of the
corresponding flow equations, \ie the Wilsonian effective action, can
be found directly in terms of renormalized quantities, that all physics
(\eg Green functions) can be extracted from this, and that
renormalizability is trivially preserved in almost any approximation
\cite{morig,rev}.

Central to many non-perturbative problems, including the examples
quoted above, is the presence of gauge invariance. (This local
invariance can be accepted as
either fundamental or, \eg in the case of high $T_c$, 
effective \cite{effectiveGauge}.) However the
introduction of a real\footnote{as opposed to \eg
analytic continuation of perturbative amplitudes in dimensional
regularisation} effective cutoff $\Lambda$, a crucial step in the definition 
of a Wilsonian RG, typically breaks this gauge invariance \cite{others}.
 
Fortunately it is possible to formulate more general exact RGs
\cite{jose}, which are gauge invariant \cite{alg,ymi,ymii,rome}.  A
wonderful extra benefit in this generalised framework is that
calculations can proceed with manifest gauge invariance preserved at
every stage \cite{alg,ymi,ymii,rome}. There is thus no need for gauge
fixing and the corresponding ghosts. The challenging non-perturbative
problem of Gribov copies \cite{Gribov},
is thus entirely avoided.\footnote{Gribov problems are known to result 
in an erroneous answer for covariant gauges \cite{neuberger, baulieu}.} Even at
the perturbative level the full power and beauty of gauge invariance
then shines through. Unlike BRST transformations, gauge transformations
are at most linear in the quantum fields and thus are not deformed at
the quantum level. For a non-Abelian gauge group, the connection in the
covariant derivative thus remains dimension one and is unrenormalized.
If the coupling is scaled out of the connection, no wave function
renormalization is possible for the gauge
field \cite{alg,ymi,ymii,rome}. Only the coupling renormalizes. The
usually na\"\i ve assumption that the
effective action is built only from gauge invariant combinations of the
covariant derivative is here true even at the quantum level; all vertices 
are subject to simple strong
constraints -- the so-called ``na\"\i ve Ward identities'' that follow
from exact gauge invariance.

In order to formulate such a gauge invariant exact RG, we need to
incorporate a gauge invariant real cutoff $\Lambda$. Moreover, this has
to appear in a way that can be naturally incorporated in the effective
action framework \cite{alg,ymii,rome}. We use the solution given in
ref. \cite{sunn}, which provides a regularisation for $SU(N)$
Yang-Mills theory in $D\le4$ spacetime dimensions, and when
$N=\infty$, in any dimension.  This is implemented by embedding the
Yang-Mills theory in a $SU(N|N)$ gauge theory, regularised by covariant
higher derivatives.  The $SU(N|N)$ gauge theory is then spontaneously
broken in the fermionic directions, at the same scale $\Lambda$, with
the resulting heavy fields playing the r\^ole of gauge invariant
Pauli-Villars fields \cite{Slavnov}.  (Actually this
corresponds to regulating a $SU(N)\times SU(N)$ Yang-Mills theory, 
but the non-unitary second copy decouples in the continuum limit 
\cite{sunn}.)
 
The work presented in this paper extends previously published results
in a number of significant ways. The flow equation in ref. \cite{ymii}
was developed intuitively from the bottom up without the author being
aware of the underlying $SU(N|N)$ structure \cite{alg}. In contrast,
the present flow equation is very simple and beautiful in its conception, its 
form being tightly constrained by the manifest invariance under the 
spontaneously broken local $SU(N|N)$. The earlier flow equation in ref.
\cite{ymii} was regularised only to one loop (and then only for
external gauge fields). Here the gauge invariant regularisation is 
complete, working to all orders in perturbation theory. The
formulation given in ref. \cite{ymii} was restricted to $N=\infty$
(again as consequence of regularisation limitations). The
present formulation makes sense also at finite $N$. 

And last but by no means least, a powerful computational technique for
working within this framework, is developed, building on the insights
gained from ref. \cite{sca}. In all realisations of the Wilsonian RG
there is an unavoidable freedom in the construction, equivalent to
regularisation scheme dependence, but especially deeply embedded
\cite{jose}. Thus for example any version of the exact RG receives its
very definition in part by specifying a cutoff function $c$. This
redundancy is magnified in the case of these gauge invariant exact RGs,
because the regularisation requires a further cutoff function $\ct$,
and the requirement of gauge invariance forces the introduction of
further choices, to a large extent arbitrary, namely the
covariantisation(s)\footnote{Different parts of the flow equation can
even have different covariantisations.} and the ``seed'' interactions.
Nevertheless, physical quantities must be independent of these choices.
Providing that we limit ourselves to controlled expansions (\eg those
in weak coupling, strong coupling, $1/N$, $1/D$ \etc), the same must be
true of the approximations.

There ought then to be a way of computing these results without having
to specify the above choices of regularisation scheme.  We uncover just
such a method. The large redundancy in the regularisation scheme is
turned to our advantage, furnishing a guide to streamlined computation
of universal quantities. Thus in the method, we are forbidden from
`looking inside' any vertices (of the the seed action or covariantised kernels
but also of the undifferentiated Wilsonian effective action). The
initial stages of the calculation are then so constrained that they can
be effectively performed diagrammatically.

Central to the method are integrated exact RG kernels which play the
r\^ole of regularised propagators, specifically by being the inverse of
the corresponding two-point vertices. For the gauge fields, since gauge
invariance is preserved, these inverses do not exist.  Instead the
integrated kernels are inverses only in the transverse space,
leaving longitudinal remainders that generate gauge transformations.

These `effective propagators' are introduced by integrating by parts
with respect to $\Lambda$, resulting in differentials of the Wilsonian
effective action. These latter are evaluated via their flow equations, after
which gauge invariance identities are used to evaluate further, where
possible.  This procedure is iterated until there are no terms left
that depend on the choice of covariantisation or seed interactions. It
is then straightforward to cast the remaining terms as total
derivatives in momentum space or otherwise show them to be universal.

Although we apply the method here only to the computation of the
one-loop $\beta$ function of $SU(N)$ Yang-Mills theory, we believe the
procedure to be of general applicability. In fact our aim is to apply
these ideas to the non-perturbative domain. As already mentioned, exact
RG equations are ideally suited for this.  It is important to note in
this context that our gauge invariant exact RG equation and the
regularisation it embodies do not require perturbation theory for their
definition.

A necessary step is to thoroughly test and understand the
framework in the perturbative domain.  For calculations at two loops
and beyond, it is helpful to augment the present flow equation. A
full report of this investigation however is left for the future
\cite{us}.

\subsection{Overview}

Most of the present paper, up to \sec{calculation},
is concerned with setting up and justifying the formalism. Specifically, we 
start in \sec{regularisation} with a review of the regularisation \cite{sunn},
and some of its novelties, adapting it to the exact RG constructed in this 
paper. Although we leave many elements of the exact RG unspecified, using
this freedom to guide the calculation and display universality, there
are a number of basic restrictions needed on the set of exact RGs we allow.
Sec. \ref{Necessary} sets out the general properties we require, and their 
consequences and interpretation in the present case. Sec. \ref{Supergauge} 
deals with a particular novelty that arises in the transformation
of the $SU(N|N)$ supergauge field functional derivative, which in turn
leads to a further constraint on the form of the exact RG. Secs. 
\ref{Covariantisation} and \ref{Decoration} spell out the restrictions placed 
on the form of supergauge covariantization and further decoration of momentum 
space kernels, and introduce the general notation used to define the 
resulting vertices. In \sec{Superfield}, we introduce the corresponding
notation for action vertices, and the form the notation takes after 
spontaneous supersymmetry breaking. 

In \sec{manifestly} we introduce the
flow equation, defining various elements and developing some of its
properties. In particular we prove its supergauge invariance, and in
\sec{Supersowing} prove a closely related property that leads to a powerful
diagrammatic incorporation of the supergauge algebra, as is explained
in \sec{DiagrammaticI}. From \sec{After} onwards we work in the spontaneously
broken theory, \sec{unbroken} containing the centrally important
resulting relations between vertices, the so-called na\"\i ve Ward identities,
both for the remaining $SU(N)\times SU(N)$ bosonic gauge invariance, but
also for the broken fermionic gauge invariances. 

In \sec{Seed}, we use
general arguments to determine the form of the classical effective
action two-point vertices. These are used to determine the kernels in
\sec{kernels}, and thus the integrated kernels in \sec{integrated}. Most
importantly, we show how these behave as effective 
propagators up to gauge remainder terms. Together with \sec{unbroken},
these provide the essential properties behind the `calculus' that follows.
These properties are seen clearly in the broken fermionic
sector, if the fermionic parts are combined into a $D+1$
dimensional vector as in \sec{Five}, a notation we then adopt for the
rest of the paper. Sec. \ref{Enforcing} explains precisely
when one can expect to get a universal result for the first two coefficients
of the Yang-Mills $\beta$ function. Although this is standard, the
universality is actually violated without the further restrictions
that are introduced in \sec{Ensuring}, a novel consequence of
Pauli-Villars regularisation in an exact RG framework \cite{alg,ymii}.

Finally, in \sec{calculation}, we set out the calculation, with 
\sec{DiagrammaticA} in particular containing the main iterative 
diagrammatic procedure, and \sec{total} the heart of the calculation
from the physics point of view. In \sec{Conclusions} we
summarise and draw our conclusions.

\section{Preliminary comments} 
\label{Preliminary}

Throughout the paper we work in Euclidean space of dimension four.
We can formulate everything in general dimension $D$, and strictly
speaking should, since the limit $D\to4$ is necessary to rigorously
define the regularisation \cite{sunn}. However, as we will show, for
the calculation of terms such as the one-loop $\beta$ function in $SU(N)$
Yang-Mills, we do not need to pay attention to this subtlety. Therefore
for simplicity of exposition we will set $D=4$ in this paper, leaving
the full generality until ref. \cite{us}.

\subsection{The regularisation}
\label{regularisation}

Instead of working just with the $SU(N)$ gauge field, which we write as
$A^1_\mu(x)\equiv A^1_{a\mu}\tau^a_1$, where $\tau^a_1$ are the $SU(N)$
generators orthonormalised to $\tr(\tau^a_1\tau^b_1)=\delta^{ab}/2$, we
embed it in a $SU(N|N)$ supergauge field \cite{sunn}:
\be
\label{defA}
\A_\mu = {\A}^{0}_{\mu} \one
+ \left( \!\! \begin{array}{cc}
                   A^{1}_{\mu} & B_{\mu} \\
                   \bar{B}_{\mu} & A^{2}_{\mu}
                   \end{array} \!\!
            \right).
\ee
Here we have written $\A$ as an element of the $SU(N|N)$ Lie
superalgebra, using the defining representation, \ie as a supermatrix
with bosonic block diagonal terms $A^i$ and fermionic block
off-diagonals $B$ and $\bar{B}$, together with the central term
$\A^0\one$. As required by $SU(N|N)$, the
supermatrix (and thus also $\A$) is supertraceless, \ie $\tr A^1 - \tr
A^2 =0$.  This excludes in particular
\be
\sigma \equiv \sigma_3 = \pmatrix{\one & 0\cr 0 & -\one},
\ee
from the Lie algebra.
The supermatrix is in addition also traceless,
the trace having been parametrised by $\A^0$.
Equivalently, we can introduce a complete set of traceless and supertraceless
generators $T_A$ (normalised as in ref. \cite{sunn})
and thus expand $\A$ as
\be
\label{expandA}
\A_\mu = {\A}^{0}_{\mu} \one + \A^A_\mu T_A.
\ee
The $B$ fields are wrong statistics gauge fields. They will be given a 
mass of order the cutoff $\Lambda$. The supergroup $SU(N|N)$ has 
$SU(N)\times SU(N)\times U(1)$ as its bosonic subgroup.
$A^2_\mu(x)\equiv A^2_{a\mu}\tau^a_2$ is the gauge field for the second 
$SU(N)$, and $\A^0$ is the $U(1)$ connection.
Interactions are built via commutators, using the covariant derivative:
\be
\label{defnabla}
\nabla_\mu = \partial_\mu -i\A_\mu,
\ee
thus the superfield strength is given by 
$\F_{\mu\nu}=i[\nabla_\mu,\nabla_\nu]$. The kinetic term will be 
regularised by higher derivatives which thus take the form:
\be
\label{keA}
\str\ \F_{\mu\nu} \left(\nabla\over\Lambda\right)^n\!\!\!\cdot\F_{\mu\nu},
\ee
(where the dot means $\nabla$ acts by commutation.
In practice we will add the higher derivatives as a power series with
coefficients determined by a cutoff function $c$.) The supertrace, which is 
necessary to ensure $SU(N|N)$ invariance, forces the kinetic term for $A^2$ 
to have wrong sign action, leading to negative norms in its Fock space 
\cite{sunn}.  

As can be seen from \eq{expandA}, $\A^0$ does not appear in the kinetic 
term. Providing the interactions can be written as $\str(\A \,\times$
commutators), $\A^0$
will not appear anywhere in the action. More generally we will need to
impose its non-appearance as a constraint, since otherwise $\A^0$ has
interactions but no kinetic term and thus acts as a 
Lagrange multiplier. This would result in a non-linear constraint on the 
theory, which does not look promising for its use as a regularisation method
for the original $SU(N)$ Yang-Mills. 

On the other hand, if the constraint is satisfied, $\A^0$ is then
protected from appearing by a local ``no-$\A^0$'' shift symmetry:
$\delta\A^0_\mu(x)=\lambda_\mu(x)$, which implies in particular that
$\A^0$ has no degrees of freedom.
Together with supergauge invariance the theory is then invariant under
\be
\label{Agauged}
\delta\A_\mu = \nabla_\mu\cdot\Omega +\lambda_\mu \one
\ee
(where the supermatrix $\Omega(x)$ is in the $SU(N|N)$ Lie algebra).
The effect of the no-$\A^0$ symmetry is to dynamically define the gauge
group as the quotient $SU'(N|N)$ $=$ $SU(N|N)/U(1)$, 
in which Lie group elements are identified modulo addition of an 
arbitrary multiple of $\one$. 

An alternative and equivalent formulation \cite{sunn} is to pick coset 
representatives, which can for example be taken to be traceless, so
that $\A^0$ is set to zero, and thus discarded. (This is the strategy
used in ref. \cite{Berkovits} to define a $SU'(N|N)$ sigma model.
Incidentally this paper contains arguments for finiteness of these
models which are similar to those given by us for $SU(N|N)$ gauge
theory \cite{sunn}.\footnote{We thank Hugh Osborn for drawing our
attention to this paper}) In this reduced representation, eq. \eq{Agauged} 
is replaced by Bars' bracket \cite{Bars}:
\be
\label{gaugeBars}
\delta\A_\mu = [\nabla_\mu,\Omega]^* \equiv [\nabla_\mu,\Omega]
-{\one\over2N}\tr[\nabla_\mu,\Omega].
\ee
The *bracket replaces the commutator as a representation of the Lie
product so in particular $\F_{\mu\nu}=i[\nabla_\mu,\nabla_\nu]^*$
\cite{sunn}. 

The lowest dimension interaction that violates no-$\A^0$ symmetry
contains four superfield strengths, for example:
\be
\label{counterX}
\str\ \left(\F_{\mu\nu}\right)^2 \!\left(\F_{\lambda\sigma}\right)^2.
\ee
Such terms are not invariant under the `Bars*' \eq{gaugeBars}, either.
Since \eq{counterX} is already irrelevant, no-$\A^0$ symmetry is 
automatic for the conventional supergauge invariant bare action
of ref. \cite{sunn}. Here there is no such bare action, and
interactions are generated by a largely unspecified exact RG, so we
need to impose no-$\A^0$ as an extra constraint.

We introduce a superscalar field
\be
\label{defC}
\C = \left( \begin{array}{cc} C^1 & D \\
			      \bar{D} & C^2       
	       \end{array}
       \right)
\ee
in the fundamental $\otimes$ its complex conjugate representation,
equivalently as a matrix in the defining representation of $U(N|N)$
\cite{sunn}. Under supergauge transformations
\be
\label{Cgauged}
\delta\C = -i\,[\C,\Omega].
\ee
In the Bars* representation we do not replace this 
by a *bracket, since commutators are necessary for powers of $\C$
(appearing in its potential) to transform covariantly \cite{sunn}. 
However, as in ref. \cite{sunn}, 
since working with the full cosets seems more elegant, we will employ
\eq{Agauged} and the full representation in this paper.

We will arrange for $\C$ to develop a vacuum expectation value along the
$\sigma$ direction, so that classically $<\C>\ =
\Lambda\sigma$.\footnote{Later however 
we will use an unconventional normalisation for $\C$.}
This spontaneously breaks the $SU(N|N)$ gauge invariance down to its 
$SU(N)\times SU(N)\times U(1)$ bosonic subgroup and provides the fermionic 
fields $B$ and $D$ with masses of order $\Lambda$. In unitary gauge, the 
Goldstone mode $D$ is eaten by $B$. Since we will not fix the $SU(N|N)$
invariance, they instead gauge transform into each other and propagate
as a composite unit (see \sec{Five}).  We arrange for the remaining `Higgs'
fields $C^i$ also to have masses of order $\Lambda$.
 
In ref.  \cite{sunn}, we proved by conventional methods that if the kinetic
term of $\A$ is supplied with covariant higher derivatives 
(parametrised by the cutoff function $c$) enhancing
its high momentum behaviour by a factor $c^{-1}\sim p^{2r}/\Lambda^{2r}$, 
and the
kinetic term of $\C$ has its high momentum behaviour similarly enhanced
by $\ct^{-1}\sim p^{2\rt}/\Lambda^{2\rt}$, then providing
\be
\label{inequalities}
r-\rt > 1\quad{\rm and}\quad \rt>1,
\ee
all amplitudes are ultraviolet finite to all orders of perturbation theory.
Since the underlying theory is renormalizable, the 
Appelquist-Carazzone theorem implies that at energies much
lower than the cutoff $\Lambda$, the remaining massless fields $A^1$
and $A^2$ decouple. In this way, we can use this framework as a regularisation
of the original $SU(N)$ Yang-Mills theory carried by $A^1$.

In brief, the reasons for the above facts are as follows. Providing
eqs.  \eq{inequalities} hold, all divergences are superficially
regularised by the covariant higher derivatives, except for some
`remainders' of one-loop graphs with only $\A$ fields as external legs
and only four or less of these legs. These remainders form a symmetric
phase contribution, in the sense that the superficially divergent
interactions between $\C$ and $\A$ are just those that come from $\C$'s
covariant higher derivative kinetic term, whilst all terms containing a
$\sigma$ from the breaking are already ultraviolet finite by power
counting.  For three or less external $\A$ legs the remainders vanish
by the supertrace mechanism:  the fact that in the unbroken theory, the
resultant terms contain $\str\A=0$ or $\str\one=0$.  By manifest gauge
invariance, the four point $\A$ remainder is then actually totally
transverse, which implies that it is already finite by power counting.
 
The decoupling of $A^1$ and $A^2$ follows from the unbroken local
$SU(N)\times SU(N)$ invariance since the lowest dimension effective
interaction
\be
\label{12ints}
{1\over \Lambda^4}\,\tr\left(F^1_{\mu\nu}\right)^2 
\tr\left(F^2_{\mu\nu}\right)^2
\ee
is already irrelevant \cite{sunn,apple}.

Actually, there are a number of differences between the treatment we
give here and that of ref. \cite{sunn}.  Since ref. \cite{sunn}
followed a conventional treatment, gauge fixing and ghosts were
introduced, with a corresponding higher derivative regularisation for
them; longitudinal parts of the four point $\A$ vertex were then
related to ghost vertices using the Lee Zinn-Justin identities, which
were separately proved to be finite. Also, a specific form of bare
action and covariantisation was chosen.

Here we do not fix the gauge and the regularisation scheme is much more
general.  As well as not specifying the covariantisation or the bare
action (see below) there is anyway much more freedom in introducing
interactions via the flow equation. We shall not here supply a rigorous
proof that up to appropriate restrictions, the flow equation is finite.
Since we never have to specify the details, we only need to {\sl
assume} that this is true for at least one choice. However, we take
care that the scheme as described above is qualitatively correctly
implemented. Where we do have to explicitly compare terms we can use
\eq{inequalities} as a guide, thus for example we do find that our final
expression for $\beta_1$ is properly regulated.
(However, it should be borne in mind that at intermediate stages
our vertices have much more freedom in their momentum dependence than that
implied by the bare action in ref. \cite{sunn}. Additionally,
cutoff functions with non-power law asymptotics, for example
exponential, could also be used.\footnote{The proof given in ref.
\cite{sunn} could also be easily extended to these cases.})  In
practice, it is easy to see at one loop that the high energy
cancellations are occurring as expected.

\subsection{Necessary properties of the exact RG and their interpretation}
\label{Necessary}

The extra fields we have added form a necessary part of the
regularisation structure. We gain an interpretation of these fields at
the effective level by imagining integrating out the heavy fields $B$,
$C$ and $D$ at some scale $\Lambda$. The result would be an effective action
containing only the unbroken gauge fields $A^i$, but it is not
well defined because it is not finite.
In particular, the one-loop determinant formed from integrating out the
heavy fields is necessarily divergent: the divergences are there to
cancel those left by the one-loop hole in the remaining covariant higher 
derivative regularisation \cite{oneloophole} of the $SU(N)\times SU(N)$ 
Yang Mills theory, in a similar way to that done
explicitly in gauge invariant Pauli-Villars regularisation
\cite{Slavnov}.
  
A gauge invariant exact RG description of gauge theory thus requires
not only a well defined finite effective action but a separate measure
term.  The measure
term is not itself finite, but can be represented by a well-defined finite
addition to the effective action, if we accept the introduction of
these auxiliary heavy fields.

Whilst this interpretation is reasonable, we nevertheless need to be
sure that we are still only representing the original quantum field
theory (here $SU(N)$ Yang-Mills). This demand is especially pertinent
in (but not restricted to) the case where there are extra regulator
fields, particularly here $A^2$ which remains massless and in this
effective description only decouples at momenta much less than
$\Lambda$. More generally, even if there are only physical fields in
the effective action, we need to be sure that locality, an important
property of quantum field theory \cite{egZinn,IZ}, is properly
incorporated.\footnote{otherwise non-physical effects or the effects of
other propagating fields, could be hidden in the vertices.} Note that
$\Lambda$ is intended to be set at the energy scales of interest, which
is why it makes sense to use the exact RG and solve for the effective
action directly in renormalized terms (see \eg ref. \cite{rev}).  Indeed, to
extract the physics (\eg correlation functions \etc) we will even want to
take $\Lambda\to0$ eventually \cite{morig,rev}.

These demands are fulfilled implicitly through the $\Lambda\to\infty$
limit, providing some very general requirements on the exact RG are
implemented, as we now explain.

Firstly, we require that all parts of the flow equation
can be expanded in external momenta to any order, so that the solutions
$S$ can also be required to have an all orders derivative expansion
\cite{ymi,ymii,rev}.\footnote{Sharp cutoff 
realisations \cite{weg} are more subtle \cite{morig,trunc} and will
not be discussed here.} This `quasilocality' requirement \cite{ymi}
is equivalent to the fundamental requirement of the Wilsonian RG
that Kadanoff blocking take place only over a localised patch \cite{Wil},
\ie here that each RG step, $\Lambda\mapsto \Lambda-\delta\Lambda$, be 
free from infrared singularities.

The flow equation is written only in terms of renormalized quantities
at scale $\Lambda$. In fact, we require that the only explicit scale
parameter that appears in the equations is the effective cutoff
$\Lambda$. Again this is so that the same can be required of $S$ where
it implements the concept of self-similar flow \cite{Shirkov}. Here
this amounts to a non-perturbative statement of renormalizability, \ie
existence of a continuum limit, equivalent to the requirement that $S$
lie on a renormalized trajectory \cite{rev}.  This is clearer if we
first scale to dimensionless quantities using the appropriate powers of
$\Lambda$. Then, $S$ is required to have no dependence on $\Lambda$ at
all except through its dependence on the running coupling(s) $g(\Lambda)$
\cite{rev}. 

Note that the $\Lambda\to\infty$ end of the renormalized trajectory,
\ie the perfect action \cite{Hasenfratz} in the neighbourhood of the
ultraviolet fixed point at $\Lambda=\infty$, amounts to our choice of
bare action. Its precise form is not determined beforehand but as a
result of solution of the exact RG, but it is constrained by choices in
the flow equation. Since these choices are however  here to a large
extent unmade, we deal implicitly with an infinite class of perfect bare 
actions.

We require that the flow of the Boltzmann measure $\exp(-S)$ is a 
total functional derivative, \ie for some generic fields $\phi$:
\be
\label{reparam}
\Lambda\partial_\Lambda \,{\rm e}^{-S} = {\delta\over\delta\phi}\left(\Psi
\,{\rm e}^{-S}\right)
\ee
(corresponding to the statement that each RG step is equivalent to an
infinitesimal field redefinition $\phi\mapsto \phi+\Psi\,
\delta\Lambda$) \cite{alg,jose}. Importantly, this ensures that the
partition function ${\cal Z}=\int\!\!{\cal D}\phi\, \exp(-S)$, and
hence the physics derived from it, is invariant under the RG flow.
Since we will solve the exact RG approximately, but by controlled
expansion in a small quantity, this property is left undisturbed.
Therefore we may use different scales $\Lambda$ at our convenience to
interpret the computation.

For example, although locality is obscured in the Wilsonian effective
action at any finite $\Lambda$, it is important to recognise that
invariance of ${\cal Z}$ together with the existence of a derivative
expansion and self-similar flow (\viz that the only explicit scale be
$\Lambda$), ensure that locality is implemented, since it is then an
automatic property of the effective action as $\Lambda\to\infty$.

Similarly, it is as $\Lambda\to\infty$ that we confirm from the
Wilsonian effective action that we are
describing $SU(N)$ Yang-Mills theory: $B$, $C$ and $D$ really are
infinitely massive, and in spacetime dimension four or less, $A^2$ is
guaranteed decoupled by the Appelquist-Carazzone theorem and
\eq{12ints}. In general strong quantum corrections might alter either
of these properties. Thus in general we would need to add appropriate
sources to the $\Lambda\to\infty$ action; compute the partition
function by computing the  $\Lambda\to0$ limit of $\exp(-S)$; and
finally explicitly test these properties by computing appropriate
correlators (formed from differentiating with respect to the sources.
This is the most general way to extract the results for
physical quantities from $S$.) However since $g$ is perturbative at
high energies (indeed $g\to0$ as $\Lambda\to\infty$), we can be sure
that the above deductions about the regulator fields, drawn at the
perturbative level, are not destroyed by quantum corrections.

As already mentioned, we require that an ultraviolet regularisation at
$\Lambda$, is implemented so that the right hand
side of the flow equation makes sense. Note that this ensures that all
further quantum corrections to $S$ (computed by solving for the flow at
scales less than $\Lambda$) are cut off (smoothly) at $\Lambda$.  Since
momentum modes $p > \Lambda$ were fully contributing to the initial
$\Lambda\to\infty$ partition function, and since ${\cal Z}$ is
invariant under the flow, we can be sure that their effect has been
incorporated $S$. In other words we can be sure that our final requirement
on the flow, namely that it corresponds to integrating out momentum
modes, has been incorporated.

(In refs. \cite{alg,ymi}, a possible further requirement on the flow
equation, called ``ultralocality'' was discussed, replacing the usual
notion of locality, although it was not clear that it was necessary
however. We have seen here that the usual concept of locality is
recovered providing the existence of a derivative expansion, invariance
of ${\cal Z}$, and self-similar flow, are implemented.  Furthermore the
successful calculations of ref. \cite{sca} and here, confirm that the 
restriction of `ultralocality' is unnecessary since they do not assume it.)

\subsection{Supergauge invariance and functional derivatives}
\label{Supergauge}

The peculiarities of $SU(N|N)$ affect functional derivatives with
respect to $\A$ and lead to some constraints on the form of the exact
RG if the flow equation is to be invariant under supergauge
transformations.  

As before \cite{ymii,sunn}, it is convenient to define the functional
derivatives of $\C$ and $\A$ so as to extract the dual from under the
supertrace. For an unconstrained field such as $\C$ we simply have
\cite{ymii,sunn}:
\be
\label{dCdef}
{\delta \over {\delta\C}} := {
\left(\!{\begin{array}{cc} {\delta / {\delta C^1}} & - {\delta /
{\delta \bar{D}}} \\ {\delta / {\delta D}} & - {\delta
/ {\delta C^2}} \end{array}} \!\!\right)},
\ee
or in components 
\be
\label{Cdumbdef}
{\delta \over {\delta\C}}^i_{\gap j} :=
{\delta \over {\delta\C}^k_{\gap i}}\sigma^k_{\gap j}.
\ee
Under supergauge transformations 
\eq{Cgauged}, the functional derivative transforms as one would hope:
\be
\label{dCgauged}
\delta \left({\delta\over\delta\C}\right) =
-i\left[{\delta\over\delta\C},\Omega\right].
\ee
Such a derivative\footnote{for simplicity, written with partial
derivatives, to neglect the irrelevant spatial dependence}
has the properties of `supersowing' \cite{ymii}:
\be
\label{sow}
{\partial\over\partial\C}\ \str \,\C Y = Y \quad\Longrightarrow\quad
\str X{\partial\over\partial\C}\ \str \,\C Y = \str XY,
\ee
and `supersplitting' \cite{ymii}:
\be
\label{split}
\str {\partial\over\partial\C}X\C Y = \str X \str Y,
\ee
\ie of sowing two supertraces together, and splitting one supertrace into
two, where $X$ and $Y$ are arbitrary constant supermatrices.  

({\it N.B.} as we will see later, it is a helpful trick to contract in 
arbitrary supermatrices at intermediate stages of the calculation: it allows 
manifestly $SU(N|N)$ invariant index-free calculations in the $SU(N|N)$ 
algebra, by permuting overall bosonic structures past each other. It also 
leads as we will show, to efficient diagrammatic techniques.  The arbitrary 
supermatrices can always be stripped off at the end, if necessary.
If we did not use this trick, we would lose manifest $SU(N|N)$ invariance at 
intermediate stages, by having to carry intermediate minus signs from 
fermionic parts of supermatrices anticommuted through each other.)

Since $\A$ is constrained to be supertraceless, its dual under
the supertrace $\str\J_\mu\A_\mu$ has without loss of generality no
$\one$ component: only 
\be
\label{realJmu}
\J_\mu -{\one\over2N}\tr \J_\mu 
\ee
really couples. The natural construction for the $\A$ functional derivative 
from \eq{expandA} \cite{sunn}:
\be
\label{dumbdef}
{\delta\over\delta\A_\mu}:=
2T_A{\delta\over\delta\A_{A\,\mu}}+{\sig\over2N}
{\delta\over\delta\A^0_\mu}
\ee
pulls out precisely this combination. However from \eq{Agauged} and
the completeness relations for the $T_A$ \cite{sunn}, under supergauge
transformations
\bea
\label{dAgauged}
\delta \left({\delta\over\delta\A_\mu}\right) &=&
-i\left[{\delta\over\delta\A_\mu},\Omega\right] 
+{i\one\over2N}\tr\left[{\delta\over\delta\A_\mu},\Omega\right]\\
&=& -i\left[{\delta\over\delta\A_\mu},\Omega\right]^*.\nonumber
\eea
The correction is to be expected since it ensures that
$\delta/\delta\A$ remains traceless, but the fact that
$\delta/\delta\A$ does not transform homogeneously means that
supergauge invariance is destroyed unless $\delta/\delta\A$
is contracted under the supertrace into something that is supertraceless
(causing the correction term to vanish). This is an extra constraint
on the form of the flow equation.

[As an alternative one might try defining $\delta/\delta\A$ as only
the $2T_A\delta/\delta\A_A$ term in \eq{dumbdef}, however one can
show from \eq{Agauged} that this does not transform into itself but into 
the full functional derivative given in \eq{dumbdef}. It works however
in the Bars* representation, where the transformation again takes the form
\eq{dAgauged}.]

Similarly there are corrections to \eq{sow} and \eq{split} that arise because 
the derivative is constrained:\footnote{ignoring the spacetime index and 
spatial dependence}
\be
\label{sowA}
\str X{\partial\over\partial\A}\ \str \,\A Y = \str XY -{1\over2N}\, \str X
\tr Y
\ee
as expected from \eq{realJmu}, and 
\be
\label{splitA}
\str {\partial\over\partial\A}X\A Y = \str X \str Y -{1\over2N}\, \tr\, Y\!X.
\ee
Since these corrections contain $\tr \cdots \equiv \str\,\sig \cdots$, 
they similarly violate $SU(N|N)$ invariance.  As
we discuss in \sec{Supersowing}, they also effectively disappear with the above
constraint that $\delta/\delta\A$ is contracted into something
supertraceless. (This is obvious in \eq{sowA} where thus $\str X=0$.)

In this way the supersplitting and supersowing rules actually become
exact for both fields, even at finite $N$ (compare \cite{ymi,ymii}). As
we will see, this leads to a very efficient diagrammatic
technique incorporated into the Feynman diagrams,
for evaluating the gauge algebra, analogous to the 't Hooft
double line notation \cite{tHooftdouble} and utilised earlier
\cite{alg,ymi,ymii}, but here applying even at finite $N$.

\subsection{Covariantisation}
\label{Covariantisation}
Given some momentum space kernel $W_p\equiv W(p,\Lambda)$, we 
write in position space
\be
\label{kdef}
W_{xy} \equiv\int\!\!{d^4\!p\over(2\pi)^4}
\,W(p,\Lambda)\,{\rm e}^{ip.(x-y)},
\ee
and introduce the shorthand:
\be
f\ker{W}g :=
\int\!\!\!\!\int\!\!d^4\!x\,d^4\!y\
f(x)\, W_{xy}\,g(y),
\ee
where $f$ and $g$ are any two functions. We define a general covariantisation
of any such kernel (the `wine' \cite{ymi,ymii}) via the supergauge invariant:
\bea
\label{wv}
&&u\,\{W\}_{\!\!{}_\A} v =\\
&&\sum_{m,n=0}^\infty\int\!\!d^4\!x\,d^4\!y\,
d^4\!x_1\cdots d^4\!x_n\,d^4\!y_1\cdots d^4\!y_m\,
W_{\mu_1\cdots\mu_n,\nu_1\cdots\nu_m}
(x_1,\cdots,x_n;y_1,\cdots,y_m;x,y) \nonumber\\
&&\phantom{\sum_{m,n=0}^\infty
    \int\!\!d^4\!x\,d^4\!y\,d^4\!x_1\cdots d^4\!x_n\,}
\str\left[\, u(x)\, \A_{\mu_1}(x_1)\cdots \A_{\mu_n}(x_n)\, 
v(y)\,\A_{\nu_1}(y_1)\cdots \A_{\nu_m}(y_m)\,\right],\nonumber
\eea
where $u$ and $v$ are any two supermatrix representations, and
where without loss of generality we may insist that 
$\{W\}_{\!\!{}_\A}$ satisfies
$u\,\{W\}_{\!\!{}_\A} v\equiv v\,\{W\}_{\!\!{}_\A} u$. We write the $m=0$
vertices (where there is no second product of gauge fields), more
compactly as
 \be
\label{compac}
W_{\mu_1\cdots\mu_n}(x_1,\cdots,x_n;x,y)
\equiv W_{\mu_1\cdots\mu_n,}(x_1,\cdots,x_n;;x,y),
\ee
whilst the $m=n=0$  term is
just the original kernel \eq{kdef}, \ie 
\be
\label{mno}
W_,(;;x,y)\equiv W_{xy}.
\ee

We leave the covariantisation general, up to
certain restrictions. One of these is already encoded into eq. \eq{wv},
namely that there is just a single supertrace in \eq{wv},
involving just two ordered products of supergauge fields.
Another is that we require that the covariantisation satisfy coincident
line identities \cite{ymi} which in particular imply that if 
$v(y)=\one g(y)$ for all $y$, \ie is in the scalar representation of the
gauge group, then the covariantisation collapses to
\be
\label{Acoline}
u\,\{W\}_{\!\!{}_\A} v = (\str\, u)\ker{W}g.
\ee
As shown in ref. \cite{ymii} (\cf sec. 5.2 of that paper), the coincident
line identities are equivalent to the requirement that the gauge fields
in \eq{wv} all act by commutation. This requirement is necessary to ensure
no-$\A^0$ remains valid and to ensure that $\delta/\delta\A$ is indeed
contracted into something supertraceless. It is this that we need
rather than the identities themselves, which are used just once,
to collect terms in the calculation.

Although we will not use it explicitly, let us remark that
these constraints are solved by the following general covariantisation
\cite{ymi,ymii}:
\be
\label{wine}
u\,\{W\}_{\!\!{}_\A} v=
\int\!\!\!\!\int\!\!d^4\!x\,d^4\!y\int\!\!\D_{\!\!{}_W}\ell_{xy}\
\str \,u(x)\,\Phi[\ell_{xy}]\,v(y)\,\Phi^{-1}[\ell_{xy}],
\ee
where 
\be
\label{line}
\Phi[\ell_{xy}]=P\exp-i\int_{\ell_{xy}}\!\!\!\!\!\!dz^\mu\A_\mu(z)
\ee
is a path ordered exponential integral, \ie a Wilson line, and the appearance
of $\Phi^{-1}[\ell_{xy}]$ means that we traverse backwards along another
coincident Wilson line. The covariantisation is determined by the
measure $\D_{\!\!{}_W}$ over configurations of the curves $\ell_{xy}$ and is
so far left unspecified except for its normalisation:
\be
\int\!\!\D_{\!\!{}_W}\ell_{xy}\ 1 = W_{xy},
\ee
as follows from \eq{wv} and \eq{mno}. It is easy to see that \eq{wine}
indeed does satisfy \eq{Acoline}.

\begin{figure}[h]
\psfrag{mu1}{$\mu_1$}
\psfrag{mu2}{$\mu_2$}
\psfrag{nu1}{$\nu_1$}
\psfrag{nu2}{$\nu_2$}
\psfrag{dots}{$\cdots$}
\psfrag{+}{$+$}
\psfrag{=}{$=$}
\begin{center}
\includegraphics[scale=.5]{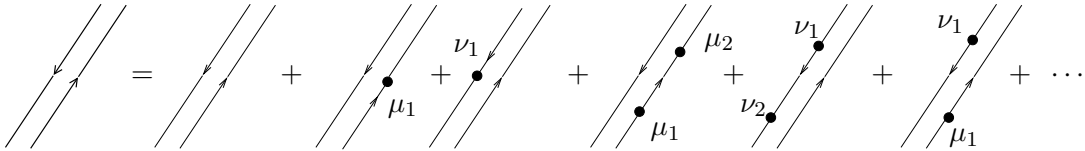}
\end{center}
\caption{Wine expansion. The thick lines are expanded into thin lines,
with the blobs representing ${\cal A}$ fields}\label{winexp}
\end{figure}

The expansion \eq{wv} can be represented as in \fig{winexp}.
As we explain later, these will act as Feynman rules, although they can
also be viewed directly as expansions of the Wilson lines \eq{line} in
the covariantised kernel \eq{wine}.

Finally, we will require that the covariantisation satisfies
\be
\label{noATailBiting}
{\delta\over\delta\A_\mu}\,\{W\}_{\!\!{}_\A} =0,
\ee
(where the functional derivative acts on all terms inside $\{W\}_{\!\!{}_\A}$
but not on the unspecified right hand attachment)
\ie that there are no diagrams  in which the wine bites its own tail 
\cite{alg,ymi,ymii}. This leads to identities for the $W$ vertices which again
we do not need in practice: as we will confirm, such terms do not in any
case contribute to the one-loop $\beta$ function. However 
such diagrams do appear in general to lead to some 
improperly regularised terms and so some restriction
is needed for consistency. We can use the representation \eq{wine} to see that
sensible solutions to \eq{noATailBiting} do exist. For example we can simply
insist that $\ell_{xy}$ is a straight Wilson line, and more generally that
the measure  $\D_{\!\!{}_W}$ has no support on curves $\ell_{xy}$ that cross
the points $x$ or $y$. The end points need defining carefully 
so that they only touch $x$ and $y$ after a limit has been taken \cite{alg}. 
However since we never specify the covariantisation, we only need to assume 
that solutions to \eq{noATailBiting} exist. In the calculation we just use 
\eq{noATailBiting} and thus just forbid all wine-biting-their-tail diagrams. 

\subsection{Decoration with $\C$}
\label{Decoration}

It will prove convenient to allow occurrences of $\C$ also on the  Wilson
lines (with the obvious corresponding extension of \fig{winexp})
although we can limit their appearance to attachments at either
end of $\ell_{xy}$. In this paper they will furthermore act only via
commutation at both ends. Precisely, we extend the definition \eq{wv}
so that
\be
\label{wev}
u\{W\}v =  u\,\{W\}_{\!\!{}_\A}v -{1\over4}
\C\!\cdot\!u\,\{W_{m}\}_{\!\!{}_\A}\C\!\cdot\!v,
\ee
where $W_m(p,\Lambda)$ is some new kernel.
In the expansion we now have vertices that come from both $\A$ and $\C$.
Typically in this case $u$ and $v$ will actually correspond to functional
differentials, with respect to, say, $Z_1$ and $Z_2$, and it will also be 
helpful to keep track of these flavours by including them as labels in the
naming convention for the kernel, {\it viz.} as $W^{Z_1Z_2}_{(m)}$.
The notation we will thus use in general is
\bea
\label{wcv}
&&{\delta\over\delta Z_1^{c}}\{W^{Z_1Z_2}\} {\delta \over\delta Z_2^{c}} =\\
&&\sum_{m,n=0}^\infty\int\!\!d^4\!x\,d^4\!y\,
d^4\!x_1\cdots d^4\!x_n\,d^4\!y_1\cdots d^4\!y_m\,
W_{\ a_1\cdots\, a_n,\ b_1\cdots b_m}^{X_1\cdots X_n, Y_1\cdots Y_m, Z_1 Z_2}
(x_1,\cdots,x_n;y_1,\cdots,y_m;x,y) \nonumber\\
&&\phantom{\sum_{m,n=0}^\infty
    \int\!\!d^4\!x\,d^4\!y\,d^4\!x_1\cdots }
\str\left[\, {\delta\over\delta Z_1^{c}(x)}\, X_1^{a_1}(x_1)\cdots 
X_n^{a_n}(x_n)\, 
{\delta\over\delta Z_2^{c}(y)}\,Y_1^{b_1}(y_1)\cdots 
Y_m^{b_m}(y_m)\,\right],\nonumber
\eea
where the superfields $X_i$, $Y_i$ and $Z_i$, are $\A$ or $\C$,
and the indices $a_i=\mu_i$, $b_i=\nu_i$ and $c=\gamma$ in the case
that the corresponding field is $\A$ and null if the field is $\C$.
In fact, as a consequence of the restricted structure \eq{wev}, the
$X_2,\cdots,X_{n-1}$ and $Y_2,\cdots,Y_{m-1}$ must be $\A$s
if they appear at all. 

We can still insist without loss of generality
that $u\{W\}v \equiv v\{W\}u$, and use the shorthand \eq{compac},
where now we keep track of flavour labels as in \eq{wcv} however.
It is still the case that with no fields on the `wine', the original
$W$ kernel is recovered as in \eq{mno}. The commutator structure in \eq{wev}
ensures that \eq{Acoline} holds for the full `wine' also:
\be
\label{coline}
v(y) = \one\, g(y)\ \ \forall\, y\quad \Rightarrow\quad 
u\,\{W\} v = (\str\, u)\ker{W}g.
\ee
Finally,  the $\C$s as further `decorations'
of the covariantised kernels are required to
partake in the restriction described below
\eq{noATailBiting}, so this equation extends to
\be
\label{noTailBiting}
{\delta\over\delta\A_\mu}\,\{W\} = {\delta\over\delta\C}\,\{W\} =0.
\ee
(In fact by $X=\one$ in \eq{split}, the
contribution from differentiating the leftmost $\C$ vanishes in any
case.)

\subsection{Superfield expansion}
\label{Superfield}

\begin{figure}[h]
\begin{center}
\psfrag{=}{$=$}
\psfrag{dot}{$\cdots$}
\psfrag{+}{$+$}
\includegraphics[scale=.5]{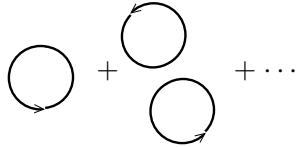}
\end{center}
\caption{Expansion of the action in products of supertraces}\label{fig:action}
\end{figure}

\begin{figure}[h]
\begin{center}
\psfrag{=}{$=$}
\psfrag{dot}{$\cdots$}
\psfrag{+}{$+$}
\includegraphics[scale=.5]{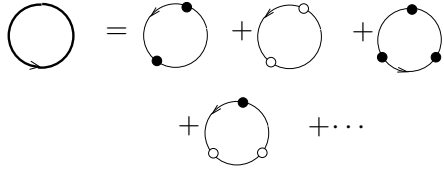}
\end{center}
\caption{Expansion of a supertrace in powers of the fields
$\A$ and $\C$.}\label{fig:fieldex}
\end{figure}

The Wilsonian effective action $S$ (and the seed action $\hS$
that we will also introduce), being supergauge invariant, has an expansion
in supertraces and products of supertraces:
\bea
\label{Sex}
S &=&\sum_{n=1}^\infty{1\over s_n}\int\!\!d^4\!x_1\cdots d^4\!x_n\,
S^{X_1\cdots X_n}_{\, a_1\cdots\,a_n}(x_1,\cdots,x_n)\ 
\str\, X_1^{a_1}(x_1)\cdots X_n^{a_n}(x_n)\nonumber\\
&+&{1\over2!}\sum_{m,n=1}^\infty{1\over s_ns_m}\int\!\!
d^4\!x_1\cdots d^4\!x_n\,d^4\!y_1\cdots d^4\!y_m\,
S_{\, a_1\cdots\, a_n,\ b_1\cdots b_m}^{X_1\cdots X_n, Y_1\cdots Y_m}
(x_1,\cdots,x_n;y_1,\cdots,y_m)\nonumber\\
&&\phantom{{1\over2!}\sum_{m,n=1}^\infty{1\over nm}\int\!\!
d^4\!x_1\cdots d^4\!x_n\,}
\str\, X_1^{a_1}(x_1)\cdots X_n^{a_n}(x_n)\
\str\, Y_1^{b_1}(y_1)\cdots Y_m^{b_m}(y_m)\nonumber\\
&+& \cdots,
\eea
where again the $X_i^{a_i}$ are $\A_{\mu_i}$ or $\C$, and $Y_j^{b_j}$
are $\A_{\nu_j}$ or $\C$. (Note that throughout this paper we discard
the vacuum energy.) Only one cyclic ordering of each list 
$X_1\cdots X_n$, $Y_1\cdots Y_m$ appears in the sum.
Furthermore, if either list is invariant under some
nontrivial cyclic permutations, then $s_n$ ($s_m$)
is the order of the cyclic subgroup, otherwise $s_n=1$ ($s_m=1$).
(For example, in the terms where every $X_i^{a_i}$ is a $\C$, $s_n=n$.)
The expansion can be represented diagrammatically, where a thick closed line 
stands for a single supertrace of any number of fields, as in 
\fig{fig:action},
and each blob represents a field in the supertrace, as in \fig{fig:fieldex}.
In a somewhat similar way to \eq{wine} and \eq{wev}, these closed lines can
be interpreted as decorated Wilson loops \cite{ymi,ymii}.

When we spontaneously break the fermionic invariance by shifting $\C$
in the $\sigma$ direction, it will prove to be better to work separately
with the bosonic and fermionic parts of the superfields. Thus we write
in the broken phase 
\be
\label{ABCD}
\A_\mu = A_\mu + B_\mu,\ins11{and} \C\mapsto \C+\sigma= C + D + \sigma.
\ee
where $A$ and $C$ are the block diagonals, 
and $B$ and $D$ are the block off-diagonals in eqs. \eq{defA} and \eq{defC}
respectively \cite{ymii}.
(We will see in the \sec{manifestly} that $\C$'s 
effective vacuum expectation value is just $\sigma$.)

Thus in the broken phase we will expand as in \eq{Sex}, but the
flavours $X$ and $Y$ are set to $A$, $B$, $C$ or $D$. There will also
be occurrences of $\sigma$. However since $\sigma$ commutes with $A$ and
$C$, and anticommutes with $B$ and $D$, to define the expansion we can
take the convention that we (anti)commute all such occurrences to the
far right in the supertrace. Upon using $\sigma^2=\one$, we are then
left with terms with either one $\sigma$ at the end of a supertrace
or none at all in that supertrace.
Since $\sigma$ has no position dependence, we put the flavour label in
the superscript, but we omit the corresponding position label (and the
term $n$-point vertex counts $n$ fields excluding $\sigma$ appearances).
Clearly, since the broken fields can still be cyclically permuted by
(anti)commutation through $\sigma$, we also omit it from the
determination of the symmetry factor, \ie $s_n$ is equal to the order
of the cyclic permutation subgroup of the fields $X_i$, ignoring the
$\sigma$ (if present). Finally note that each
supertrace term  must separately hold only totally bosonic combinations
since if $X_1\cdots X_n$ (or $X_1\cdots X_n\sigma$) is fermionic, it is block 
off-diagonal and has vanishing supertrace. 

Similarly, in \eq{wcv}, in the broken phase, $X$, $Y$ and $Z$ will be
$A$, $B$, $C$ or $D$. Note that $Z_1$ can be the opposite statistic
partner from $Z_2$. Since it is a single supertrace, again each
contribution in \eq{wcv} is overall bosonic however.  Single occurrences
of $\sigma$ can also appear at the ends of the Wilson lines, after
taking into account that these can also (anti)commute through the $Z$
functional derivatives.

Finally, the momentum space vertices are written as
\be
S^{X_1\cdots X_n}_{\, a_1\cdots\,a_n}(p_1,\cdots,p_n)\
(2\pi)^4\delta( 
\sum_{i=1}^np_i)  
=\int\!\!d^4\!x_1\cdots d^4\!x_n\,{\rm e}^{-i\sum_ix_i\cdot p_i}
S^{X_1\cdots X_n}_{\, a_1\cdots\,a_n}(x_1,\cdots,x_n),
\ee
where all momenta are taken pointing into the vertex, and similarly for
all the other vertices including  \eq{wcv}. 
We use the short hand $S^{XY}_{ab}(p)\equiv S^{XY}_{ab}(p,-p)$
and $S^{XY\sigma}_{ab}(p)\equiv S^{XY\sigma}_{ab}(p,-p)$
for action two-point vertices.

\section{A manifestly $SU(N|N)$ gauge invariant exact RG}
\label{manifestly}

Our strategy is to write down a manifestly supergauge invariant flow
equation, obeying the rules outlined in \sec{Preliminary}, and then
spontaneously break it. Defining $\Sigma_g=g^2S-2\hS$, we simply set 
\be
\label{sunnfl}
\ldl S  =
- a_0[S,\Sigma_g]+a_1[\Sigma_g],
\ee
where
\be 
\label{a0}
 a_0[S,\Sigma_g] ={1\over2}\,\frac{\delta S}{\delta {\cal
A}_{\mu}}\{\dDelta^{\!\A\A}\}\frac{\delta \Sigma_g}{\delta {\cal
A}_{\mu}}+{1\over2}\,\frac{\delta S}{\delta {\cal C}}\{\dDelta^{\C\C}\}
\frac{\delta \Sigma_g}{\delta {\cal C}}, 
\ee
and
\be
\label{a1}
a_1[\Sigma_g] = {1\over2}\,\frac{\delta }{\delta {\cal
A}_{\mu}}\{\dDelta^{\!\A\A}\}\frac{\delta \Sigma_g}{\delta {\cal
A}_{\mu}} + {1\over2}\,\frac{\delta }{\delta {\cal C}}\{\dDelta^{\C\C}\}
\frac{\delta \Sigma_g}{\delta {\cal C}}.
\ee
In the rest of this section we explain the meaning of the various
components, at the same time developing some of the properties of this 
exact RG. 

The definition of $\Sigma_g$ and the form of the flow equation
\eq{sunnfl} are the same as in refs. \cite{ymi,ymii}. In contrast to
ref. \cite{ymii} however, the exact RG is very simple in conception.
The basic structure is inherited from the Wilson exact RG
\cite{Wil,Pol,alg}: the bilinear functional -$a_0$ generates the
classical corrections, whilst the linear functional $a_1$ generates
quantum corrections. As in refs.  \cite{ymi,ymii}, $a_1$ has exactly
the same structure as $a_0$ except that the leftmost functional
derivatives differentiate everything to their right. Consequently we
have
\be
\ldl \,{\rm e}^{-S} = a_1[\Sigma_g \,{\rm e}^{-S}],
\ee
which shows that condition \eq{reparam} is fulfilled. 

As before, $g(\Lambda)$ is the renormalized coupling of the $SU(N)$
Yang-Mills theory carried by $A^1$. It is defined through the
renormalization condition:
\be
\label{defg}
S[\A=A^1, \C={\bar\C}] ={1\over2g^2}\,{\rm tr}\!\int\!\!d^4\!x\, 
\left(F^1_{\mu\nu}\right)^2+\cdots,
\ee
where the ellipsis stands for higher dimension operators (and the 
ignored vacuum energy), and ${\bar\C}$ is the effective vacuum expectation 
value defined so as to minimise the effective potential $V(\C)$ in $S$:
\be
\label{defCbar}
\left.{\partial V\over\partial\C}\right|_{\C={\bar\C}} =0.
\ee
${\bar\C}$ is spacetime independent and generically contains terms
proportional to $\sigma$ and $\one$ \cite{sunn}. We will see later that for 
the purposes of this paper we can simply set ${\bar\C}=\sigma$.

The similarities mean that the general structure of the perturbative
expansion is the same as in refs. \cite{ymi,ymii}:
we see from \eq{sunnfl}, that $S\sim 1/g^2$ at the classical 
level [consistent with \eq{defg}], and by iteration, using 
\eq{sunnfl}, that $S$ has as expected the weak coupling expansion 
\be
\label{Sloope}
S={1\over g^2} S_0+S_1+g^2 S_2 +\cdots,
\ee
where $S_0$ is the classical effective action, $S_1$ the one-loop
correction, and so on.
Substituting this expansion in \eq{sunnfl}, we see that
the $\beta$ function must also take the standard form
\be
\label{betafn}
\beta:=\Lambda{
\partial g\over\partial\Lambda}=\beta_1g^3+\beta_2g^5+\cdots,
\ee
with coefficients to be determined. From \eq{Sloope} and \eq{betafn}, we 
obtain the loopwise expansion of \eq{sunnfl}:
\bea
\Lambda{\partial\over\partial\Lambda}S_0 &=& -a_0[S_0,S_0-2\hS],
\label{ergcl}\\
\Lambda{\partial\over\partial\Lambda}S_1&=&2\beta_1S_0-2a_0[S_0-\hS,S_1]
+a_1[S_0-2\hS], \label{ergone}
\eea
\etc Actually, we will find it convenient to add some simple quantum
corrections to the supergauge invariant seed action $\hS$, giving it a
$g$ dependence (as we outline below).  We also need to take account of
the flow of $g_2$, the coupling for the second $SU(N)$ carried by
$A_2$.  However, neither of these complications has an effect on the
one-loop $\beta$ function computation, so will be largely ignored here, and
developed fully only when we consider multi-loops \cite{us}.

$\hS$ is used to determine the form of the classical effective kinetic
terms and the kernels $\dDelta(p,\Lambda)$.  It therefore has to
incorporate the covariant higher derivative regularisation and allow
the spontaneous symmetry breaking we require. Unlike previously
\cite{alg,ymi,ymii}, we will see that we otherwise leave it almost
entirely unspecified. The kernels $\dDelta$ are determined 
by the requirement that after spontaneous
symmetry breaking, the two-point vertices of the classical effective
action $S_0$ and $\hS$ can be set equal.  As previously
\cite{alg,ymi,ymii}, this is imposed as a useful technical device,
since it allows classical vertices to be immediately solved in terms of
already known quantities.  It also means that the integral of the
kernels defined via
\be
\label{defprop}
\ldl\Delta = - \dDelta
\ee
will play a closely similar r\^ole to that of propagators, in particular
being the inverse of these two-point vertices up to gauge
transformations.

The $\C$ commutator terms in \eq{wev}, yield $\sigma$
commutators on spontaneous symmetry breaking. Since $\sigma$ commutes
with $A$ and $C$ but anticommutes with $B$ and $D$, $\Delta^{\A\A}_m$
and $\Delta^{\C\C}_m$ allow for the addition of spontaneous mass creation
for $B$ and $D$ whilst still allowing the solution that the two-point vertices 
of $\hS$ and $S_0$ are equal. The appearance of the $\C$ commutator on both sides 
allows us to insist that $\C\leftrightarrow-\C$ is an invariance of the 
symmetric phase. The form \eqs{a0}{a1} preserves charge conjugation symmetry 
$\C\mapsto \C^T$, $\A\mapsto-\A^T$ (using the definition of the supermatrix 
transpose in ref. \cite{sunn}. Note that here the transformation for
$\C$ is as given so that its vacuum expectation value
is invariant under charge conjugation. Since charge conjugation reverses
the order of terms in a supertrace, diagrammatically it corresponds to
reversing the arrows in figs. \ref{winexp} -- \ref{fig:fieldex}, \ie to 
mirror reflection \cite{ymi,ymii}.)

From \eq{dCgauged} and \sec{Decoration},
it is trivial to see that the $\delta/\delta\C$ terms are supergauge 
invariant. Under a supergauge 
transformation we have by \eq{dAgauged} and \eq{coline},
\be \label{tlgaugetr}
\delta\left(\frac{\delta S}{\delta {\cal
A}_{\mu}}\{\dDelta^{\!\A\A}\}\frac{\delta \Sigma_g}{\delta {\cal
A}_{\mu}}\right) = {i\over2N}\,
\tr\!\left[{\delta S\over\delta\A_\mu},\Omega\right]\!\cdot
\dDelta^{\!\A\A}\!\cdot\str{\delta\Sigma_g\over\delta\A_\mu}
+ (S\leftrightarrow\Sigma_g),
\ee
where $S\leftrightarrow\Sigma_g$ stands for the same term with $S$ and
$\Sigma_g$ interchanged. But by \eq{dumbdef} and no-$\A^0$,
\be
\str{\delta\Sigma_g\over\delta\A_\mu}={\delta\Sigma_g\over\delta\A^0_\mu}=0,
\ee
similarly for $S$, and thus the tree-level terms are supergauge invariant.
Similarly, the quantum terms are $SU(N|N)$ gauge invariant, since
\be \label{qcgaugetr}
\delta\left(\frac{\delta }{\delta {\cal
A}_{\mu}}\{\dDelta^{\!\A\A}\}\frac{\delta }{\delta {\cal
A}_{\mu}}\Sigma_g\right) = {i\over N}\,\tr\!\left[{\delta \over\delta\A_\mu},
\Omega\right]\!\cdot\dDelta^{\!\A\A}\!\cdot
\str{\delta\Sigma_g\over\delta\A_\mu} =0.
\ee
This completes the proof that the exact RG  is supergauge
invariant!

Note that there is no point in incorporating longitudinal terms into
the exact RG (as was done in ref. \cite{ymii}) because here the
manifest supergauge invariance means that they can be exchanged for
$\C$ commutators:
\be
\label{gaugeS}
\nabla_\mu\!\cdot\!{\delta S\over\delta\A_\mu}\, 
= \,i\,\C\!\cdot\!{\delta S\over\delta\C}
\ee
(as holds for any supergauge invariant functional) and thus absorbed into the $\dDelta^{\C\C}_m$ term. 

It is important for the working of the $SU(N|N)$
regularisation that the effective scale of spontaneous symmetry
breaking is tied to the higher derivative regularisation scale, which
thus both flow with $\Lambda$. This is not the typical situation, but
can be arranged to happen here by constraining $\hS$ appropriately. 
However, as we
now show, the constraint is straightforward only if we take $\C$
to be dimensionless in \eq{sunnfl} -- \eq{a1}.

Contracting an arbitrary constant supermatrix $X$ into
\eq{defCbar} (for convenience, \cf \sec{Supergauge}) and
differentiating with respect to $\Lambda$, we have:
\be \label{minimum}
\left[\str\,{\partial{\bar\C}\over\partial\Lambda}
{\partial\over\partial\C}\,\str\, X{\partial V\over\partial\C}
+ \str\, X{\partial \over\partial\C}{\partial V\over\partial\Lambda}
\right]_{\C={\bar\C}}  =0.
\ee
We can compute the flow ${\partial V/\partial\Lambda}$ by setting
$\A=0$ and $\C={\bar\C}$ in \eq{sunnfl}. Taking the classical limit 
$V\to V_0$, we find that the resulting equation simplifies dramatically.
Using eqns. \eq{ergcl}, \eq{a0}, \eq{defCbar}, \eq{wv}, \eq{mno},
the fact that vertices in the actions with only one $\A_\mu$, 
vanish at zero momentum (by Lorentz invariance), and
\be
[{\bar\C},\left.{\partial{\hat V}\over\partial\C}]\ \right|_{\C={\bar\C}}\!\!=0,
\ee
which follows from global $SU(N|N)$ invariance
(where ${\hat V}$ is the potential in $\hS$), we get
\be
\label{minCondn}
\str\,\left[\left(\Lambda{\partial{\bar\C}\over\partial\Lambda}
+\dDelta^{\C\C}(0,\Lambda){\partial{\hat V}\over\partial\C}\right)
{\partial\over\partial\C}\,\str\, X{\partial V_0\over\partial\C}
\right]_{\C={\bar\C}}  =0.
\ee

With $\C$ dimensionless, we can and will insist that the classical vacuum
expectation value ${\bar\C}=\sigma$.  Eq. \eq{minCondn} is then
satisfied if and only if\footnote{We will see that the requirement that
$C$ has a mass in the broken phase forces
$\dDelta^{\C\C}(0,\Lambda)\ne0$.} 
${\hat V}$ also has a minimum at ${\C}=\sigma$. This is
delightful since it ensures that at the classical level at least, 
neither action has one-point $\C$
vertices in the broken phase.  We
will thus impose 
\be
\left.{\partial {\hat V}\over\partial\C}\right|_{\C=\sigma} =0
\ee
as a constraint on $\hS$.

Had we not taken $\C$ to be dimensionless, we would have had
to require that ${\bar\C}$ depend on $\Lambda$, in order that the
effective breaking scale flows with $\Lambda$. Since $X$ is general,
\eq{minCondn} would then imply that ${\hat V}$ {\sl cannot} have a
minimum also at $\C={\bar\C}$. Further analysis shows that ${\hat V}$
is then forced to violate $\C\leftrightarrow-\C$ symmetry in the
symmetric phase. 

Although conventionally $\C$ would have dimension one, for these reasons
we will take it to be dimensionless from now on. (It is intriguing that
the conclusion that $\C$ [actually $C$] must be dimensionless was reached
for very different reasons in refs. \cite{alg,ymii} which are no longer 
necessarily applicable, now that \eq{gaugeS} is a symmetry.)

At the quantum level, ${\bar\C}=\sigma$ can be expected to receive loop
corrections.  Since $SU(N)\times SU(N)$ invariance is left unbroken,
these corrections can only be proportional to $\sigma$ or $\one$.
Corrections proportional to the latter do not affect the breaking (but
presumably through \eq{defg} give important contributions at higher
loops), however corrections proportional to $\sigma$ would result,
through \eq{gaugeS}, in broken gauge invariance identities that
explicitly involve $g$ and thus mix different loop orders. We can avoid
this by again using the freedom in our choice of $\hS$ to design things
appropriately. We can constrain the appearance of ${\hat V}$ one-point 
vertices in the broken phase  
\be
\label{defv}
v^{C}\, \str\,\C\ +v^{C\sigma}\,\str\,\C\sigma
\ee
by imposing ${\bar\C}=\sigma$ as a renormalization condition. 
Each $v$ is then a non-vanishing function of $g$, but from
the analysis above, only from one-loop onwards:
\be
\label{expv}
v^{C}(g) = v^C_1\,g^2+v^C_2\,g^4+\cdots\ins11{and}
v^{C\sigma}(g) = v^{C\sigma}_1\,g^2+v^{C\sigma}_2\,g^4+\cdots.
\ee
However the fact that these corrections start only at one loop,
makes them already too high an order to
affect the one-loop $\beta$ function calculation. (This is particularly
clear from the perspective of higher loop calculations \cite{us}.)

\subsection{Supersowing and supersplitting in the $\A$ sector}
\label{Supersowing}

The inherent supersymmetry has a remarkable effect on the gauge
algebra: one can replace the usual manipulation of structure constants
and reduction to Casimirs, which becomes increasingly involved 
at higher loops, by simple steps \eq{sow} and \eq{split} which always
either just sow together supertraces or split them open. These have an
immediate diagrammatic interpretation.  The apparent violations present
in \eq{sowA} and \eq{splitA} must somehow disappear since they would
violate even global $SU(N|N)$. We first prove that this indeed the case.

For the case where the action contains just a single supertrace, which
will turn out to be all we need here, we could adapt the proof given in
sec. 6.2 of ref. \cite{sunn}. However, in preparation for future work,
we will give a more sophisticated proof which is applicable when
working with multiple supertrace contributions. Indeed we will see that
there is then one special case, where the corrections in
\eqs{sowA}{splitA} do survive, and result in a simple supergauge
invariant correction.

The corrections present in \eqs{sowA}{splitA} arise because $\A$ is
constrained to be supertraceless. To compare their effect to the
unconstrained case \eqs{sow}{split}, we momentarily `lift' $\A$ to a full
superfield $\A^e$ by adding a $\sigma$ part:
\be
\A_\mu\mapsto \A^e_\mu := \A_\mu+\sigma\A^\sigma_\mu.
\ee
$\A^\sigma_\mu$ is taken arbitrary so the map is not at all unique. We
similarly extend all functionals of $\A$ to the full space, simply
by replacing $\A$ with $\A^e$, \eg
\be
S^e[\A^e,\C] := S[\A\mapsto\A^e,\C].
\ee
Again, this is a not unique procedure, as can be seen for example in
the fact that  $\str\,\A$ vanishes, but the promoted functional
$\str\,\A^e$ does not.  We also introduce the projection back onto the
supertraceless space:
\be
\label{defpi}
\pi \A^e_\mu = \A_\mu, \quad \pi S^e=S, \quad\etc,
\ee
which of course is unique. 
Functional derivatives with respect to $\A^e$ can be written as
\be
\label{dAedef}
{\delta\over\delta\A^e_\mu} = {\delta\over\delta\A_\mu} 
+ {\one\over2N} {\delta\over\delta\A^\sigma_\mu},
\ee
using \eq{dumbdef}, or equivalently defined as in \eq{Cdumbdef}. 
$\delta/\delta\A^e$ thus satisfies
the exact supersowing and supersplitting relations \eqs{sow}{split}.
In the extended space, the constrained derivative
\eq{dumbdef} can now be written in terms of an unconstrained 
derivative:
\be
\label{dAdef}
{\delta\over\delta\A_\mu} = {\delta\over\delta\A^e_\mu} 
-{\one\over2N}\,\tr{\delta\over\delta\A^e_\mu}.
\ee
Of course $\pi$ and $\delta/\delta\A^\sigma$ do not
commute, however
\be
\label{prepre}
\frac{\delta S}{\delta {\cal
A}_{\mu}}\{\dDelta^{\!\A\A}\}\frac{\delta \Sigma_g}{\delta {\cal
A}_{\mu}} = \pi\left\{
\frac{\delta S^e}{\delta {\cal
A}_{\mu}}\{\dDelta^{\!\A\A}\}^e\frac{\delta \Sigma^e_g}{\delta {\cal
A}_{\mu}} \right\},
\ee
since $\A^\sigma$ is not differentiated on the right hand side.
Substituting \eq{dAdef} or \eq{dAedef}, and using \eq{coline} and \eq{dumbdef}, 
the term in big curly braces becomes
\be
\label{preAAsow}
\frac{\delta S^e}{\delta {\cal
A}^e_{\mu}}\{\dDelta^{\!\A\A}\}^e\frac{\delta \Sigma^e_g}{\delta {\cal
A}^e_{\mu}}
-{1\over2N}{\delta\S^e\over\delta\A^0_\mu}\cdot\dDelta^{\!\A\A}\cdot
{\delta\Sigma_g^e\over\delta\A^\sigma_\mu}
-{1\over2N}{\delta\Sigma_g^e\over\delta\A^0_\mu}\cdot\dDelta^{\!\A\A}\cdot
{\delta\S^e\over\delta\A^\sigma_\mu}.
\ee
Now, as we explain below, no-$\A^0$ symmetry is violated in the
extended space. However the $\A^0$ derivatives in \eq{preAAsow} do
vanish after the projection.  Thus \eq{prepre} becomes
\be
\label{sowAe}
\frac{\delta S}{\delta {\cal
A}_{\mu}}\{\dDelta^{\!\A\A}\}\frac{\delta \Sigma_g}{\delta {\cal
A}_{\mu}} = \pi\left\{
\frac{\delta S^e}{\delta {\cal
A}^e_{\mu}}\{\dDelta^{\!\A\A}\}^e\frac{\delta \Sigma^e_g}{\delta {\cal
A}^e_{\mu}} \right\},
\ee
which says precisely that the corrections in \eq{sowA} can be ignored: 
exactly the same result is obtained if exact supersowing is used.

However, performing the same analysis on the corresponding quantum term
in \eq{a1}, we get a correction to exact supersplitting, consisting
of an attachment of the (zero-point) kernel $\dDelta^{\A\A}(p,\Lambda)$
to two $\A$ points in $\Sigma_g$:
\be
\label{splitAe}
\frac{\delta }{\delta {\cal
A}_{\mu}}\{\dDelta^{\!\A\A}\}\frac{\delta \Sigma_g}{\delta {\cal
A}_{\mu}} = \pi\left\{
\frac{\delta }{\delta {\cal
A}^e_{\mu}}\{\dDelta^{\!\A\A}\}^e\frac{\delta \Sigma^e_g}{\delta {\cal
A}^e_{\mu}} \right\} -{1\over N} \pi {\delta\over\delta\A^\sigma_\mu}\cdot
\dDelta^{\!\A\A}\cdot{\delta\Sigma_g\over\delta\A^0_\mu}.
\ee
To understand when this correction is non-vanishing, we need briefly to
analyse the consequences of no-$\A^0$ symmetry in more detail.
Considering the transformation\footnote{there are higher order
constraints from separating out higher powers of $\A^0$ but from
\eq{splitAe} we only need the first order} $\delta\A_\mu=
\lambda_\mu\one$ in \eq{Sex}, we see that the result must vanish either
via the supergroup algebra because the corresponding vertex contains a
factor $\str\A\A$, thus generating $\str\A=0$ (but $\str\A^e\ne0$ in
the extended space), or because a non-trivial constraint exists on the
corresponding vertex function.  (This is simply that the sum over all
possible valid placings of $\A^0$s associated position and Lorentz
argument inside a vertex function leaving other arguments alone, yields 
zero, \cf \eq{noAreln} and refs. \cite{antonio,us} for more detail.) 
This non-trivial constraint then causes the
coefficient to vanish whether or not the remaining supergauge fields
are extended by $\A^\sigma\sigma$.  Thus the correction in \eq{splitAe}
vanishes in all cases except where the zero-point $\dDelta^{\A\A}$
kernel attaches each end to a $\str\A\A$ factor. Comparing the result to the
computation assuming exact supersplitting, \ie the first term in
\eq{splitAe}, we see that instead of getting a supergroup factor
$(\str\one)^2=0$ we get $-{1\over N}\str\sigma$ \ie a supergroup factor of
$-2$.

(Note that in deriving this rule we have assumed that vertices in $\Sigma_g$ 
with factors $\str\A$ have been set to zero from the beginning [as would
follow immediately from the $SU(N|N)$ group theory]. If for some reason
this was not done then the first term in \eq{splitAe} can get a non-zero
contribution from the kernel attaching to this $\str\A=2N\A^\sigma$ point. 
However it then also appears in the correction with precisely equal
and opposite coefficient.) 

This supergroup factor should have been expected since the algebra
part of the attachment of a zero-point kernel to a two-point vertex simply
counts the number of bosonic degrees of freedom in the algebra
minus the number of fermionic degrees of freedom. There are $N^2$ fermionic
such terms in $B$, but only $N^2-2$ in $A$, since both $\A^\sigma$ and,
by no-$\A^0$ symmetry, $\A^0$, are missing. 

Since the correction in \eq{splitAe} is non-vanishing only when using up
a separate $\str\A\A$ factor, it is clear that the result is still supergauge
invariant in the remaining external superfields. Furthermore in the
present case where we will be able to work with actions with only a single 
supertrace, the entire effect of the correction is a just vacuum energy 
contribution, which from now on we ignore.

\subsection{Diagrammatic interpretation}
\label{DiagrammaticI}

\begin{figure}[hh]
\psfrag{=}{$=$}
\psfrag{-}{$-$}
\psfrag{+}{$+$}
\psfrag{ldl}{$\ldl$}
\psfrag{S}{$S$}
\psfrag{si}{$\Sigma_g$}
\psfrag{1/2}{$\displaystyle \, \frac{1}{2}$}
\psfrag{xi}{\tiny $f$}
\psfrag{sumi}{$\displaystyle\sum_{f=\A,\C}$}
\begin{center}
\includegraphics[scale=.5]{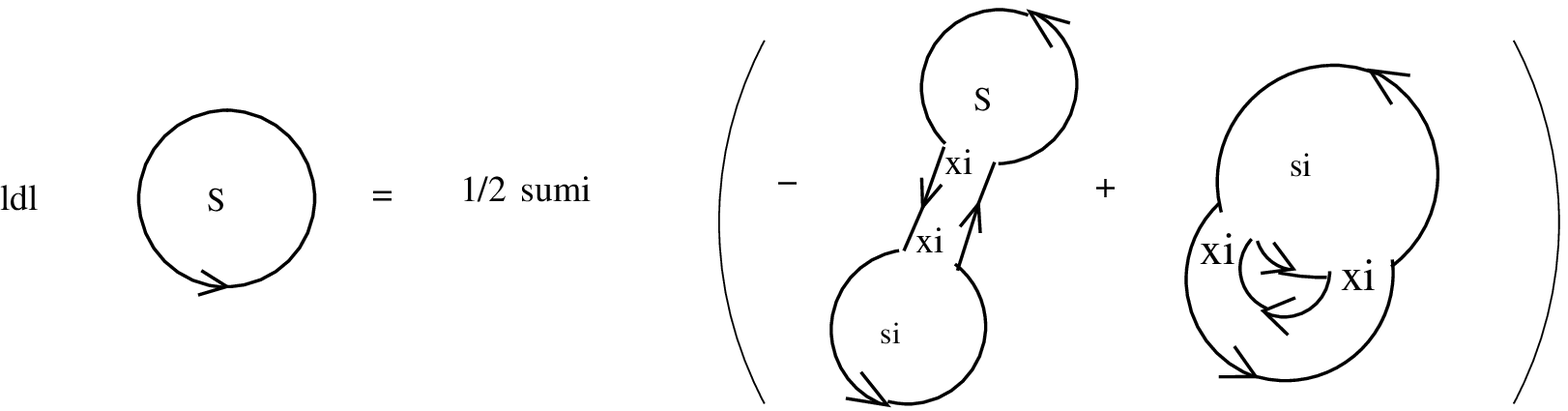}
\end{center}
\caption{Graphical representation of the exact RG, when $S$ and $\hS$ 
contain only single supertraces.}\label{fig:floeq}
\end{figure}

$\A$ thus also effectively satisfies the exact supersowing and
supersplitting relations \eq{sow} and \eq{split}. By using these
equations when the covariantised kernels \eq{wcv} act on
the actions \eq{Sex}, and comparing the result to
the diagrammatic interpretation of the covariantised kernels 
and actions, \fig{winexp} and figs. \ref{fig:action},\ref{fig:fieldex}, 
it is clear that the exact RG is given diagrammatically as in \fig{fig:floeq}.

Here we have specialised to the case of interest in this paper, where 
$S$ and $\hS$ can be assumed to have only a single supertrace. (The
extension to the more general contributions \fig{fig:action} is obvious.)
Expanding the thick lines (representing any number of fields) into a
power series in the fields, we translate the figure into individual
Feynman diagrams, whose Feynman rules are given by the momentum space
versions of the vertices in \eq{wcv} and \eq{Sex} (without the symmetry
factors).\footnote{This part of the analysis is the same as in
ref. \cite{ymii}, except that here we make explicit the factor of
$1/2$ from \eq{a0} and \eq{a1}, in \fig{fig:floeq} and the Feynman diagrams,
and the factor of $1/\Lambda^2$ is now incorporated in the definition of
the kernels $\dDelta$.} The points representing individual fields 
and their associated momenta and Lorentz indices,
appear in all places on a composite
loop with equal weight, whilst respecting the cyclic order.
Of course if one of the corresponding vertices
does not appear in the expansions \eq{wcv} and \eq{Sex}, the corresponding
Feynman rule is zero. 

It can be seen from \fig{fig:floeq} that the tree-level corrections preserve
the assumption that there is only a single supertrace in $S$, but that
each quantum correction results in
an extra supertrace factor. Thus in general $S$ has terms with any number
of supertraces, and already a product of two supertraces at 
one-loop. However for the computation of the $\beta$ function, we need only
look at contributions to the $AA$ two-point vertex (see \eq{defg} and later,
or refs. \cite{alg,ymi,ymii}). Since $A$ is both traceless and supertraceless,
to get a non-vanishing answer both $A$s must lie in the same
supertrace, leaving the other one empty of fields. In this way, $S$ 
effectively contains only a single supertrace to the order in which
we are working.

\subsection{After spontaneous breaking}
\label{After}

We substitute $\C\mapsto\C+\sigma$, and from now on work in the
spontaneously broken phase. Working with fields appropriate for the
remaining $SU(N)\times SU(N)$ symmetry, we break $\A$ and $\C$ down
to their bosonic and fermionic parts $A$, $B$, $C$ and $D$ as in
\eq{ABCD}. 

\begin{figure}[h]
\begin{center}
\includegraphics[scale=.7]{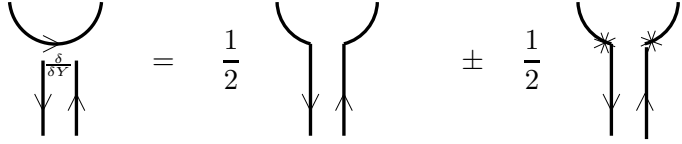}
\end{center}
\caption{Feynman diagram representation of attachment via a
partial supermatrix ${\rm d}_\pm Y = \pm Y$.}\label{partialAttach}
\end{figure}

The diagrammatic interpretation is still the same, except that
we now have the four flavours to scatter around the composite loops,
and appearances of $\sigma$, which can be simplified as explained in 
\sec{Superfield}. (Some terms are then related, for example 
$\dDelta^{A,AA}_\mu =$ $\dDelta^{B,AB}_\mu$,
although we never need to use this explicitly.) In addition, we
must recall the corrections to supersplitting and supersowing arising
from differentiating only partial supermatrices \cite{ymii}. These 
lead to further appearances of $\sigma$ which are
easily computed by expressing the partial supermatrices in terms
of full supermatrices via the projectors ${\rm d}_\pm$ onto the block
(off)diagonal components
\be
{\rm d}_\pm X = {1\over2}(X\pm\sigma X\sigma),
\ee
(hence $C={\rm d}_+\C$, $D={\rm d}_-\C$, \etc).
Diagrammatically this simply amounts to
corrections involving a pair of $\sigma$s inserted either side of the 
attachment as in \fig{partialAttach} \cite{ymii}.
\psfrag{+}{$\pm$}
\psfrag{=}{$=$}
\psfrag{d}{\tiny $\frac{\delta}{\delta Y}$}
\psfrag{1/2}{${\ds \frac{1}{2}}$}

For tree-level type attachments as in \eq{sow}, the corrections merely
ensure that the coefficient supermatrices ($X$ and $Y$) have the appropriate
statistics to make each supertrace term totally bosonic (\cf 
\sec{Superfield}), but this has already been taken into account in the 
Feynman rules. Thus these corrections have no effect at tree level 
\cite{ymii}. 

Since the classical action $S_0$ (similarly $\hS$) has only a single
supertrace and respects $\C\leftrightarrow-\C$ invariance in the symmetric
phase (\cf \sec{manifestly}), upon spontaneous breaking
we have the `theory space' symmetry 
\bea
\label{z2}
C &\leftrightarrow& -C,\nonumber\\ 
D &\leftrightarrow& -D,\nonumber\\
\sigma &\leftrightarrow& -\sigma.
\eea

The single supertrace part of the one-loop effective action $S_1$ 
has a single supertrace because it also has
a supertrace void of fields (\cf \sec{DiagrammaticI}). 
In order for this not to vanish it must `trap' a $\sigma$ (so that we
get $\str\,\sigma=2N$ rather than $\str\,\one=0$). Therefore, the non-trivial
supertrace has one less $\sigma$ (mod two) and is thus odd under the symmetry
\eq{z2}. 

These observations, which can be easily extended to multiple loops
and supertraces, are useful in limiting the possible vertices.

\subsection{(Un)broken gauge invariance}
\label{unbroken}

Splitting $\Omega$ into its fermionic and bosonic parts: 
$\tau={\rm d}_-\Omega$ and $\omega={\rm d}_+\Omega$, we obtain from 
\eq{defnabla}, \eq{Agauged} and \eq{Cgauged} the unbroken 
$SU(N)\times SU(N)$ transformations
\bea
\label{bgauged}
\delta A_\mu &=& D_\mu \cdot\omega \nonumber\\
\delta B_\mu &=& -i\, B_\mu \cdot\omega \nonumber\\
\delta C &=& -i\, C \cdot\omega\nonumber\\
\delta D &=& -i\, D \cdot\omega,
\eea
where $D_\mu = \partial_\mu -iA_\mu$ is the covariant derivative
for the $SU(N)\times SU(N)$  (the dot again means action by commutation,
and we have used the fact that $[\sigma,\omega]=0$) and the broken
fermionic gauge transformations
\bea
\label{fgauged}
\delta B_\mu &=& D_\mu\cdot\tau\nonumber\\
\delta A_\mu &=& -i\, B_\mu\cdot\tau \nonumber\\
\delta D &=& -i\, C\cdot\tau +2i\tau\sigma \nonumber\\
\delta C &=& -i\, D\cdot\tau.
\eea
From the first of \eq{bgauged} we see that $A$ can have no
wavefunction renormalization because if it did, then replacing $A$
by $Z^{1/2}A$, this becomes $\delta A_\mu 
=D_\mu\cdot\omega+(Z^{-1/2}-1)\partial_\mu\omega$, \ie the gauge symmetry
is violated \cite{alg,ymi,ymii}. (Clearly this is true whether or not
one tries to reparametrize $\omega$ also.) This is the reason for scaling the
coupling $g$ out of the connection \eq{defnabla}: $A$ then has no
anomalous dimension and only $g$ renormalizes. Since the 
corresponding marginal term to \eq{defg} exists for $A_2$, there is also
a coupling $g_2$. Its renormalization is important for calculations in
the physical sector beyond one loop \cite{us}. 

In a similar way the first two relations of \eq{fgauged} imply
that $B$ can have no wavefunction renormalization, whilst at first
surprisingly the last
two relations imply that $C$ and $D$ cannot have any wavefunction 
renormalization either. However, these last two relations are
a consequence of the vacuum expectation value ${\bar\C}=\sigma$ being
protected by the introduction of the terms \eqs{defv}{expv}.

Applying these relations to the field expansions in momentum
space, we get the `trivial'
Ward identities which relate vertices via the manifest bosonic and broken 
fermionic gauge invariance. These identities provide a powerful check
on solutions for $S$, and play a crucial r\^ole in the calculation to follow.

Applying the unbroken gauge symmetries \eq{bgauged}:
\be
q^{\nu}U^{\cdots X A Y\cdots}_{\cdots\ a\, \nu\, b\,\cdots}(\cdots,
p,q,r,\cdots)=U^{\cdots X Y\cdots}_{\cdots\ a\,  b\,\cdots}(\cdots,
p,q+r,\cdots)-U^{\cdots X Y\cdots}_{\cdots\ a\,  b\,\cdots}(\cdots,
p+q,r,\cdots),
\label{bosonwi}
\ee
where $U$ stands for any element, \ie a vertex from a covariantised
kernel or from an action. $X_a$ and $Y_b$ are the fields $A$, $B$, $C$
or $D$, with $a$ and $b$ Lorentz indices or null as appropriate. Geometrically, 
the momentum of
the gauge field is pushed forward along the direction of the Wilson line to
the next `obstruction' (with a plus) or pulled back against the
direction of the Wilson line to the previous obstruction (and given a
minus sign) \cite{alg}. If $A$ is at the end of a line in a wine
vertex, then either $X$ or $Y$ is $Z_1$ or $Z_2$ in the expansion
\eq{wcv} as appropriate, and the momentum is pushed forward (pulled
back) onto this \cite{ymi,ymii}. Since $\sigma$ commutes with $\omega$,
any $\sigma$ insertion is `invisible' in this process and the momentum
$q$ is pushed forward (pulled back) through the $\sigma$ position to
the next `real' obstruction. [This is also clear by temporarily
(anti)commuting the $\sigma$ out of the way and then applying
\eq{bosonwi}.]

Similarly, applying the broken supergauge symmetries \eq{fgauged}, we get
\bea
\label{fermionwi}
&& q^{\nu}U^{\cdots X B Y\cdots}_{\cdots\ a\, \nu\, b\,\cdots}(\cdots,
p,q,r,\cdots) - 2 U^{\cdots X D\sigma Y\cdots}_{\cdots\ a\,\phantom{D\sigma} 
b\,\cdots}(\cdots, p,q,r,\cdots) \nonumber\\
&&\qquad\qquad = U^{\cdots X \hat{Y}\cdots}_{\cdots\ a\,b\,\cdots}(\cdots,
p,q+r,\cdots)-\, U^{\cdots \hat{X} Y\cdots}_{\cdots\ a\,  b\,\cdots}(\cdots,
p+q,r,\cdots). 
\eea
$X_a$ and $Y_b$ have the same interpretation as before. $\hat{X}$ and
$\hat{Y}$ are the opposite statistics partners (thus $\hat{A}_\mu=B_\mu$,
\etc). This time, since $\tau$ anticommutes with $\sigma$, if the momentum
$q$ is pushed back (pulled forward) through a $\sigma$ then the 
corresponding term on the right hand side of \eq{fermionwi} changes sign.

\section{Seed action two-point vertices}
\label{Seed}

As we have already emphasised, we do not restrict the seed action $\hS$
to a particular choice. We will restrict the set of possible choices however, 
for technical reasons. For example in this paper we insist that $\hS$
has only a single supertrace. We could have taken
the form of the bare action from ref. \cite{sunn} as one choice of seed
action, but it is very helpful
to add more general interactions than this, in
order to avoid the appearance of certain flowing classical couplings 
(\cf sec. \ref{Ensuring}). More importantly we now realise that it is to our
advantage to keep $\hS$, and the forms of covariantisation, general, since
this guides us to an efficient procedure for calculation. Providing we
are computing a well defined physical quantity, we are guaranteed that
the result is independent of the detailed choices.

Consider first the $DD$ two-point vertex.\footnote{There is no
$DD\sigma$ vertex since by $\{D,\sigma\}=0$ and cyclicity,
$\str DD\sigma = -\str D\sigma D = -\str DD\sigma$.} By Goldstone's theorem 
\cite{Goldstone}, $D$ must be massless, thus by Lorentz invariance and
dimensions its kinetic term takes the form
\be
\label{hSDD}
\hS^{DD}(p)=\Lambda^2p^2/\ct_p,
\ee
where $\ct_p=\ct(p^2/\Lambda^2)$ is a dimensionless smooth strictly positive 
function. (Recall from \sec{Necessary} 
that $\Lambda$ is the only explicit scale that can appear, and smoothness is
a requirement for all vertices.) Although it is not necessary \cite{ymii},
we set the kinetic term to be conventionally normalised, and so
restrict our choices to $\ct(0)=1$. 

Proceeding similarly, we have that in general there are two types of 
$AA$ vertex, however by \eq{z2}, $\hS^{AA\sigma}_{\mu\nu}=0$. 
From \eq{bosonwi}, the $AA$ vertex is totally transverse
\be
\label{transverse}
p^\mu \hS^{AA}_{\mu\nu}(p)=0.
\ee
(Since $\str A=0$, single point $A$ 
vertices do not exist.\footnote{But also for many other reasons: 
Poincar\'e invariance, charge conjugation invariance, \etc})
By dimensions and Lorentz invariance, it therefore takes the form
\be
\label{hSAA}
\hS^{AA}_{\mu\nu}(p) = 2\Box_{\mu\nu}(p)/c_p,
\ee
where $\Box_{\mu\nu}(p)\equiv p^2\delta_{\mu\nu}-p_\mu p_\nu$ is the usual 
transverse kinetic term, which will appear often, and $c_p=c(p^2/\Lambda^2)$ 
is another dimensionless smooth strictly positive function. 
Recall from \sec{manifestly} that we set the classical 
two-point vertex equal to this:
\be
\label{SAA}
S_{0\,\mu\nu}^{\ph0AA}(p)=\hS_{\mu\nu}^{AA}(p).
\ee
This implies from the renormalization condition \eq{defg}, and
\eq{Sloope}, that $c(0)=1$. In order to maintain finiteness, we must limit
the large momentum behaviour of $c$ and $\ct$, for power law large momentum
behaviour as in \eq{inequalities}.

This analysis shows that so far, apart from some very basic restrictions, the 
introduction of the cutoff functions merely parametrise the most general 
two-point vertices.

Now consider the $BB$ vertex. (Like $DD\sigma$, a $BB\sigma$ vertex
cannot exist.) It has a transverse part, which by dimensions and Lorentz
invariance has the same form as \eq{hSAA} but with a possibly different
cutoff function. However for $SU(N|N)$ invariance to be recovered at
high energies, these cutoff functions must agree at high energies.
For simplicity we just set them equal. (In the symmetric phase, $B$ and
$A$ of course have the same kinetic term. In the broken phase they 
of course differ, for example from
$\hS^{\A\C\A\C}\mapsto\hS^{\A\sigma\A\sigma}$.
For the two kinetic terms to 
disagree also at high energies, these 
higher point interactions with $\C$ would have to have 
momentum dependence in the ultraviolet
so violent as to destroy the higher derivative part of the
regularisation of ref. \cite{sunn}.) The longitudinal part of the $BB$
vertex is already determined by two applications of
broken fermionic gauge invariance \eq{fermionwi}:
\bea
p^{\mu}\hS^{BB}_{\mu\nu}(p)&=&-2\hS^{BD\sigma}_{\nu}(p)\nonumber\\
p^{\mu}\hS^{BD\sigma}_{\mu}(p)&=&-2\hS^{DD}(p).\label{hSBDwi}
\eea
[In the first line we use cyclicity and then Lorentz invariance:
$\hS^{BD\sigma}_\nu(-p,p)=-\hS^{BD\sigma}_\nu(p)$.] Thus
\be
\label{hSBB}
\hS^{BB}_{\mu\nu}(p)=2\,\c_p\Box_{\mu\nu}(p)
+4\Lambda^2\,\ctil_p\delta_{\mu\nu},
\ee
and using \eq{hSBDwi},
\be
\label{hSBDs}
\hS^{BD\sigma}_{\mu}(p) = -2\Lambda^2p_\mu/\ct_p.
\ee
By using the fact that the vertex must be overall bosonic, and using
charge conjugation [or the (broken super)gauge symmetries, or sometimes
just \eq{z2}] one 
may readily show that all other mixed two-point vertices are disallowed.

Finally, we know from \eq{z2} that there is no $CC\sigma$ vertex. The
difference between the $CC$ and $DD$ vertex, amounts
to the addition of a new cutoff function that serves
to give $C$ a mass, and thus must not vanish at $p=0$. In addition at
high momentum it must be subleading compared to the $DD$ part
in order that the symmetric phase be regained (as with $BB$
versus $AA$). For simplicity we simply choose it to be constant and thus,
\be
\label{hSCC}
\hS^{CC}(p)= \Lambda^2p^2/\ct_p + 2\lambda\Lambda^4,
\ee
where $\lambda>0$ is a constant parameter that is left undetermined.

This completes the parametrisation of the seed action two-point vertices. 
In point of fact they are the ones that would be obtained by setting
the seed action to have the same form as the bare action of ref. \cite{sunn},
however we emphasise that the higher-point $\hS$ vertices will not agree with
the bare ones from \cite{sunn}. These higher point vertices are constrained
by the symmetries of the theory and most powerfully by 
\eqs{bosonwi}{fermionwi}. By iterative use of these identities and the flow
equations \eqs{ergcl}{ergone}, we will be able to reduce the complete 
calculation of $\beta_1$ to a dependence only on the two-point vertices
above. In this way, although we for simplicity set a number of 
restrictions (equality of cutoff functions for $B$ and $A$, $\ct(0)=1$,
and that the mass term for $C$ is simply a constant), these come into
play only at the end of the computation and could easily be lifted.

\section{The kernels}
\label{kernels}
\begin{figure}[h]
\psfrag{sim}{}
\psfrag{=}{$=$}
\psfrag{Wp}{} 
\psfrag{Kp}{} 
\psfrag{Hp}{} 
\psfrag{Gp}{} 
\psfrag{Am}{$A_{\mu}$}
\psfrag{Bm}{$B_{\mu}$}
\psfrag{C}{$C$}
\psfrag{D}{$D$}
\begin{center}
\includegraphics[scale=.5]{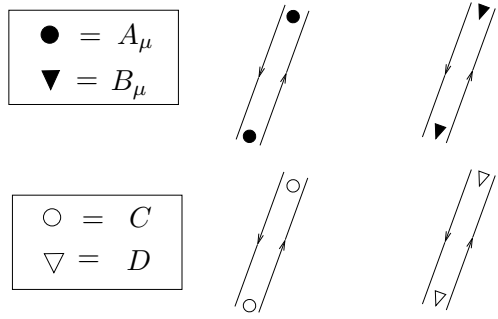}
\end{center}
\caption{Graphical representation of 0-point wines}\label{fig:wines}
\end{figure}

\begin{figure}[h]
\psfrag{S}{$S_0$}
\psfrag{Si}{$\Sigma_0$}
\psfrag{=}{$=$}
\psfrag{-1/2l2}{$-\displaystyle\frac{1}{2}$}
\psfrag{mu}{$\mu$}
\psfrag{nu}{$\nu$}
\psfrag{ldl}{$\ldl$}
\psfrag{+}{$+$}
\psfrag{<->}{$\leftrightarrow$}
\psfrag{p}{$p$}
\psfrag{-}{$-$}
\psfrag{pa}{$\left(\right.$}
\psfrag{sis}{$\left.\right)$}
\begin{center}
\includegraphics[scale=.55]{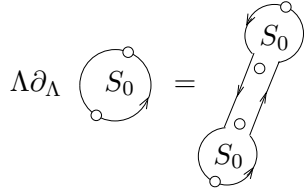}
\end{center}
\caption{$CC$ tree-level equation. One $C$ has momentum $p$
and the other $-p$.}\label{CC}
\end{figure}

From \eq{a0} and \eq{wev},
the zero-point kernels in the broken phase take the form
\be
\label{bzpt}
\dDelta^{AA} = \dDelta^{\A\A},\quad
\dDelta^{BB} = \dDelta^{\A\A}+\dDelta^{\A\A}_m,\quad
\dDelta^{CC} = \dDelta^{\C\C}\quad{\rm and}\quad
\dDelta^{DD} = \dDelta^{\C\C}+\dDelta^{\C\C}_m.
\ee
They will be represented graphically as in \fig{fig:wines}, and
are determined by the requirement that the two-point vertices of
classical effective
action $S_0$ can be set equal to those of $\hS$ (\cf \sec{manifestly}).
After setting $\hS^{CC}=S^{CC}_0$, the flow equation for $S^{CC}_0$ takes the 
simple form given in \fig{CC}, giving
\be
\label{SCCfl}
\ldl S^{CC}_0(p)=S^{CC}_0(-p)\dDelta^{CC}_pS^{CC}_0(p),
\ee
thus, since $\hS^{CC}$ is an even function of $p$,
\bea
\label{invCC}
\dDelta^{CC}_p &=& -\ldl (\hS^{CC}_p)^{-1}\\
&=& {1\over\Lambda^4}\ {1\over x}\left({2x^2\ct\over x+2\lambda\ct}\right)',
\label{kCC}
\eea
where here and later we use the notation, $x=p^2/\Lambda^2$, 
the cutoff terms being functions of this ratio,
and prime denotes differentiation with respect to this. 

Similarly, using \eq{SAA}, the classical $AA$-point flow equation is 
\be
\label{SAAfl}
\ldl \hS^{AA}_{\mu\nu}(p) = \hS^{AA}_{\mu\alpha}(p)
\dDelta^{AA}_p \hS^{AA}_{\alpha\nu}(p),
\ee
If \eq{hSAA} were invertible, $\dDelta^{AA}$
would also take the form \eq{invCC}. Instead, substituting \eq{hSAA}
we get:
\be
\label{kAA}
\dDelta^{AA}={c'_p/\Lambda^2}
\ee

\begin{figure}[h!]
\psfrag{S}{$S_0$}
\psfrag{Si}{$\Sigma_0$}
\psfrag{=}{$=$}
\psfrag{mu}{$\mu$}
\psfrag{nu}{$\nu$}
\psfrag{ldl}{$\ldl$}
\psfrag{+}{$+$}   
\begin{center}
\includegraphics[scale=.5]{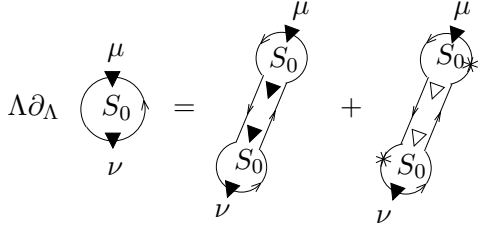}
\end{center}
\caption{$BB$ tree-level equation.}
\label{BB}
\end{figure}

The $BB$, $DD$ and $BD\sigma$ classical flow equations are coupled,
\cf \eg \fig{BB}: 
\bea
\label{SBDfl}
\ldl\hS^{BB}_{\mu\nu} &= \hS^{BB}_{\mu\alpha}\dDelta^{BB}\hS^{BB}_{\alpha\nu}
+\hS^{BD\sigma}_\mu\dDelta^{DD}\hS^{BD\sigma}_\nu\nonumber\\
\ldl\hS^{BD\sigma}_\mu &= \hS^{BB}_{\mu\alpha}\dDelta^{BB}\hS^{BD\sigma}_\alpha
+\hS^{BD\sigma}_\mu\dDelta^{DD}\hS^{DD}\nonumber\\
\ldl\hS^{DD} &= \hS^{BD\sigma}_\alpha\dDelta^{BB}\hS^{BD\sigma}_\alpha
+\hS^{DD}\dDelta^{DD}\hS^{DD},
\eea
where $p$ is the momentum argument in all the above terms. These three 
equations are of course not independent: the last two are readily seen to 
follow from the first, on using the broken gauge transformations \eq{hSBDwi}.
By substituting \eq{hSBB} and \eq{hSBDs} in the first (and isolating
the transverse part or otherwise), it is straightforward to solve for
the kernels:
\bea
\label{kBB}
\dDelta^{BB} &=& -\ldl (2p^2/c\, +4\Lambda^2/\ct)^{-1}\ 
= {1\over\Lambda^2}\left({xc\ct\over x\ct+2c}\right)'\\
\label{kDD}
\dDelta^{DD} &=& -\ldl \left({\ct\over\Lambda^2p^2}\right)
-{4\over p^2}\,\dDelta^{BB}\
= {1\over\Lambda^4}\ {1\over x}\left({2x^2\ct^2\over x\ct+2c}\right)'.
\eea

From \eq{bzpt}, the original kernels for \eqs{a0}{a1} and \eq{wev} are thus
given by \eq{kCC}, \eq{kAA}, and
\bea
\dDelta^{\A\A}_m &=& -{1\over\Lambda^2}\left({2c^2\over x\ct+2c}\right)'\\
\dDelta^{\C\C}_m &=& {1\over\Lambda^4}\ {1\over x}
\left({4x^2\ct\,(\lambda\ct^2-c)\over(x\ct+2c)(x+2\lambda\ct)}\right)'.
\eea
Importantly, these mass-term type corrections to $\dDelta^{\A\A}$ and 
$\dDelta^{\C\C}$, which behave as expected from \eq{hSBB} and \eq{hSCC},
decay much faster than \eq{kAA} and \eq{kCC}, thus ensuring that at high
momentum $p$, the exact RG \eq{sunnfl} goes over to one appropriate for
the symmetric phase of the $SU(N|N)$ theory. These corrections thus
behave as required by the discussion below \eq{inequalities}, in particular 
the $\C$ (+$\sigma$) decorations of \eq{wev}, which destroy the supertrace 
mechanism, here can be taken to be regularised by the covariant higher 
derivatives alone.

\section{The integrated kernels}
\label{integrated}

\begin{figure}[h!]
\psfrag{-}{$-$}   
\psfrag{=}{$=$}   
\psfrag{ldl}{$\ldl$}
\psfrag{1}{$1$}   
\psfrag{ar}{$\Rightarrow$}
\begin{center}
\includegraphics[scale=.5]{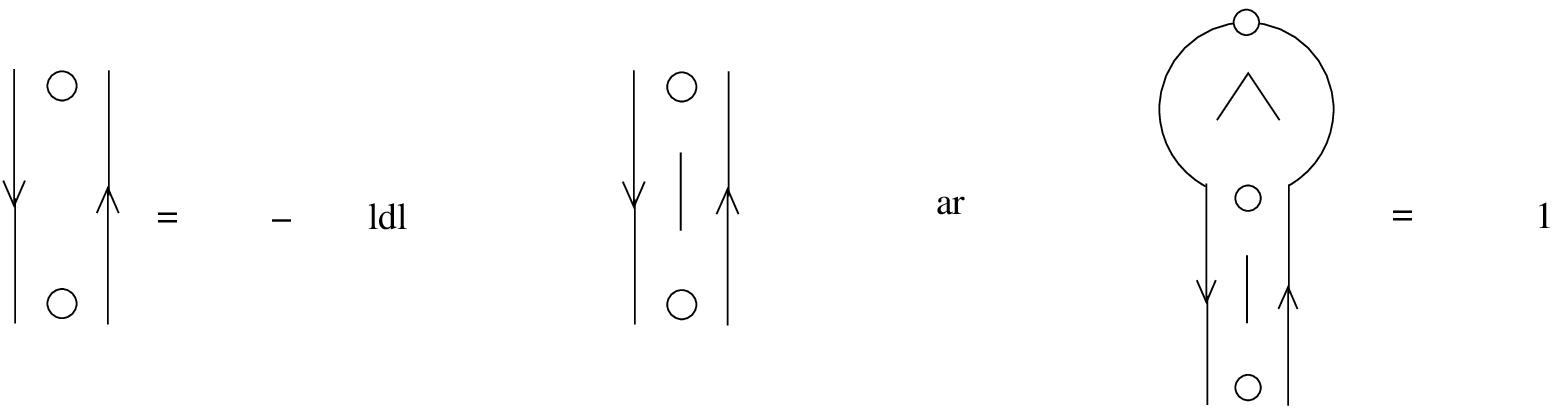}
\end{center}
\caption{$C$ integrated wine. In this case there is no remainder.}
\label{inverse}
\end{figure}

By \eq{defprop} and \eq{invCC},
we immediately see that the $CC$ integrated kernel is just the inverse 
kinetic term:
\be
\label{iCC}
\Delta^{CC} S^{CC}_0 =1\quad\Rightarrow\quad
\Delta^{CC}={1\over\Lambda^4}\,{\ct\over x+2\lambda\ct}
\ee
(choosing the integration constant here and later, so that the 
`effective propagator' vanishes as $p\to\infty$). We represent the 
integrated wine as in \fig{fig:wines}, but with a line down its spine,
and thus \eq{iCC} is represented 
diagrammatically as in \fig{inverse}. 

Integrating \eq{kAA}, we have
\be
\label{ikAA}
\Delta^{AA}_p = {c_p\over 2p^2}.
\ee
Despite its similarity to a regularised Feynman propagator, we 
stress that no gauge fixing has taken place. Indeed this
`effective propagator' is the inverse of the classical $AA$ kinetic
term only in the transverse space:
\be 
\label{iAA}
\Delta^{AA}_p S^{\ph0AA}_{0\,\mu\nu}(p) = \delta_{\mu\nu} -
{p_\mu p_\nu\over p^2}.
\ee
Since in practice $\Delta^{AA}$ will be connected to an $A$ point on some 
other vertex, the remainder term above
will simply generate gauge transformations 
via \eq{bosonwi}. This observation proves crucial in the `magic' 
of the calculation to follow.

The integrals via \eq{defprop} of \eq{kBB} and \eq{kDD} are also immediate,
and thus we find
\bea
\label{iBD}
\Delta^{BB} &=& {1\over2\Lambda^2}\,{c\ct\over x\ct+2c} \nonumber\\
\Delta^{DD} &=& {1\over\Lambda^4}\,{\ct^2\over x\ct+2c}.
\eea
Note that despite the classical $D$ kinetic term being that of a massless
(Goldstone) field, the $D$ effective propagator like that of $C$ and $B$
(but unlike $A$) has no massless pole. Of course this is nothing but the
Higgs mechanism, arising here from
the $B$ and $D$ two-point vertices being intimately entangled via
\eq{SBDfl}. Similarly to the above reasoning, the pair of effective
propagators \eq{iBD} would form the inverse of the {\sl matrix} of
these fermionic two-point vertices, if this matrix were invertible. It is
not, for the same reason that these flows are necessarily entangled: $B$ and
$D\sigma$ rotate into each other under the broken supergauge transformations
\eq{fermionwi}. 

\subsection{Five dimensions in the fermionic sector} 
\label{Five}

\begin{figure}[h!]   
\psfrag{=}{$ = (-)^{1+f_\sigma}$}   
\begin{center}
\includegraphics[scale=.6]{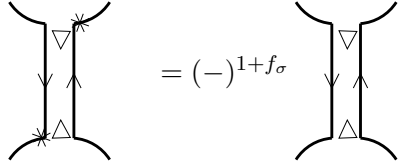}
\end{center}
\caption{Wine attachment to $D\sigma$.}
\label{wineDs}
\end{figure}

We thus need to consider $B$ and $D\sigma$ together. We write $B$ and 
$D\sigma$ as elements of a Euclidean 5 dimensional vector\footnote{Note
that this respects charge conjugation symmetry since $F_M\mapsto -F^T_M$.}
\be
\label{defF}
F_M=(B_\mu,\, D\sigma).
\ee
Introducing the `5-momentum' 
\be
q_M=(q_\mu,-2),
\ee 
we see from \eq{fermionwi} and \eq{hSBDwi}, that the matrix 
$\hS^{FF}_{MN}(p)\equiv \hS^{FF}_{MN}(p,-p)$ is going to be transverse. 
Indeed, defining
\be
\label{hSFF}
\hS^{FF}_{MN}(p) =
\pmatrix{\hS^{BB}_{\mu\nu}(p) &\ \hS^{BD\sigma}_\mu(p) \cr
-\hS^{BD\sigma}_\nu(p) & -\hS^{DD}(p)},
\ee
where we have used $\hS^{D\sigma B}_\nu(p)=-\hS^{BD\sigma}_\nu(p)$
and $\hS^{D\sigma D\sigma}=-\hS^{DD}$, we have:
\be
\label{ftransverse}
p_M\hS^{FF}_{MN}(p)= \hS^{FF}_{MN}(p) (-p)_N =0.
\ee
Note that the argument of the 5-momentum is that of the 4-momentum inflow to the 
corresponding point,\footnote{{\it N.B.} $(-p)_M\ne -p_M$.} and by cyclicity 
the matrix is of course symmetric in this sense:
\be
\label{Ftranspose}
\hS^{FF}_{MN}(p)=\hS^{FF}_{NM}(-p).
\ee
Wine attachments to $D$ must now attach to $D\sigma$. The result is an
extra factor of $(-)^{1+f_\sigma}$, where $f_\sigma=0(1)$ if either side of 
the wine is bosonic (fermionic), as is clear from 
\fig{wineDs}.

Thus the fermionic effective propagators \eq{iBD} collect into:
\be
\label{ikFF}
\Delta^{FF}_{MN}(p) =\pmatrix{ \Delta^{BB}_p\delta_{\mu\nu} & 0\cr
0 & -\Delta^{DD}_p},
\ee
and $\dDelta^{FF}_{MN}$ is simply the differential of this, according to
\eq{defprop}. Apart from these extra factors (and the need to add
5-indices $M$ and $N$ under the flavour labels) 
the tensorial expansions \eq{wcv}
and \eq{Sex} map unchanged to five dimensional notation.
In this way, all equations relating to the fermionic sector
can be written in this language, where they take a compact form
very similar in appearance to the corresponding expressions for $A$ and $C$.
For example from \eq{SBDfl}, 
$\ldl\hS^{FF}_{MN} =\hS^{FF}_{MA}\dDelta^{FF}_{AB}\hS^{FF}_{BN}$,
\cf \eq{SCCfl} and \eq{SAAfl}, and from \eq{fermionwi}, broken fermionic
gauge transformations now map exclusively to lower point vertices as [compare
\eq{bosonwi}]:
\be
q^{N}U^{\cdots X F Y\cdots}_{\cdots\,a\, N\, b\,\cdots}(\cdots,
p,q,r,\cdots)=U^{\cdots X \ra{Y}\cdots}_{\cdots\ a\,  b\,\cdots}
(\cdots,p,q+r,\cdots)
-U^{\cdots \la{X} Y\cdots}_{\cdots\ a\,  b\,\cdots}
(\cdots, p+q,r,\cdots),
\label{Fwi}
\ee
where on the elementary fields $X=A, B, C, D$, $\La{X}=\Ra{X}={\hat X}$
is just the opposite statistics partner. Similarly, 
$\Ra{F}_M=(A_\mu,\,C\sigma)$, but $\Ra{F}_M=(A_\mu,-C\sigma)$, the extra sign 
being picked up by $\tau$ anticommuting through $\sigma$. As for 
\eq{fermionwi}, there are compensating signs on the right hand side for any 
other $\sigma$ that the momentum is pushed through. 

Finally, multiplying \eq{hSFF} and \eq{ikFF} one readily finds the
pair [related by transposition as in \eq{Ftranspose}]:
\bea
\label{iFFf}
\hS^{FF}_{RS}(p)\Delta^{FF}_{ST}(p)&=&\delta_{RT}-p'_R\, p_{\,T} \\
\label{iFFb}
\Delta^{FF}_{RS}(p)\hS^{FF}_{ST}(p)&=&\delta_{RT}-(-p)_R (-p)'_{\,T},
\eea
where introducing the useful shorthands
\be
\label{deffg}
f={\ct\over x\ct+2c}\qquad{\rm and}\qquad g={c\over x\ct+2c},
\ee
we define a dual 5-momentum $p'_M$ as follows
\be \label{defdual}
p'_M = (fp_\mu/\Lambda^2,\ -g).
\ee
Note that since $2g+xf=1$, we have $p'_Mp_M=1$, and thus \eq{iFFf} and
\eq{iFFb} are projectors onto the appropriate transverse space, just as
in \eq{iAA}. Furthermore, we see that since in practice $\Delta^{FF}$ will 
be connected to an $F$ point on some other vertex, the 
remainder from unity in \eqs{iFFf}{iFFb} always generates 
supergauge transformations via \eq{Fwi}.

\section{Guaranteeing universality of $\beta_1$ (and $\beta_2$)}
\label{Enforcing}

We now review in this context, the standard argument for why we should
expect to get the same value for $\beta_1$, and indeed $\beta_2$, in
the $\beta$ function \eq{betafn} as in other methods, despite the
fact that our renormalization scheme for $g(\Lambda)$ differs from that of 
the corresponding coupling $\gt(\mu\mapsto\Lambda)$ defined by these other 
methods. (We note that the
Gribov problem \cite{Gribov} which in truth invalidates these
methods since they proceed by gauge fixing, is not expected to alter
purely perturbative results.) 

In principle we can extract from \eq{Sloope}, 
by computing quantum corrections, the value of the other coupling as a 
function of ours, and thus match the two couplings perturbatively:
\be
\label{gt}
{1/\gt^2} = {1/g^2} +\gamma +O(g^2),
\ee
where the classical agreement is guaranteed by the standard normalisations
of the fields and kinetic term in \eq{defg}, after scaling $g$ back to its
usual position, and $\gamma$ is a one-loop matching coefficient. 
Differentiating with respect to $\Lambda$ and using \eq{betafn},
the corresponding $\beta$ function for $\gt$,  and \eq{gt}, we have
\be
\label{betas}
\bt_1 +\bt_2\, g^2 = \beta_1 +\beta_2 \,g^2 +\ldl\gamma +O(g^4).
\ee
Since $\gamma$ is dimensionless, it cannot depend upon $\Lambda$, there being
no other scale to form the necessary dimensionless combination. Thus
$\ldl\gamma=0$ in \eq{betas},  and we immediately recover the standard
facts that $\bt_1=\beta_1$ and $\bt_2=\beta_2$.

Clearly this argument fails if some other scale has been introduced,
for example the standard arbitrary finite physical scale $\mu$, or if
other running couplings get introduced. (After solving for their flows, \ie
solving their corresponding $\beta$ functions, this becomes equivalent
to the first failure since by dimensional transmutation a new finite
physical scale has been introduced.) Importantly, $\ldl\gamma$ can then have
an $O(g^2)$ one-loop contribution or in extreme cases even a tree-level
$O(g^0)$ contribution. From \eq{betas} one sees that a one-loop contribution
to the running of $\gamma$ destroys $\beta_2$ agreement, whilst a 
tree-level running would even modify $\beta_1$.

As we will see shortly, a generic $\hS$, including the simple form used for
the bare action in ref. \cite{sunn}, can lead to such tree-level
corrections. Fortunately, there is also an infinite class of seed actions
that cannot. As with the earlier constraints discussed,
since we never specify $\hS$,
it is not the solution that matters, only knowing that one exists.

To get agreement with the standard $\beta$ function at the two-loop level,
one needs to confirm that there are no further couplings hidden, that
run at one loop, and to take into account contributions from 
$g_2(\Lambda)$. This can be done \cite{us}.

Even with a non-vanishing $\ldl\gamma$, one could still recover 
the usual $\beta$ function coefficients, by defining a standard low 
energy --or infrared-- coupling $\gt(\mu)$ at some scale $\mu<\Lambda$,  
this coupling being distinguished from the `ultraviolet' coupling 
$g(\Lambda)$ in the effective action 
$S_\Lambda$ \cite{bonini,litim}. We want to avoid this because the
introduction of $\mu$  would destroy, or at least obscure, the
power and elegance of self-similarity \cite{Shirkov} (\cf \sec{Necessary}).

\subsection{Ensuring no running couplings at tree level}
\label{Ensuring}

The incorporation of Pauli-Villars type fields directly into an exact RG
causes some novel classical divergences in the $\Lambda$ integrals defining
the classical vertices,
just as they did in ref. \cite{ymii,alg}. They can be 
cancelled by appropriate choices of integration constant. However, 
generically this results in introducing another finite scale $\mu$, even at
the classical level, again just as it did in ref. \cite{ymii}. The 
resulting loss of self-similarity leads to non-universal contributions
creeping in at a particular point in the calculation of $\beta_1$ that 
follows. Indeed, we will see there that it is precisely the classical
dependence on $\mu$ that causes the problem.

To show how this arises, and how we can avoid it, consider the classical
$CCA$ vertex. This is one of a number of affected vertices that form
part of the $\beta_1$ calculation. The flow follows from eq. \eq{ergcl},
and is given by the first four diagrams on the right hand side of \fig{aff}
after replacing the star by an open circle. The resulting equation,
up to changes of notation, 
is precisely equivalent to the $CCA$ vertex in ref.\cite{ymii}:
\bea
\label{SACC}
S^{\ph0ACC}_{0\,\mu}(p,q,r) =
&&- \int^{\Lambda_0}_\Lambda{d\Lambda_1\over\Lambda_1}\,\left\{ 
\hS^{ACC}_\mu(p,q,r)
\left[\dDelta^{CC}_q\hS^{CC}_q+\dDelta^{CC}_r\hS^{CC}_r\right]
\right.\nonumber\\
&&\left.\ph{- \int^{\Lambda_0}_\Lambda{d\Lambda_1\over\Lambda_1}\quad}
+\hS^{ACC}_\alpha(p,q,r)\dDelta^{AA}_p\hS^{AA}_{\alpha\mu}(p)
+\hS^{CC}_r\dDelta^{A,CC}_\mu(p;r,q)\hS^{CC}_q\right\}\nonumber\\
&&\ph{- \int^{\Lambda_0}_\Lambda}
+\Lambda_0^2(q-r)_\mu+\ct'_0(q^2+r^2)(r-q)_\mu
+\gamma^{ACC}q_\alpha\Box_{\alpha\mu}(p),
\eea
where all the
terms inside the curly brackets are to be understood as being functions
of $\Lambda_1$ (not $\Lambda$), and $\ct'_0\equiv\ct'(0)$.   
Here we have recognized that we can integrate the flow equation immediately
with respect to $\Lambda$, but to make explicit the divergences we have
replaced the upper limit by $\Lambda_0$. The integration constant ensures
however that the complete expression is finite, so the continuum limit
$\Lambda_0\to\infty$ can actually be safely taken.

The first two terms in the integration constant, are forced by 
gauge invariance \cite{ymii}. Indeed, setting $\Lambda=\Lambda_0$ in 
\eq{SACC}, we see that the integration constant is nothing but the $ACC$
vertex of the classical bare action $S_0 |_{\Lambda=\Lambda_0}$. Its
longitudinal part follows from \eq{bosonwi} and \eq{hSCC} which implies, 
\be
p^\mu S^{\ph0ACC}_{0\,\mu}(p,q,r) = \hS^{CC}_r -\hS^{CC}_q.
\ee
This equation is readily solved 
at $\Lambda=\Lambda_0$, by expanding both sides as a power series in 
$\Lambda_0$, and noting that all negative powers can be 
discarded.\footnote{The coefficients are
purely local, \ie polynomials in momenta, with determined dimension, 
\cf \sec{Necessary}.}
Equivalently and more simply, the longitudinal terms follow from any 
covariantisation of \eq{hSCC}, \eg
\be
{1\over2}\,\str\!\int\!\!d^4x\ C\,
\{\,2\lambda\Lambda_0^4 -\Lambda_0^2\, D_\mu^2+\ct'_0 [D_\mu^2]^2
+O(1/\Lambda_0^2)\,\}\,C \quad +\cdots
\ee
(Recall that $D_\mu$ is $\partial_\mu-iA_\mu$ and acts by commutation.
The ellipsis refers to terms not containing $CCA$ vertices.)

The final term in the integration constant in \eq{SACC}, 
$q_\alpha\Box_{\alpha\mu}(p)$, is the unique
transverse combination that is allowed by dimensions (\ie is not accompanied
by a negative power of $\Lambda_0$, and importantly satisfies all the other
symmetries specifically charge conjugation and no-$\A^0$ symmetry), and as 
such has an undetermined (dimensionless)
coefficient: $\gamma^{ACC}$. The fact that it is undetermined does not
matter: the whole calculation is independent of such details. 
However, by the same token $\zeta^{ACC}$, the coefficient of this momentum 
term in the small momentum (\aka derivative) expansion of the 
integrand in \eq{SACC}, is also dimensionless, thus independent of 
$\Lambda_1$, and thus yields $-\zeta^{ACC}\ln(\Lambda_0/\Lambda)$ on 
integration. In order to ensure the finiteness of \eq{SACC}, we are forced 
to introduce a new finite physical scale by including 
$\zeta^{ACC}\ln(\Lambda_0/\mu)$ in the integration constant $\gamma^{ACC}$.

An alternative to introducing $\mu$ directly, is to allow $\gamma^{ACC}$
to become a logarithmically running coupling $\gamma^{ACC}(\Lambda)$ 
at the classical level, correcting the flow to take account of its 
$\beta$ function. However as we explained in the previous section, this is in 
the end equivalent and still results in the loss of universality for 
$\beta_1$.

In general we see that we may be forced to introduce $\mu$ at the classical
level wherever a purely local vertex with dimensionless coupling can be 
constructed, which is transverse, in the sense that it is unrelated to lower 
point vertices via  either bosonic gauge invariance \eq{bosonwi}, or the 
broken fermionic gauge invariance \eq{fermionwi}. 

We note that the problem is associated
with the Pauli-Villars sector because these terms necessarily have a
divergent {classical} action as $\Lambda\to\infty$, at least in so far
as they have divergent masses. There is a problem with this only for
the generated {\sl logarithmic} divergences along the marginal directions
however, whose cancellation 
necessarily requires introducing a new finite physical scale.
There are however infinitely many of these directions because we can have
any number of $C$ points, since $C$ is dimensionless.
 
The solution is to tune the corresponding terms in $\hS$. 
Indeed by noting from \eq{invCC}, that $\dDelta^{CC}\Delta^{CC}$ $=$
$\ldl\ln\hS^{CC}$ and thus equals 4 at zero momentum, we get from \eq{SACC}
that the shift 
\be
\hS^{ACC}_\mu(p,q,r)\mapsto\hS^{ACC}_\mu-{1\over8}\,
\zeta^{ACC}q_\alpha\Box_{\alpha\mu}(p)
\ee
precisely cancels the coefficient of $q_\alpha\Box_{\alpha\mu}(p)$ 
in the integrand, thus removing the logarithmic divergence from the
$\Lambda_1$ integral.

Since the structure of the classical flow equations \eq{ergcl} 
is such, that the flow of every vertex 
$S^{\ph0X_1\cdots X_n}_{0\,a_1\,\cdots\,a_n}$ has the corresponding
$\hS^{X_1\cdots X_n}_{\,a_1\,\cdots\,a_n}$ as its highest-point $\hS$
contribution,
contracted via kernels into all the 
appropriate two point vertices (\viz $\dDelta^{X_iX_i}\hS^{X_iX_i}$) 
\cite{ymi,ymii}, and since these $\dDelta^{X_iX_i}\hS^{X_iX_i}$ terms are 
non-vanishing at zero momentum precisely when $X_i$ is a 
massive Pauli-Villars field, 
it follows that we can always remove the divergence associated with 
these marginal directions by tuning 
$\hS^{X_1\cdots X_n}_{\,a_1\,\cdots\,a_n}$ in the same direction.

In this way, we completely avoid introducing $\mu$ (equivalently
marginal running couplings) at the classical level.

\section{The calculation}
\label{calculation}
The whole of the paper up to this point has been concerned with setting
up and justifying the formalism we will now use. We can now finally
turn to computation itself.

\begin{figure}[h!]
\begin{center}
\includegraphics[scale=.4]{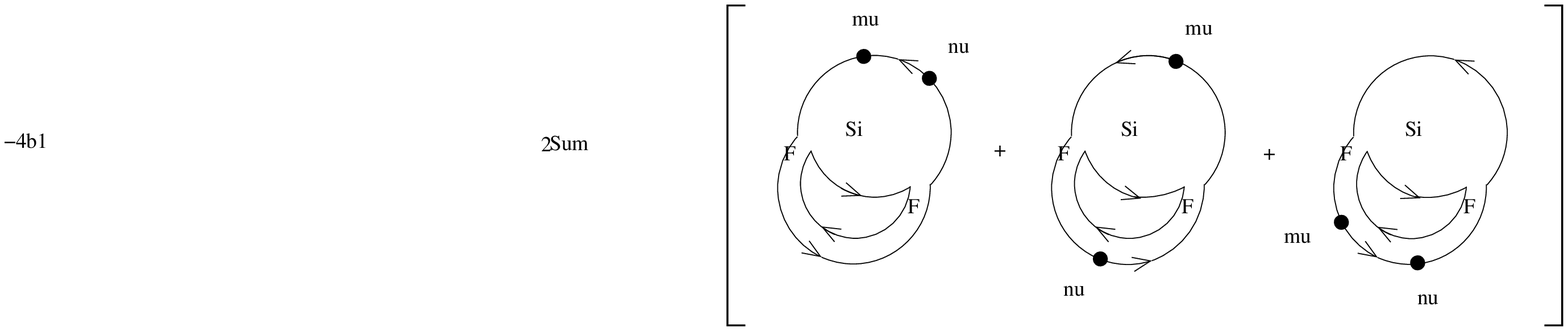}
\caption{Graphical representation of the equation for 
$\beta_1$}\label{beta1}
\end{center}
\end{figure}

The renormalization condition \eq{defg} constrains the two-$A$ point
vertices, and by \eq{hSAA} and \eq{SAA}, this constraint is already 
saturated at tree level:
\be
S^{AA}_{\mu\nu}(p) + S^{AA\sigma}_{\mu\nu}(p) =
{2\over g^2}\Box_{\mu\nu}(p)+O(p^3) =
{1\over g^2}S^{\ph0AA}_{0\,\mu\nu}(p)+O(p^3).
\ee
It follows that all higher loop contibutions, 
$S^{\ph{n}AA}_{n\,\mu\nu}(p) + S^{\ph{n}AA\sigma}_{n\,\mu\nu}(p)$,
must vanish at $O(p^2)$. From the discussion below \eq{z2}, the one-loop
contribution is purely of form $S^{\ph1AA\sigma}_{1\,\mu\nu}$.
(Thus it is already clear that $g_2$ has the opposite sign $\beta_1$,
consistent with the wrong sign action in \eq{defg} \cite{ymii}. 
This will be fully developed in ref. \cite{us}.)
Specialising the one-loop flow equation
\eq{ergone} to two $A$s and $O(p^2)$, and using \eq{SAA}, we see that
it collapses to the purely algebraic relation \cite{ymi,ymii}:
\be
-2\beta_1S^{\ph0AA}_{0\,\mu\nu}(p)+O(p^3) =
a_1[\Sigma_0]^{AA\sigma}_{\mu\nu}(p).
\ee
($\Sigma_0=S_0-2\hS$.) Diagrammatically, this takes the form 
of \fig{beta1}, after including the factor $2$ from 
the two different supertraces that $AA$ can go into
(\cf end of \sec{DiagrammaticI}), and a factor $2$
from adding the contribution $p_\mu\leftrightarrow -p_\nu$ (which is
equal by Lorentz invariance. Whenever such terms arise we will typically
combine them.)

\psfrag{O(p3)}{}
\psfrag{=}{}
\psfrag{mu}{$\mu$}
\psfrag{nu}{$\nu$}
\psfrag{Si}{$\Sigma_0$} 
\psfrag{F}{\tiny$f$}
\psfrag{+}{$+$}
\psfrag{S0}{$S_0$}
\psfrag{Sum}{$\displaystyle \sum_{f=A,C,F}$}  
\psfrag{-4b1}{$-4\beta_1 \Box_{\mu \nu} (p) + O(p^3) = 2$}
\psfrag{(p)}{}
\psfrag{2}{}  

Performing the $\sigma$ algebra as in \fig{partialAttach}, we thus find
\bea 
\label{oneloop}
-4\beta_1\Box_{\mu\nu}(p) &&+O(p^3) = 
2N \int{d^4k\over(2\pi)^4}\Big\{  \nonumber\\
&&\ph+\dDelta^{CC}_k\Sigma^{\ph0AACC}_{0\,\mu\nu}(p,-p,k,-k)
+\dDelta^{A,CC}_\mu(p;-k-p,k)\Sigma^{\ph0ACC}_{0\,\nu}(-p,p+k,-k)\nonumber\\
&&\qquad+\dDelta^{AA,CC}_{\,\mu\,\nu}(p,-p;-k,k)\Sigma^{CC}_0(k)\nonumber\\
&&+\dDelta^{AA}_k\Sigma^{\ph0AAAA}_{0\,\alpha\,\alpha\,\mu\,\nu}(k,-k,p,-p)
+\dDelta^{A,AA}_\mu(p;-k-p,k)
  \Sigma^{\ph0AAA}_{0\,\nu\,\alpha\,\alpha}(-p,p+k,-k)\nonumber\\
&&\qquad+\dDelta^{AA,AA}_{\,\mu\,\nu}(p,-p;-k,k)
   \Sigma^{\ph0AA}_{0\,\alpha\,\alpha}(k)\nonumber\\
&&-\dDelta^{FF}_{SR}(k)\Sigma^{\ph0AAFF}_{0\,\mu\nu RS}(p,-p,k,-k)
-\dDelta^{A,FF}_{\mu,SR}(p;-k-p,k)
  \Sigma^{\ph0AFF}_{0\,\nu RS}(-p,p+k,-k)\nonumber\\
&&\qquad-\dDelta^{AA,FF}_{\,\mu\,\nu,SR}(p,-p;-k,k)
   \,\Sigma^{\ph0\,FF}_{0\,RS}(k)\Big\}.
\eea

Although extra $\sigma$s appear via \fig{partialAttach}, the part that 
contributes from the $\sigma$ trapped in the empty supertrace (thus giving 
$\str\,\sigma=2N$) ultimately comes from the breaking of $SU(N|N)$,
\ie from shifting $\C$,
since otherwise the fermionic and bosonic attachments would just give
equal and opposite $\sigma$ contributions in \fig{partialAttach},
combining to give a full supermatrix differential.
Indeed at high momentum $k$, 
exact $SU(N|N)$ invariance is recovered, resulting in regularisation
of \eq{oneloop}, since
the $F$ sector then cancels the $A,C$ sector. 

Recall that we exclude diagrams where the wine bites its own tail, as in
\eq{noTailBiting}. In fact such terms vanish for $\beta_1$ in any case,
since the attachments are via a full $\A$ or $\C$ in \eq{wev} with no
possibility of trapping an extra $\sigma$, thus yielding $\str\,\one=0$.

The $\beta_1$ computation splits into one-loop contributions from the
three sectors $C,A,$ and $F$, each of which appears in \eq{oneloop} in
almost identical form. Thus apart from the sign, we get the $A$ sector
terms from the $F$ sector simply by replacing $F$ by $A$, and $R,S$
by $\alpha,\beta$, recognizing that the wines just have
$\delta_{\alpha\beta}$ as a factor. Similarly, we get the $C$ contribution
from the $F$ contribution by $F\mapsto C$, dropping $S$ and $R$ altogether.

\subsection{Diagrammatic analysis}
\label{DiagrammaticA}

\begin{figure}[h!]
\begin{center}
\includegraphics[scale=.5]{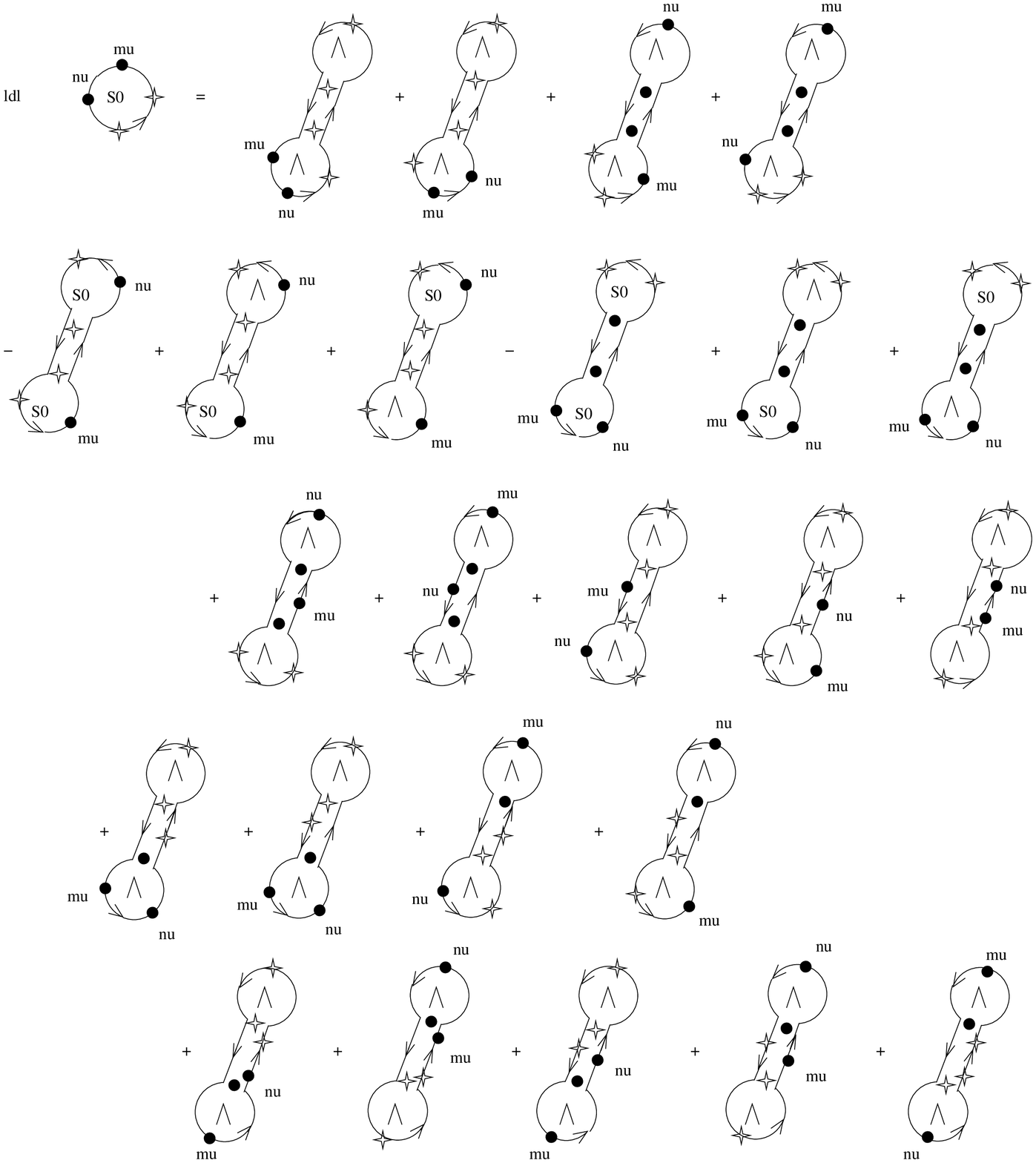}
\caption{Flow of the tree-level $AAFF$ vertex.
$F$ is represented by the star.}
\label{aaff}
\end{center}
\end{figure}

\begin{figure}[h!]
\begin{center}
\includegraphics[scale=.37]{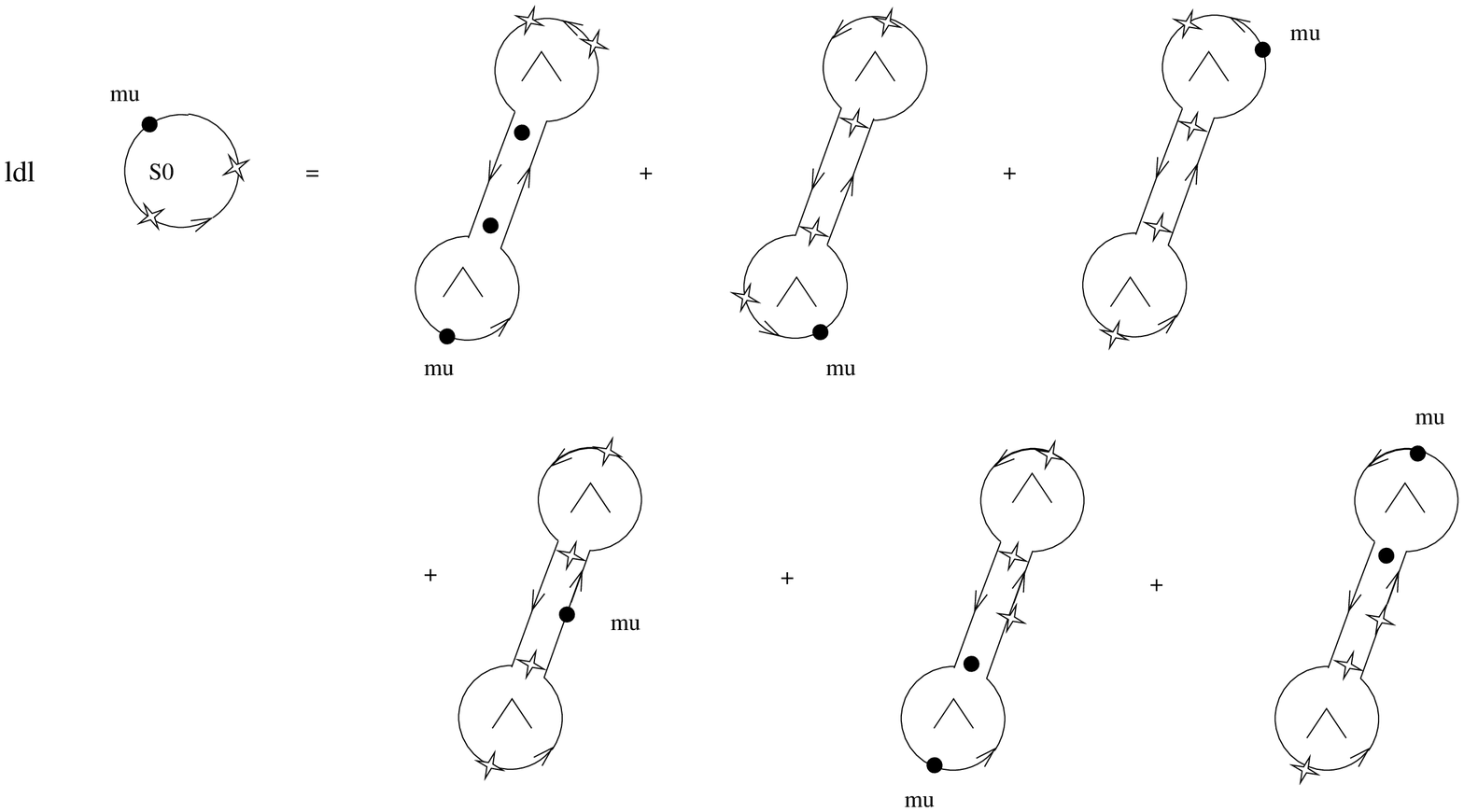}
\caption{Flow equation for the tree-level $FFA$ vertex}
\label{aff}
\end{center}
\end{figure}

The similarity goes deeper. The $F$ type
classical four-point vertex in \eq{oneloop}
is determined by the flow equation given diagrammatically in 
\fig{aaff}, whilst the $F$ type classical three-point vertex (appearing in 
\eq{oneloop} and \fig{aaff})  is expressed through the flow equation of 
\fig{aff}. The corresponding $A$ ($C$) sector diagrams are given simply by 
replacing the star with a filled (empty) circle! 
\psfrag{=}{=}
\psfrag{mu}{$\mu$}
\psfrag{nu}{$\nu$}
\psfrag{+}{$+$}
\psfrag{-}{$-$}
\psfrag{S0}{$S_0$}
\psfrag{ldl}{$\Lambda\partial_{\Lambda}$}

This works 
because all cyclically allowed configurations of the external fields appear 
in \fig{aaff} and \fig{aff}, after which the flavour of the point the
wine attaches to, is uniquely determined by the requirement that each
component vertex has an even number of stars, so as to be overall bosonic. 
For the $A$ sector the fact that there is just one wine attachment 
for each external configuration is enough to ensure the mapping works. 
For the $C$ sector,
the symmetry \eq{z2} ensures that each vertex also has an even number
of open circles, which thus go uniquely where the stars had gone before.
There are a couple of provisos however. Firstly, all the wines which attach
via an $F$ at one end and an $A$ at the other, map under $F\mapsto C$,
to wines that do not exist in \eq{a0}. This does not matter: we can simply 
assign them a zero value, and carry these vanishing
terms through the computation. (There is no fundamental reason for their 
non-appearance: recall that we exchanged such terms for simpler terms using 
\eq{gaugeS}. Also note that actually the $F$ attachment here is exclusively
a $B$. When we take this into account, it will be clear that the term only
has a mapping to the $A$ sector.) Secondly \eq{z2} in fact
also allows an odd number of $C$'s per classical vertex in the $C$ sector, 
providing that the vertex also has a $\sigma$. But in the case of
the wines, these terms vanish by \eq{wev} since the $\sigma$ commutes with the 
bosonic $A$ or $C$ derivative, whilst the only action vertex that could 
contribute is $AAC\sigma$ ($AC\sigma$ being already excluded by symmetries 
\cf \sec{Seed}, and all other possibilities being too high order). However, by 
charge conjugation invariance
\be
S^{AAC\sigma}_{\mu\nu}(p,q,r)=S^{AAC\sigma}_{\nu\mu}(q,p,r),
\ee
while so long as we insist on a single supertrace,
by no-$\A^0$ symmetry:
\be
\label{noAreln}
S^{AAC\sigma}_{\mu\nu}(p,q,r)+S^{AAC\sigma}_{\nu\mu}(q,p,r)=0,
\ee
thus the vertex actually vanishes. (Allowing multiple supertrace
terms, it can be shown that the part contributing to $\beta_1$
vanishes after ensuring no running classical couplings as in \sec{Ensuring}.)

We can map the effective propagator relations \eqs{iFFf}{iFFb} in an obvious 
way to the corresponding relation for $A$, \viz \eq{iAA}, and $C$,
\viz \eq{iCC}, leaving the gauge remainder terms in the case of $F$ and $A$
till last (where the $A$ sector expression follows from the map
$k_N\mapsto k_\nu$, $k'_M\mapsto k_\mu/k^2$, and of course in the $C$
sector these terms map to zero).
This all means that we have an added bonus: except for
these final stages, we need only present the $F$ sector computation and
then just map the result to the other two sectors.

\begin{figure}[h!]
\begin{center}
\includegraphics[scale=.4]{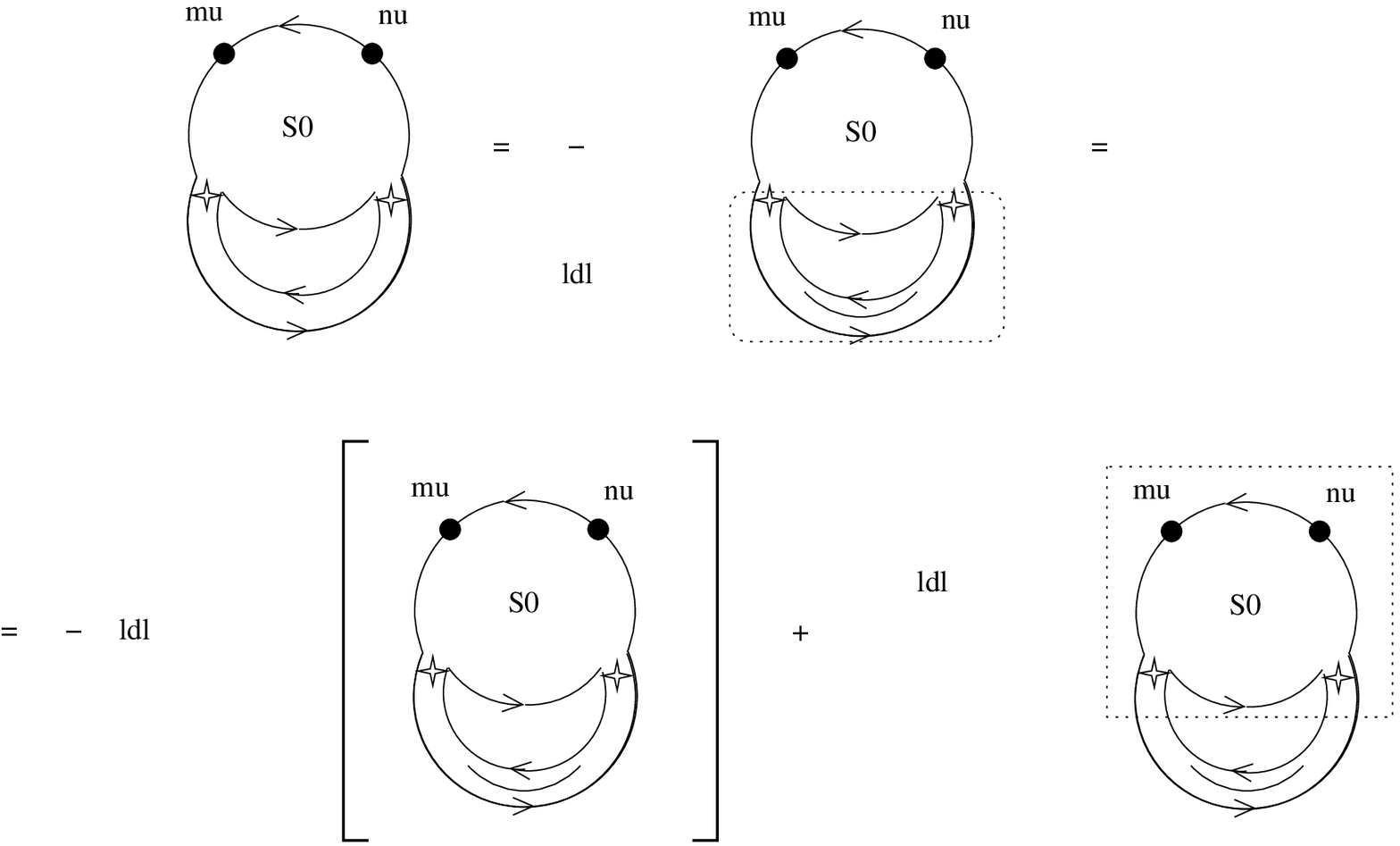}
\caption{The first step in the calculation of $\beta_1$.}
\label{trick1f}
\end{center}
\end{figure}

Substituting $\Sigma_0=S_0-2\hS$ in \fig{beta1}, we start with the highest 
point vertex and convert the $S_0$ part to a total $\Lambda$ derivative and 
remainder, using 
\eq{defprop}:
\bea
\label{trick1}
\dDelta^{FF}_{SR}(k)\,S^{\ph0AAFF}_{0\,\mu\nu RS}(p,-p,k,-k)
=&&-\ldl\left[\Delta^{FF}_{SR}(k)
\,S^{\ph0AAFF}_{0\,\mu\nu RS}(p,-p,k,-k)\right]\nonumber\\
&&+\Delta^{FF}_{SR}(k)\,\ldl S^{\ph0AAFF}_{0\,\mu\nu RS}(p,-p,k,-k),
\eea
as shown diagrammatically in \fig{trick1f}. 

\psfrag{S0}{$\!\! S_0$}
\psfrag{Si}{$\Sigma_0$}
\psfrag{=}{$=$}
\psfrag{mu}{$\mu$}
\psfrag{nu}{$\nu$}
\psfrag{ldl}{$\ldl$}
\psfrag{+}{$+$}  
\psfrag{-}{$-$}   

In the first term, we will put the $\ldl$ outside the $k$ integral. If the
integral were convergent, the part we want, namely the $O(p^2)$ part, would then
vanish since it is a dimensionless function of the only scale $\Lambda$. After 
adding in the $A$ and $C$ sectors, the integral will only have an infrared 
divergence, which $\ldl$ converts to a universal contribution. 

\begin{figure}[h!]
\begin{center}
\includegraphics[scale=.45]{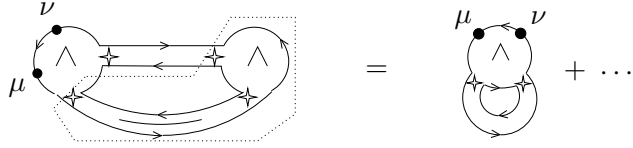}
\caption{One of the pair cancelling $-2\hS^{AAFF}$ in fig. 11. The ellipsis are the gauge remainder terms.}\label{twof}
\end{center}
\end{figure}

In the second 
term we can now exchange the four-point vertex for lower point vertices via 
\fig{aaff}: the generated four-point $\hS$ vertices must cancel the 
$-2\hS^{AAFF}_{\mu\nu\, RS}$ in \eq{oneloop} in order for the result to be 
universal. This must be so because $\hS^{AAFF}$ can contain
arbitrary transverse terms, which thus have no relation to lower point 
vertices. Such a cancellation is precisely what we find from the first two 
diagrams in \fig{aaff}, on using \eq{iFFb}, 
as we can see from \fig{twof}. 
\psfrag{=}{$=$}
\psfrag{mu}{$\mu$}
\psfrag{nu}{$\nu$}
\psfrag{ldl}{$\ldl$}
\psfrag{+}{$+$}  
\psfrag{-}{$-$}   
\psfrag{R34}{$\cdots$}
The remainder term from \eq{iFFb} generates a gauge transformation via
\eq{Fwi} and thus also maps to lower point vertices. We will return to
these gauge remainder terms later.

\begin{figure}[h!]
\begin{center}
\includegraphics[scale=.35]{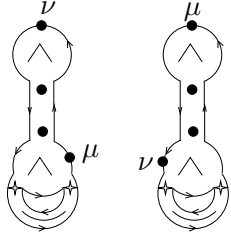}
\caption{The two remaining terms containing a four-point vertex.}
\label{onef}
\end{center}
\end{figure}

We are still left with two terms containing a four-point vertex, that
of \fig{onef}, as generated by the next two diagrams in \fig{aaff}. 
However now note,
as we will frequently, that the 
two-point $A$ vertex is already $O(p^2)$, by gauge invariance \cf \eq{hSAA}. 
Therefore the only 
part that can contribute to \eq{oneloop} is where $p$ is set to zero in
the rest of the expression. After using Lorentz invariance to exchange
$p_\mu$ and $-p_\nu$ in the second diagram, this 
results in a contribution
\be
\label{here}
-\Delta^{FF}_{SR}(k)\left[\hS^{AAFF}_{\mu\,\alpha\,RS}(0,0,k,-k)+
\hS^{AAFF}_{\alpha\,\mu\,RS}(0,0,k,-k)\right]\dDelta^{AA}_0
\hS^{AA}_{\alpha\nu}(p)
\ee
to the integrand in \eq{oneloop}. As soon as we have an $A$-point with zero
momentum, we can simplify via gauge invariance. Using \eq{bosonwi}
twice over, we have:
\be
\hS^{AAFF}_{\mu\,\nu\,RS}(-\epsilon,\epsilon,k,-k)\epsilon^\mu\epsilon^\nu
=\hS^{FF}_{RS}(k+\epsilon)-\hS^{AFF}_{\mu\,RS}(0,k,-k)\epsilon^\mu
-\hS^{FF}_{RS}(k).
\ee
Thus, Taylor expanding to $O(\epsilon^2)$, we determine the symmetric part:
\be
\label{00wi}
\hS^{AAFF}_{\mu\,\nu\,RS}(0,0,k,-k)+ \hS^{AAFF}_{\nu\,\mu\,RS}(0,0,k,-k)
= \partial^k_\mu\partial^k_\nu\hS^{FF}_{RS}(k).
\ee
Substituting this in \eq{here} we reduce \fig{onef} to an expression
depending only on seed action two-point vertices and their associated
zero-point kernels (integrated or otherwise). We will refer to such terms
as `potentially universal',
since the seed action two-point vertices and the kernels derived from them
are the only things that we have 
explicitly prescribed. For the result to be universal,
it must be that we can reduce everything to such potentially universal
terms or to total $\Lambda$ derivatives as in \eq{trick1}. In turn,
potentially universal terms must, and do, collect into total $k$
derivatives, whose boundary terms on integration,
are universal as a result of 
restrictions on the large momentum behaviour, \eg \eq{inequalities},
and the renormalization condition \eq{defg}. (Actually, 
since $\dDelta^{AA}_0\propto c'_0$, by \eq{kAA}, and $1/c'_0$ is
never produced, terms such as \eq{here} are universal only because they
combine to give boundary terms that vanish.)

Proceeding with the remaining terms in \fig{aaff},
we generate many further reductions similar to the ones above.

\begin{figure}[h!]
\begin{center}
\includegraphics[scale=.4]{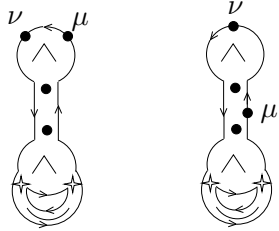}
\caption{These diagrams do not contribute at $O(p^2)$}
\label{threef}
\end{center}
\end{figure}

There are eight terms that immediately have vanishing $O(p^2)$ component.
Two examples are shown in \fig{threef}. By Lorentz invariance,
the first diagram has only odd powers of $p$, its dependence coming from
$\hS^{AAA}_{\mu\nu\alpha}(p,-p,0)$, whilst the second
diagram is also too high order since $\dDelta^{A,AA}_\mu(0;0,0)=0$.
Finally, the last term from \fig{aaff} is obviously $O(p^4)$.

\begin{figure}[h!]
\begin{center}
\includegraphics[scale=.45]{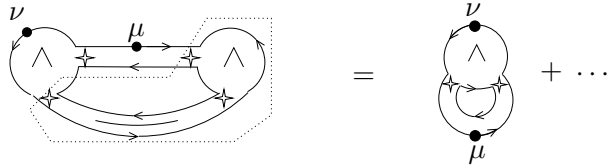}
\caption{One of the pair cancelling $-2\hS^{AFF}$ in fig. 11.}
\label{fourf}
\end{center}
\end{figure}

\begin{figure}[h!]
\begin{center}
\includegraphics[scale=.4]{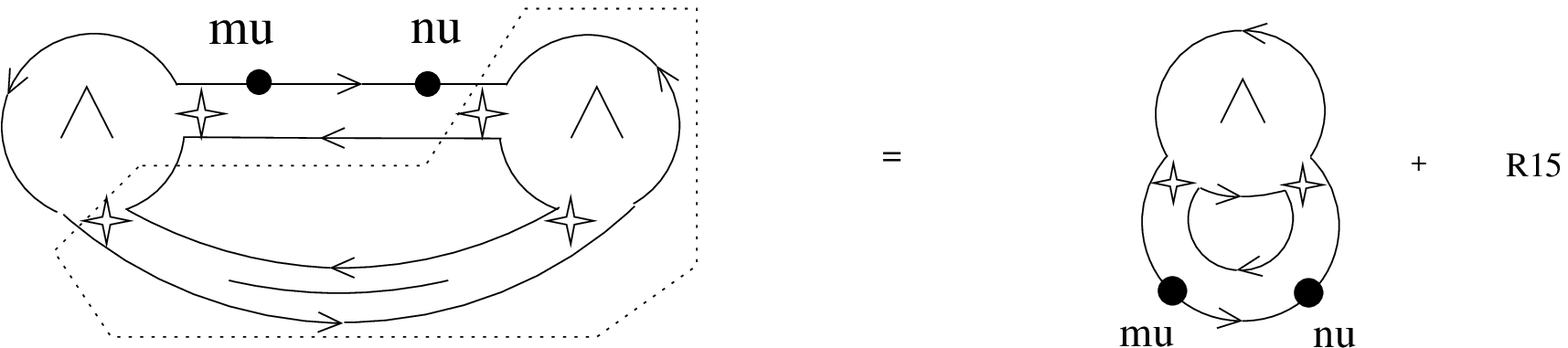}
\caption{Cancellation of $-\hS^{FF}$ in fig. 11.}
\label{fivef}
\end{center}
\end{figure}

After using \eqs{iFFf}{iFFb}, a mirror pair (\ie related by mirror
reflection, equivalent to charge conjugation) cancel the seed action part of 
$\Sigma_0^{AFF}$, as shown in \fig{fourf}, generating further gauge remainder
terms.
\psfrag{=}{$=$}
\psfrag{mu}{$\nu$} 
\psfrag{nu}{$\mu$} 
\psfrag{ldl}{$\ldl$}
\psfrag{+}{$+$}  
\psfrag{-}{$-$}   
\psfrag{R1314}{$\cdots$}
One term cancels the final $\Sigma_0^{FF}=-\hS^{FF}$ 
term in \eq{oneloop}, as shown in \fig{fivef}. 
Actually, here there is no point in carrying forward
the gauge remainder, since it is clear by \eq{iFFb} and \eq{ftransverse},
and by \eq{iAA} and \eq{transverse}, that it vanishes for both $F$ and $A$
sectors.
\psfrag{=}{$=$}
\psfrag{mu}{$\mu$}
\psfrag{nu}{$\nu$} 
\psfrag{+}{$+$}  
\psfrag{-}{$-$}   
\psfrag{R15}{$\cdots$}

\begin{figure}[h!]
\begin{center}
\includegraphics[scale=.42]{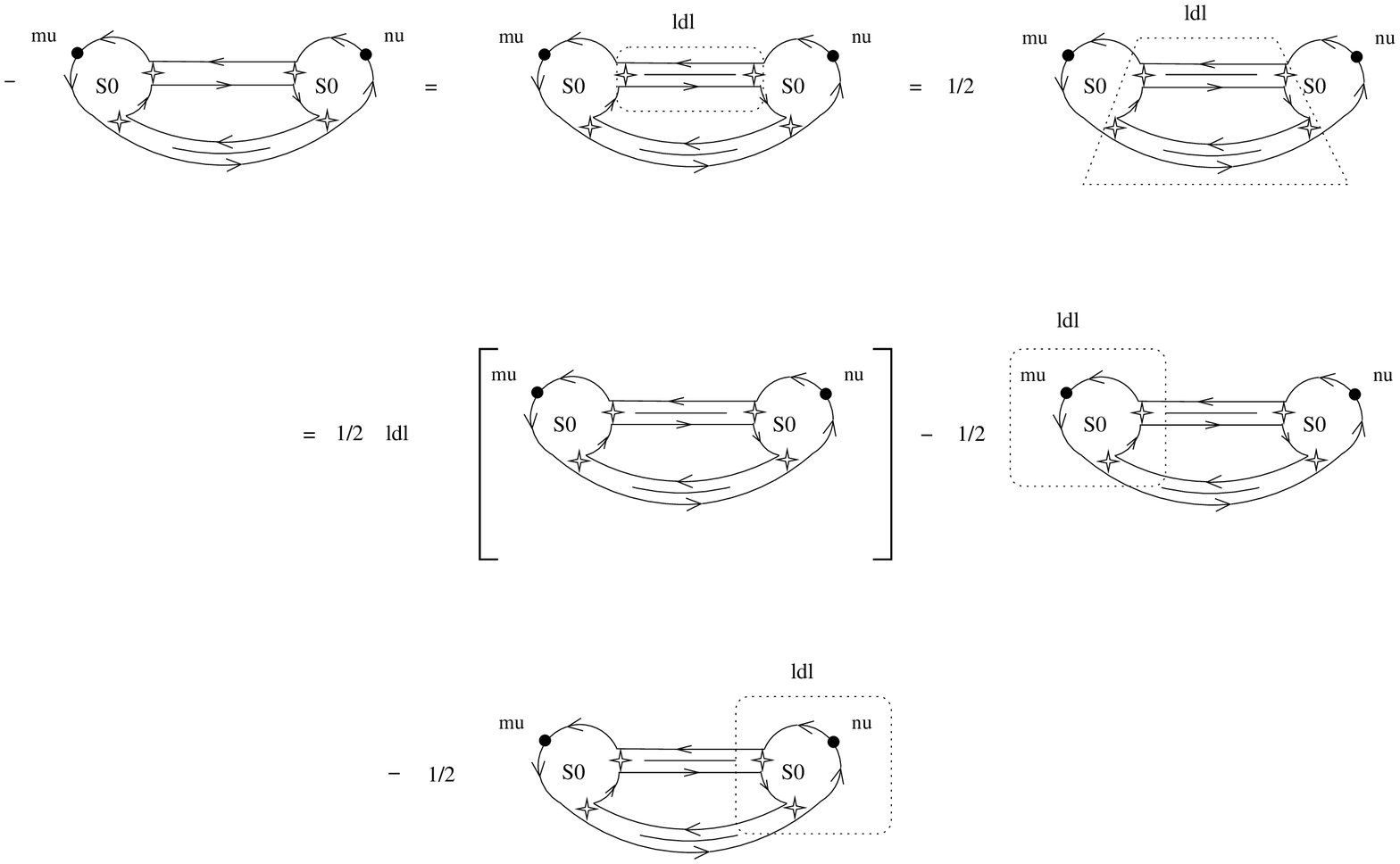}
\caption{Evaluating the `smiling frog' diagrams.}\label{trick2f}
\end{center}
\end{figure}

From the second line of \fig{aaff}, there are three terms generated where the
integrated kernel attaches to two different three-point vertices.
These are either both $S_0$ vertices, or one $S_0$ vertex and one $\hS$ vertex.
They simplify after introducing a second integrated kernel into the
$S_0$--$S_0$ term, as in \fig{trick2f} (the third step following
after using $p_\mu\leftrightarrow-p_\nu$ and relabelling the loop momentum).

The total $\Lambda$ derivative will be
considered along with that of \fig{trick1f}. The other two terms are 
evaluated by substituting \fig{aff} and its $p_\mu\mapsto-p_\nu$ partner.
\psfrag{=}{$=$}
\psfrag{mu}{$\nu$} 
\psfrag{nu}{$\mu$} 
\psfrag{+}{$+$}  
\psfrag{-}{$-$} 
\psfrag{S0}{\small $\! S_0$}
\psfrag{ldl}{$\ldl$}
\psfrag{1/2}{$\, \frac{1}{2}$}

\begin{figure}[h!]
\begin{center}
\includegraphics[scale=.5]{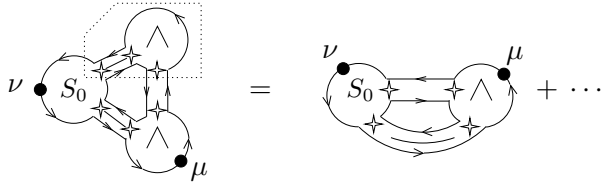}
\caption{Cancellation of the `winking smiling frog' diagrams.}
\label{term1}
\end{center}
\end{figure}

\begin{figure}[h!]
\begin{center}
\includegraphics[scale=.5]{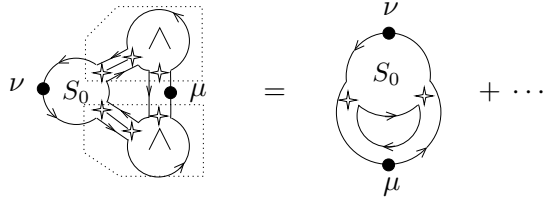}
\caption{One of a pair cancelling the $S_0^{AFF}$ term in fig. 11.}
\label{term2}
\end{center}
\end{figure}

\begin{figure}[h!]
\begin{center}
\includegraphics[scale=.45]{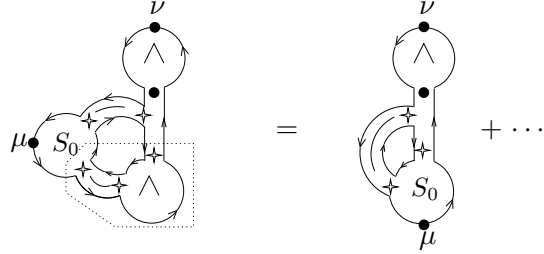}
\caption{Further diagrams from fig. 13.} 
\label{term3}
\end{center}
\end{figure}

\begin{figure}[h!]
\begin{center}
\includegraphics[scale=.45]{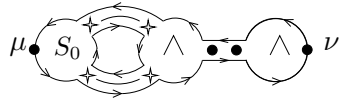}
\caption{A potentially universal term generated by fig. 13.}
\label{terms}
\end{center}
\end{figure}

Again, making use of the effective
propagator relations \eqs{iFFf}{iFFb}, many cancellations occur.
Thus the last two terms on the first line of \fig{aff} (and its 
$p_\mu\mapsto-p_\nu$ partner) result in
cancelling the mixed $\hS$--$S_0$ terms as illustrated in \fig{term1}.
Moreover, the $S_0$ part of $\Sigma_0^{AFF}$, the last term left
in \eq{oneloop}, is cancelled by the first term on the second line
of \fig{aff}, as can be seen from \fig{term2} (after adding the 
$p_\mu\mapsto-p_\nu$ partner). 
\psfrag{=}{$=$}
\psfrag{mu}{$\nu$} 
\psfrag{nu}{$\mu$} 
\psfrag{+}{$+$}  
\psfrag{-}{$-$} 
\psfrag{S0}{\small $\! S_0$}
\psfrag{ldl}{$\ldl$}
\psfrag{R1}{$\,\,\cdots$}
\psfrag{=}{$=$}
\psfrag{mu}{$\nu$} 
\psfrag{nu}{$\mu$} 
\psfrag{+}{$+$}  
\psfrag{-}{$-$} 
\psfrag{S0}{\small $\! S_0$}
\psfrag{ldl}{$\ldl$}
\psfrag{R2}{$\,\,\cdots$}
The last two terms in \fig{aff} result in diagrams of the form shown in
\fig{term3}. These cancel two corresponding terms generated by \fig{aaff},
where the $S_0$ three-point vertex is replaced by the
$\hS$ three-point vertex. The cancellation occurs because the top lobe
in \fig{term3} is already $O(p^2)$, and thus in the bottom lobe, by
\eq{bosonwi} and equality of $S_0$ and $\hS$ two-points, 
what actually counts is 
\be
\label{gthree}
S^{\ph0AFF}_{0\,\mu RS}(0,k,-k)=\partial^k_\mu\hS^{FF}_{RS}(k)
=\hS^{AFF}_{\mu\,RS}(0,k,-k).
\ee
\psfrag{=}{$=$}
\psfrag{mu}{$\mu$} 
\psfrag{nu}{$\nu$}
\psfrag{+}{$+$}  
\psfrag{-}{$-$} 
\psfrag{S0}{\small $\! S_0$}
\psfrag{ldl}{$\ldl$}
\psfrag{R3}{$\,\,\cdots$}
After combining via $p_\mu\leftrightarrow-p_\nu$,
we are left with just the diagram generated by the first term in \fig{aff},
namely \fig{terms}, all the other terms from \fig{aff} having been dealt with. 
It is easy to see by gauge invariance that this is potentially universal,
\ie depends only on two-point seed action vertices and the zero-point
kernels. (In fact, just as in \fig{onef}, this contains $c'_0$ and must
combine to give a vanishing contribution.)

The only remaining terms to be processed are those generated by the
first four diagrams on the last line of \fig{aaff}. We easily see however
that up to the gauge remainders which as above we set aside, 
all these correspond to wine-biting-their-tail
diagrams and are thus annihilated by \eq{noTailBiting}.
If we ignore this constraint then, on tidying up using 
the coincident line identities \cite{ymi,ymii}
\bea
\dDelta^{A,F,FA}_{\mu,R,A\alpha}(p;k;-k,-p) &=&
-\dDelta^{AF,FA}_{\mu R,A\alpha}(p,k;-k,-p)
-\dDelta^{FA,FA}_{R\mu,A\alpha}(k,p;-k,-p),\nonumber\\
\dDelta^{,AF,FA}_{,\mu R,A\alpha}(;p,k;-k,-p)&=&
\dDelta^{FA,FA}_{R\mu,A\alpha}(k,p;-k,-p),
\eea
\etc, we get a contribution to the integrand in \eq{oneloop} of the form:
\be
2\hS^{AA}_{\nu\alpha}(p)\dDelta^{AF,FA}_{\mu R, R\alpha}(0,k;-k,0).
\ee
Now, since there is no $\dDelta^{\C\A}$ kernel, we have by \eq{wev}:
\bea
\dDelta^{AF,FA}_{\mu R, R\alpha}(0,k;-k,0) 
&=& \dDelta^{AB,BA}_{\mu\alpha}(0,k;-k,0)\\
&=& \dDelta^{AA,AA}_{\mu\alpha}(0,k;-k,0).\label{dDAAAA}
\eea
By replacing $F$ as described below \eq{oneloop}, we see by the same argument
that there is no corresponding $C$ sector contribution, 
but by \eq{dDAAAA}, there is an equal and opposite $A$ sector contribution.
Thus the wine-biting-their-tail diagrams exactly cancel here in any case, 
as already explained below \eq{oneloop}. 

Collecting the results, we have
\bea
&&-4\beta_1\Box_{\mu\nu}(p)+O(p^3) =
N \int{d^4k\over(2\pi)^4}\Bigg\{ 
\ldl\Big[ 
2\Delta^{FF}_{SR}(k) \,
S^{\ph0AAFF}_{0\,\mu\,\nu RS}(p,-p,k,-k)\nonumber\\ &&
\qquad-\Delta^{FF}_{SR}(k)\Delta^{FF}_{TU}(k-p)
S^{\ph0AFF}_{0\,\mu TS}(p,k-p,-k)
S^{\ph0AFF}_{0\,\nu RU}(-p,k,p-k)
\Big]\nonumber\\ &&
\quad-\frac{4}{\Lambda^2}c'_0\Box_{\nu\alpha}(p)\left[
\Delta^{FF}_{SR}(k)\partial_{\mu}^k\partial_{\alpha}^k
\hS^{FF}_{RS}(k)
-\Delta^{FF}_{RU}(k)\Delta^{FF}_{ST}(k)
\partial^k_{\mu}\hS^{FF}_{RS}(k)
\partial^k_{\alpha}\hS^{FF}_{TU}(k)\right]\nonumber\\ &&
\quad+4\dDelta^{FF}_{TU}(k)k'_Uk_R\hS^{AAFF}_{\mu\,\nu\, RT}(p,-p,k,-k)
+4\dDelta^{A,FF}_{\mu,TU}(p;-k,k-p)k'_Tk_R\hS^{AFF}_{\nu\,RU}(-p,k,p-k)
\nonumber\\ &&
\quad-4S^{\ph0AFF}_{0\,\nu RU}(-p,k,p-k)\hS^{AFF}_{\mu\,TM}(p,k-p,-k)
\Delta^{FF}_{TU}(p-k)\dDelta^{FF}_{MN}(k)k'_Nk_R
\nonumber\\ &&
\quad-2S^{\ph0AFF}_{0\,\mu RT}(p,k-p,-k)\left[
2\dDelta^{A,FF}_{\nu,UR}(-p;p-k,k)(-k)'_U(-k)_T\right.\nonumber\\ &&
\qquad\left.-\dDelta^{A,FF}_{\nu,UV}(-p;p-k,k)(-k)'_U(-k)_T(k-p)'_V(k-p)_R
\right]\nonumber\\ &&
\quad-8\Box_{\nu\alpha}(p)\dDelta^{F,AB}_U(k;0,-k)k'_\alpha k_R\partial^k_{\mu}
\hS^{FF}_{RT}(k)\Delta^{FF}_{TU}(k)
\nonumber\\ &&
\quad-8\Box_{\nu\alpha}(p) (-k)'_\alpha (-k)_R
\dDelta^{AF,BA}_{\mu R}(0,k;-k,0)\ 
\Bigg\}
\label{final}
\eea
The first two lines contain the total $\Lambda$ derivative terms. The next
line contains the only potentially universal terms so far. There then follows
in the order generated above, all the gauge remainder terms,\footnote{Terms
related by charge conjugation, $p_\mu\leftrightarrow-p_\nu$, and relabelling
the loop momentum, have been combined.} in particular 
the last term is all that is left from the diagrams above that generated
wine-biting-their-tail diagrams.

Note that up until now, all the essential steps have been performed at a 
totally diagrammatic level.\footnote{See ref. \cite{antonio} for some of the
algebraic expressions.} Indeed it is far more efficient and elegant to
do so, not the least because the diagrams make very clear the relations
due to cyclicity, and charge conjugation (\aka mirror symmetry), and
are already identical if the algebraic expressions are equal after
relabelling the loop momentum.  It is possible to evaluate the gauge
transformations in \eq{final} also at this level \cite{ymi,ymii}, so
it may be possible to push these diagrammatic techniques further.
However we now have to take account of the differences between the $A$,
$C$ and $F$ sectors. In this paper, we will pursue the remainder of the 
analysis at the level of equations.

\subsection{The total $\Lambda$ derivative contribution}
\label{total}

We start by computing the $O(p^2)$ part of the
first two lines in \eq{final}. These two terms
seem to have the clearest physical interpretation. They both
depend only on effective propagators and the effective action $S$
(which, unlike the seed action, can be expected to contain the real
physics \cite{sca}).  Their diagrams, \cf \fig{trick1f} and
\fig{trick2f}, are actually just the usual tadpole and self-energy
Feynman graphs respectively. From the translations below \eq{oneloop} we see 
that the $A$ sector looks as though it is in Feynman gauge (although ghosts
are missing):
\bea
\label{Aphysics} 
N\ldl \int\!\!{d^4k\over(2\pi)^4}\Big[ &&
-2\,\Delta^{AA}_k\,S^{\ph0AAAA}_{0\,\mu\nu\alpha\alpha}(p,-p,k,-k)\\
&&\qquad
+\Delta^{AA}_k\Delta^{AA}_{k-p}\ S^{\ph0AAA}_{0\,\mu\alpha\beta}(p,k-p,-k)
\,S^{\ph0AAA}_{0\,\nu\beta\alpha}(-p,k,p-k)\ \Big]_{O(p^2)}\ ,
\nonumber
\eea
whilst the $F$ and $C$ sectors via \eq{iCC} and \eq{iBD}
give copies with $\sim\Lambda$ massive effective propagators, that are
the expected Pauli-Villars regularising terms. Indeed if the five dimensional
notation is expanded using \eq{defF}, we see that apart from the 
relative minus sign, the $B$ sector looks identical in form to \eq{Aphysics}.
Similarly, since the sign in \eq{ikFF} cancels on translating the vertices 
back to $AADD$ and $ADD$ form, the $D$ sector has identical appearance, but
opposite sign, to the $C$ sector. Since we have ensured
that at high momentum $k$, the $B$ ($D$) terms do actually equal the $A$ ($C$)
terms, we see that the regularisation is incorporated as required. Finally,
the overall $\Lambda$ derivative is just as one would expect to convert the 
$\ln\Lambda$ divergent result to a contribution to the $\beta$ function 
\cite{sca}.

As shown in \eq{Aphysics}, we now exchange the order of 
$\ldl$ and the $k$ integral. Pulling out the $O(p^2)$
part, the integral is actually dimensionless. Since the only explicit scale
is $\Lambda$, if the integral were convergent it would have to be a constant,
thus \eq{Aphysics} combined with the $B,C,D$ sectors, would
vanish. In fact the momentum integral is not well defined 
only because the second term in the $A$ sector has an infrared divergence. 
We can still keep $\ldl$ outside if 
we use the standard trick of introducing an infrared cutoff 
$k>\epsilon$ in this term, taking $\epsilon\to0$ at the
end. Now this term, and this term only, depends on $\Lambda$, through 
$\Lambda/\epsilon$. Therefore it suffices to analyse its infrared
behaviour, replacing $\ldl$ with $-\epsilon\partial_\epsilon$.

Note that the $S_0$ vertices
have Taylor expansions in small momenta, \cf \sec{Necessary}, and the $B$,
$C$ and $D$ effective propagators \eq{iCC}, \eq{iBD}, are regular as $k\to0$,
whilst the $A$ effective propagator \eq{ikAA} has an infrared double pole.
Thus only in the $A$ sector is there a problem.
The result is integrable in the first term in \eq{Aphysics} but not in the 
second, once expanded to order $p^2$. However,
whilst $\partial_\Lambda$ is inside the integral
there is actually no infrared divergence, it being ameliorated via
\eq{kAA} or the flow equations for the three-point vertices, equivalently
as we derive below, because the only terms in the three-point vertices that
contribute are independent of $\Lambda$. One can then confirm
that the above limit $\epsilon\to0$ gives 
the same answer. Of course the same answer is also
obtained by using dimensional regularisation or by converting this
part to a total derivative in $k^2$ as \eg in ref. \cite{sca}.

If we had not taken care to exclude the logarithmic classical divergences in 
\sec{Ensuring}, it is at this point that we would have picked up 
extra contributions
since the $B,C,D$ parts of the integral can then be functions of 
$\Lambda/\mu$. Worse, we have confirmed that if these parts are
evaluated first, using here 
$\partial/\partial\Lambda\equiv-\partial/\partial\mu$,
the resulting $k$ integrals are non-universal, 
as may be expected from the general  arguments in \sec{Enforcing}.
However, in ref. \cite{ymii}, where very similar classical $\mu$ dependent 
terms were not excluded, the right answer was obtained by keeping an upper
limit $\Lambda_0$ in the $\Lambda$ integrals as in \eq{SACC}, and in fact
for these terms sending $\Lambda_0\to\infty$ only after the $k$ integral
had been performed. It thus appears that if the logarithmic classical 
divergences are not excluded, then the result actually depends on the order 
in which the ultraviolet limits on the $\Lambda$ and $k$ integrals are 
performed.

Returning to the main analysis, we note that we are interested in expanding
in $p$ and then in $k$. $S^{\ph0AAA}_{0\,\mu\,\nu\,\lambda}$ is 
regular in small momenta and the lowest order term in its momentum 
expansion, with one momentum, is fixed 
uniquely by \eq{defg} to be the standard Feynman gluon vertex.
If we take the $p$ part in both three-points in
\eq{Aphysics}, this already saturates the order $p^2$ required. Furthermore,
we then must take only the $1/2k^2$ parts of 
the effective propagators in order to get an infrared divergence.
If we take a $k$ part from the three-point vertices, then in order to
maintain an infrared divergence we must use one of the more divergent terms 
from the $p$ expansion of $1/2(k-p)^2$, thus again saturating the expression
at $O(p^2)$. Thus we see that the only part that contributes to the total
$\Lambda$ derivative terms in \eq{final}, is universal
[as a consequence of \eq{defg}], and precisely of the same form as the 
standard gluonic self-energy term in Feynman gauge. We have
\bea
&& N\epsilon\partial_\epsilon \int_\epsilon{d^4k\over(2\pi)^4}\,\Big\{ 
{1\over k^4}\partial^p_\beta\Box_{\mu\alpha}(p)
\partial^p_\beta\Box_{\nu\alpha}(p)
+\left[4{(k\cdot p)^2\over k^8}-{p^2\over k^6}\right]
\partial^k_\mu\Box_{\alpha\beta}(k)\partial^k_\nu\Box_{\alpha\beta}(k)
\nonumber\\
&&\ph{N\epsilon\partial_\epsilon \int_\epsilon{d^4k\over(2\pi)^4}\,\Big\{ 
{1\over k^4}}
+2{k\cdot p\over k^6}\left[
\partial^k_\mu\Box_{\alpha\beta}(k)\partial^p_\beta\Box_{\alpha\nu}(p)
+\partial^p_\beta\Box_{\mu\alpha}(p)\partial^k_\nu\Box_{\alpha\beta}(k)
\right]\ \Big\}\nonumber\\
=&& {N\over (4\pi)^2}\left\{ {19\over3}\, p^2\delta_{\mu\nu}
-{22\over3}\, p_\mu p_\nu\right\},
\label{totalD}
\eea
where on the left hand side we used \eq{gthree} to evaluate 
$S^{\ph0AAA}_{0\,\mu\alpha\beta}(p,-p,0)$, 
$S^{\ph0AAA}_{0\,\mu\alpha\beta}(0,k,-k)$, \etc, and substituted
$S^{\ph0AA}_{0\,\mu\nu}\sim2\Box_{\mu\nu}$, \cf \eqs{hSAA}{SAA}, and the 
right hand side follows after averaging over $k$ directions and expressing
as a radial $k$ integral.

As expected of a universal term, \eq{totalD} does not, apparently, 
depend on the  regulating $C$ and $F$ sectors. But despite this, and
the apparently clear physical interpretation, it is not transverse on its 
own as would be 
required by gauge invariance. This result seems all the more surprising
once one notes that actually at a formal level
the $A$ sector {\sl is} gauge invariant on its own \cite{ymi}
(and as we will confirm, the other $A$ sector terms in \eq{final} are
already transverse). Indeed, taking care to keep the $\Lambda$ derivative
inside the \eq{Aphysics} integral (so there is no problem with infrared 
divergences), contracting with $p_\mu$ and $-p_\nu$, then using
\eq{bosonwi}, \eqs{hSAA}{SAA} and \eq{ikAA}, and shifting $k$ in some terms
(discarding one odd in $k$),\footnote{See refs. \cite{antonio,ymi,ymii} 
for further comments pertinent to here and above.}
one finds that the longitudinal part of \eq{Aphysics} is
\be
\label{AphysL}
2N\int\!\!{d^4k\over(2\pi)^4}\ \ldl {p_\beta\Box_{\beta\nu}(k) p_\nu
\over k^2(k-p)^2},
\ee
which obviously vanishes since the remaining terms have no $\Lambda$
dependence. Of course these manipulations do not make sense without
ultraviolet regularisation. The apparent independence of \eq{totalD} 
on the $C$ and $F$ sectors
is illusory since the derivation of \eq{totalD} from \eq{Aphysics}
is only legitimate if such sectors exist with the property
that they cancel its ultraviolet divergences whilst not adding any new
infrared ones. Taking into account the other sectors, 
one finds that the $C$ sector also gives a purely transverse contribution, 
but the $F$ sector yields a longitudinal part, which is precisely the one in
\eq{totalD}.

The full understanding of the apparent Feynman gauge in the total derivative 
term \eq{totalD} is closely related. It comes from the absence of
longitudinal terms in the RG equation, which were not included, as
by supergauge invariance they could be exchanged for $\C$ commutators,
\cf \eq{gaugeS}. Had we introduced them,  \eg as the term
\be
\label{longAdd}
{1\over2}\nabla_\mu\!\cdot\!{\delta S\over\delta\A_\mu}
\ \{\dDelta_L\}_{\!\!{}_\A}
\nabla_\nu\!\cdot\!{\delta \Sigma_g\over\delta\A_\nu}
\ee
in \eq{a0} [and correspondingly in \eq{a1}],
the $A$ effective propagator would 
change to include the characteristic longitudinal dependence in general
gauge, with\footnote{The $x=p^2/\Lambda^2$ term is missing by \eq{defprop},
because \eq{longAdd} must be analytic in $p$ \cf \sec{Necessary}.}
\be
\Delta_L(p,\Lambda) ={1\over p^4}\left[{1\over\xi}-1 +O(x^2)\right]
\ee 
parametrising it, generalising the usual gauge fixing parameter dependence.
Such a change would of course be compensated for by similar changes
in the $B$ and $D$ effective propagators, via \eq{gaugeS} or \eq{hSBDwi}.
However, unlike in a gauge fixed theory, the two-point classical action
vertices would remain as \eq{hSAA}, \eq{hSBB} and \eq{hSBDs},
gauge invariant and completely independent of the
introduction of $\Delta_L$. The calculation would simply have been 
rearranged, by finessing some longitudinal parts into the $F$ sector.

Finally, let us make a trivial, but important, observation. Although
\eq{totalD} is precisely the $\ln(\Lambda/\epsilon)$ contribution
from the standard gluonic self-energy term in Feynman gauge, it is here
a contribution to the Wilsonian effective action, not a contribution
to the $S$ matrix. Feynman's unitarity arguments for 
the existence of ghosts \cite{Feynman} cannot be directly
applied to vertices of the Wilsonian effective action. Of course,
if we wished to talk about on-shell gluons, we would have to gauge
fix and introduce ghosts, but that is not what we are talking about.
As already discussed, here the ghostly contributions are replaced by
shadows from the regularisation sector.

\subsection{The gauge remainders}
\label{gauge}
We next turn to the gauge remainders, the fourth line onwards in \eq{final}.
As well as the $F$ sector shown, we also have the $A$ sector contributions
which we consider first.
Recall that they come with opposite sign and are translated as described
at the beginning of this section. Making use of \eq{bosonwi} and
the same sort of simplifications as in the previous subsections,
many terms cancel out, including some that would otherwise not be
potentially universal. We are left with purely transverse contributions:
an ultraviolet divergent contribution to the 
integrand\footnote{Clearly, by Lorentz invariance 
of the $k$ integration, this is transverse on $\nu$ also.}
\be
\label{cp0A}
{8c'_0\over \Lambda^2 k^2}\Box_{\mu\alpha}(p)\left(\delta_{\alpha\nu}
-k_\alpha k_\nu/k^2\right)
\ee
which will be combined later with the potentially universal terms
on the third line of \eq{final} for the reason given below \eq{00wi},
and after averaging over $k$ directions, a term
\be
\label{grA}
-{2N\over\Lambda^4}\Box_{\mu\nu}(p)\int\!\!{d^4k\over(2\pi)^4}\  
\Big(5 c'_k\Lambda^2/k^2+c''_k \Big) = {N\over2\pi^2}\Box_{\mu\nu}(p),
\ee
where we have used the fact that,
after casting as an $x=k^2/\Lambda^2$ integral, the integrand is a total
derivative, vanishing at $x=\infty$ and universal at $x=0$ by $c(0)=1$.

Evaluating the  $F$ sector using \eq{Fwi} and similar simplifications
to above, but keeping the compact 5 dimensional notation, already many
terms cancel out, again in particular all terms that would otherwise
not be potentially universal. This time the remaining terms 
are all clearly transverse except one:
\be
4\,\Delta^{FF}_{SN}(k)k'_N(k-p)'_\nu\hS^{BF}_{\mu S}(k)\Big|_{O(p^2)}.
\ee
After some algebra, and including the integral in \eq{final},
this takes the form
\be \label{nontrans}
-{2\over3}{N\over(4\pi)^2}\left(p^2\delta_{\mu\nu}+2p_\mu p_\nu\right)
\int^{\infty}_0\!\!\!\!dx\
 g'(x)\left(x^3f'(x)\right)',
\ee
where $f$ and $g$ were defined in \eq{deffg}. Substituting $g$ in terms
of $f$ and integrating by parts this may be cast as a total derivative.
Using $\ct(0)=c(0)=1$ and for example \eq{inequalities}, it evaluates to
\be
\label{grF}
{1\over3}{N\over(4\pi)^2}\left(p^2\delta_{\mu\nu}+2p_\mu p_\nu\right).
\ee
Although it is not transverse,
added to \eq{totalD}, it results in a transverse contribution, namely
$20N/3(4\pi)^2\ \Box_{\mu\nu}(p)$. In view of the comments below \eq{AphysL},
we should expect to find such a non-transverse correction lurking in 
the $F$ sector. 

The transverse supergauge remainder terms yield a contribution to the
integrand of \eq{final} of
\be
\label{cp0F}
-{8\over\Lambda^2}c'_0\Box_{\mu\alpha}(p)\left[k'_\alpha k'_\nu
+\partial_\nu(k_\alpha/k^2)\right]
\ee
to be compared with \eq{cp0A} and similarly saved for later, and a number of 
terms which after some algebra turn into a total derivative in $x$, yielding
\be
-{2N\over(4\pi)^2}\Box_{\mu\nu}(p)\ (xc'-4c)\,x^2f^2\Big|^\infty_0,
\ee
(where $f$ and $c$ are evaluated at $x$). Considering again the behaviour
of $c$ and $\ct$ at the boundaries, we see that the result vanishes.

We are left only to evaluate the terms proportional to $c'_0$. Before we
do so, we note that the total contribution to \eq{final} from the other
terms, namely \eq{totalD}, \eq{grA} and \eq{grF} sum to 
$44N/3(4\pi)^2\ \Box_{\mu\nu}(p)$, \ie already yielding the famous result
\be \label{betaone}
\beta_1 = -{11\over3}{N\over(4\pi)^2}.
\ee

\subsection{The $c'_0$ terms}
\label{c0terms}
As we have already anticipated, all the $c'_0$ terms must sum to zero.
Together with \eq{cp0A} and \eq{cp0F}, we also have the three sectors
contributions from the third line of \eq{final}. This term can be
simplified in a way that can be mapped also to the $C$ and $A$ sectors.
Thus, suppressing the $F$ superscripts and the $k$ momentum arguments:
\bea
&&\Delta_{SR}\partial_\mu\partial_\alpha\hS_{RS}-\Delta_{RU}\Delta_{ST}
\partial_\mu\hS_{RS}\partial_\alpha\hS_{TU}\nonumber\\
&&=\partial_\mu\!\left(\Delta_{RS}\partial_\alpha\hS_{RS}\right)
-\partial_\mu\Delta_{RS}\partial_\alpha\hS_{RS}
-\Delta_{RU}\Delta_{TS}\partial_\mu\hS_{RS}\partial_\alpha\hS_{TU}\nonumber\\
&&\equiv\partial_\mu\!\left(\Delta_{RS}\partial_\alpha\hS_{RS}\right)
-\Delta_{RU}\partial_\alpha\hS_{RS}\,\partial_\mu\!\left(k'_Uk_S\right)
-k'_Tk_R\,\partial_\mu\Delta_{TS}\partial_\alpha\hS_{RS}\nonumber\\
&&=\partial_\mu\!\left(\Delta_{RS}\partial_\alpha\hS_{RS}\right)
-2k'_\alpha k'_\mu-2\partial_\alpha k'_\mu
\label{c0all}
\eea
This holds in all sectors using the maps at the beginning of this section.
(To see this, rewrite the third term of the second line to contain
$\partial_\mu(\Delta_{RU}\hS_{RS})$
plus remainder, and use \eqs{iFFf}{iFFb}, and $k\mapsto-k$. 
In the third line, third term, transfer the $\partial_\alpha$ to $k_R$,
using \eq{ftransverse}, equivalently \eq{transverse}, and
use $\partial_\alpha k_R=\delta_{R\alpha}$, which
holds in both $F$ and $A$ sectors. Finally in this term form  
$\partial_\mu(S_{RS}\Delta_{ST})$ and simplify similarly, 
noting that since it is a function only of $k$ it is automatically $\alpha
\leftrightarrow\nu$ symmetric.)

By dimensions and Lorentz invariance 
$\Delta_{RS}\partial_\alpha\hS_{RS}=2k_\alpha F(x)/\Lambda^2$ for some
function $F(x=k^2/\Lambda^2)$, with corresponding expressions in
the $A$ and $C$ sectors. Indeed by \eq{iCC}, we have for the $C$ sector,
$F={d\over dx}\ln(2\lambda+x/\ct)$; by \eq{ikAA} and \eq{hSAA}, we have
$F=3{d\over dx}\ln(x/c)$ in the $A$ sector; and finally in the $F$
sector, using \eq{hSDD}, \eq{hSBB}, \eq{ikFF} and \eq{iBD},
\be
F={d\over dx}\ln(x/c+2/\ct)+{d\over dx}\ln(x/\ct+2c/\ct^2)
-{c\over x\ct+2c}-{\ct^2\over x\ct+2c}{d\over dx}{c\over\ct^2}.
\ee
Including the multiplicative factors in \eq{final}, the first term of 
\eq{c0all} thus gives the surface contribution:
\be
\label{firstc0all}
-{N\over4\pi^2}c'_0\Box_{\mu\nu}(p)\ x^2F\Big|^\infty_0.
\ee
Summing this contribution across all sectors (remembering the relative
minus sign for $A$ and $C$), one
finds no contribution from the $x=0$ boundary in any sector,
whilst the otherwise divergent $x=\infty$ terms cancel exactly between the 
fermionic and bosonic sectors.

This leaves just the second and third terms of \eq{c0all}. Incorporating the
extra factors in \eq{final}, we see that the second term cancels the first
in \eq{cp0F}, whilst the remaining terms combine to an ultraviolet finite 
total derivative $2\partial_\mu(k_\alpha/k^2-k'_\alpha)$. Comparing with
the above structure and using \eq{defdual}, we see trivially that
this integrates to \eq{firstc0all} with $F=1/x-f$, and thus vanishes.
Finally, recalling the relative minus sign, we see that 
the $A$ sector translation of the second and third terms in \eq{c0all}
exactly cancel \eq{cp0A}. We have thus shown that all the $c'_0$ terms 
disappear, as expected.

\section{Conclusions}
\label{Conclusions}

We have proposed a technique which allows manifestly gauge
invariant and universal computations to be performed, directly in the 
continuum. No gauge fixing or ghosts are required, thus avoiding completely
the problem of Gribov copies~\cite{Gribov}. The full power and beauty of
local gauge invariance is clear and central to the whole calculation. Moreover,
renormalization group properties are built in from the beginning. The
calculation proceeds very economically,
by confirming and exploiting the independence 
of physical quantities on the details of the regularisation.

At its heart, lies the successful combination of gauge invariance with the 
introduction of a real effective cutoff \cite{sunn}. This has long been an
outstanding problem, as any straightforward division of momenta into
large and small, according to some effective scale $\Lam$, is not 
preserved by gauge transformations. (Explicitly,
under a gauge transformation of some matter field
$\phi(x) \rightarrow \Omega(x) \, \phi(x)$, momentum modes $\phi(p)$ are
mapped to a convolution with the modes from $\Omega$.) 
Its solution allows us to use Wilson's insight~\cite{Wil,jose} making
renormalization properties, normally subtle and complicated, trivially clear
and straightforward. 

This continuum cutoff is simply
spontaneously broken $SU(N|N)$ gauge theory with covariant
higher derivatives. Sketched at the start
of sec.~\ref{Preliminary} (for a complete analysis, see \cite{sunn}),  
it includes two copies of the $SU(N)$ gauge field, $A_1$ and $A_2$,
and a pair of wrong statistics gauge fields $B$, $\bar{B}$. 
A superscalar field is added to cause spontaneous symmetry
breaking down to the bosonic $SU(N)\times SU(N)$, giving
masses of order the cutoff to all fields but
$A_1$ and $A_2$. Depending on the representation chosen, 
one also encounters a
$U(1)$ connection, $\A^0$. This latter, however, does not appear anywhere
in the action provided all interactions are of the form $\str (\A \times
\mbox{commutators})$, more generally provided a no-$\A^0$ shift symmetry
$\delta\A_\mu=\lambda_\mu\one$ is respected.
Such a shift symmetry, which amounts to dynamically defining the coset space
$SU(N|N)/U(1)$, plays an important r\^ole, 
ensuring that our flow equation, \eq{sunnfl}, is indeed gauge
invariant [\cf eqs.~(\ref{tlgaugetr})-(\ref{qcgaugetr})].
We can already see that it will play an even more central r\^ole 
in the future \cite{us}.

This regularisation fits very well within the 
effective action framework.
A manifestly gauge invariant flow equation for the Wilsonian effective
action, $S$, has been written down,
\eq{sunnfl}, which leaves the partition function - and hence the physics
derived from it -- unchanged. This is achieved by demanding that the
Boltzmann measure, $\exp -S$, flow into a total derivative. 

Solutions for $S$ can then be found directly in terms 
of renormalized quantities, without specifying a bare action.
Note that almost any  approximation can be considered without 
spoiling this property \cite{rev}.

As discussed in \sec{Necessary}, since
the partition function is left unchanged, it is possible to 
interpret our results at different scales $\Lam$. 
The fact that locality is correctly incorporated, for example, 
is best understood through the $\Lambda=\infty$ limit, 
recalling that the
effective action is Taylor expandable to all orders in momenta (this
requirement of ``quasilocality''~\cite{ymi}, is tantamount to demanding that
each infinitesimal RG step be free from infrared singularities) and
that $\Lam$ is the only explicit scale parameter to appear in the action,
thus implementing the concept of a self-similar flow~\cite{Shirkov}.  
Again it is at $\Lam \to \infty$ that we see we are indeed describing
$SU(N)$ gauge theory: massive fields are actually infinitely 
massive and $A_2$ decouples, as
ensured by the Appelquist-Carazzone theorem \cite{apple}.

Besides manifest gauge invariance, another important step is the 
exploitation of the freedom coming from scheme
independence~\cite{jose}: not only did the seed action $\hS$
incorporate the gauge invariant regularisation and allow the spontaneous
symmetry breaking [see comments below \eq{ergone}], but also we could
choose it such that the minimum of the potential would flow with $\Lam$
(\cf\eq{minimum} and below)
and such that logarithmic classical divergences would be completely 
absent (see sec.~\ref{Ensuring}). This latter property guarantees
the universality of the
one-loop Yang-Mills $\beta$ function (\cf \sec{Enforcing}). 
By further fashioning $\hS$,
these properties extend to two-loop order  and higher \cite{us}.
In a sense the generalised exact RG framework \cite{alg,jose} gives
us the freedom to create `designer field theories' in the regularisation
sector, tailored to our purpose.

The seed action (not bare action) therefore represents the details put in by hand;
the physics is naturally encoded in the effective action. 
Hence, physical quantities must be independent of the choice of $\hS$ 
\cite{jose}, as our tests here, and earlier \cite{sca}, have shown.
Note: we do not pick a specific $\hS$. Apart from the constraints above, 
it turns out to be very advantageous to keep it as general as possible. 
Since the final result must be 
independent of its detailed form, any simplification has to occur before 
we look into any of its vertices.
Similarly, we do not specify what covariantisation we use. All we need 
is to name the vertices (\cf sec.~\ref{Covariantisation}). 

We do introduce some further restrictions purely for convenience. Thus
we determine the RG kernels
$\dot{\Delta}$ in such a way that after spontaneous symmetry breaking,
the two-point classical effective action vertices can be set
equal to those of the seed action. This greatly simplifies the
flow equations for higher point interactions, \cf sec.~\ref{manifestly} 
\cite{alg,ymi}. Actually, the
two-point seed action vertices are not as general as they can be, \cf
sec.~\ref{Seed}. It would not be difficult to work with completely general
versions however, leaving
no restrictions except those placed by smoothness, symmetries
(including broken supergauge invariance) and limits on their 
ultraviolet behaviour, \cf \eg \eq{inequalities}.
Very little would change in the calculation right up to the final
stages in secs.~\ref{gauge} and \ref{c0terms}.
 
Last but not least, the diagrammatic interpretation introduced in
sec.~\ref{DiagrammaticI} (generalising that of \cite{alg,ymi,ymii}
to finite $N$)
turned out to be very useful. Already at the
level of diagrams the big potential of the method comes out, \eg making
cancellations among higher order seed action vertices self evident. 

In view of the novelty of the construction, it is desirable to test the
formalism first. We have computed for the first time at finite
$N$ without fixing the gauge,
the one-loop $\beta$ function for $SU(N)$
Yang-Mills theory, verifying the standard result,
and confirming its universality with respect to regularisation
scheme.

Let us briefly recapitulate the simple steps that form the calculation, 
as discussed in detail in sec.~\ref{calculation}.
From the renormalization condition, eq.~\eq{defg}, we can write down an
{\it algebraic} equation for the one-loop beta
function, $\beta_1$ (\cf Fig.~\ref{beta1}), which receives 
contributions, equal in form, from all the sectors of the theory, $A$, $C$, 
and $F$. Introducing integrated
kernels and integrating by parts the diagram containing the
effective action vertex with the highest number of points,
leaves us with a total $\Lam$-derivative plus the $\Lam$-derivative of
this vertex (see fig.~\ref{trick1f}). Then using
the flow equation and the relation between integrated
kernels and two-point functions, as in \eq{iCC} and \eq{iAA},  to simplify the
diagrams obtained so far, we can eliminate the higher point vertices. 
Repeating the above steps several times over, we
get to the point where there is no dependence left on the details of 
the covariantisation or on the seed action.\footnote{The calculation
in ref.~\cite{sca} can also be organised according to this iterative
method.}

This last step follows either because we are left with expressions containing 
only two-point vertices and the corresponding zero point kernels, which are 
then shown to be universal by casting them as total derivatives in momentum 
space, where they depend only on ultraviolet limits or in the infrared on the
renormalization condition \eq{defg}, or, 
as in the total $\Lambda$-derivative term, because the result
can be seen explicitly to depend only on vanishing momenta where again the
dependence becomes universal as a consequence of \eq{defg}.

Some comments are in order.
As emphasised earlier, central to the method is the use of manifest gauge
invariance. It results in two sets of ``trivial'' Ward identities, derived
in sec.~\ref{unbroken}, corresponding to the (bosonic) fermionic parts of 
the (un)broken gauge invariance.
The broken Ward identities, in particular, cause $B$ and $D\sigma$ to
rotate into one another. This led to the important
realisation that technically they should be
tied together as elements of a five dimensional vector, $F_M$,
\cf \eq{defF}. As well as simplifying greatly the calculations in this sector,
it also casts them in a form which is virtually isomorphic with
the computations in the $A$ and $C$ sectors.

The calculation has been carried out in dimension $D=4$, even though
to rigorously define the regularisation to all orders, one 
should preregularise, \eg by working in general dimension $D$ and taking
the limit $D\to 4$ at the very end \cite{sunn}. This is necessary
in general \cite{alg} because finiteness is achieved by adding separately 
divergent pieces together, namely the $F$ versus $A$ and $C$ sectors, as is 
typical of Pauli-Villars type regularisations. However
at one loop, it is only necessary to preregularise the four-point $\A$
vertex; the two-point vertex (which is what we compute) is well defined 
without preregularisation, providing only that global
$SU(N|N)$ invariance is kept manifest throughout the computation \cite{sunn}. 

We did not do this, since we broke the fields $\A$ and $\C$ into their
fermionic and bosonic parts. (Had we worked with the full superfields
we would not have been able to (anti)commute $\sigma$. This leads to
a substantial increase in the number of different vertices required for
the computation.) But, by performing the calculation in the $F$
sector and then simply mapping the results to the $A$ and $C$ sectors
using the insight above, we performed the calculation in the same way
for all sectors, achieving the same effect as would be obtained by keeping
the global $SU(N|N)$ invariance manifest.\footnote{We have also checked 
explicitly that there are no 
ambiguities related to the limit $D\to 4$ \cite{antonio}.}

The final result, \eq{betaone}, comes almost entirely from the $A$
sector, except for the non-transverse term, \eq{grF}, which however
is needed to make the whole thing transverse. On the one hand,
we should not be
surprised if the {\sl whole} contribution had come from just the $A$ sector.
After all, the one-loop $\beta$
function can be cast as the derivative of the cutoff on an infrared
divergent integral. If the fields are massive (\viz $B$, $C$, $D$), 
their masses act instead as the infrared regulator
and thus they give vanishing contribution [as in the discussion below 
\eq{Aphysics}]. On the other hand, the
pure gauge sector yields \eq{totalD} plus \eq{grA},
apart from contributions that must vanish eventually
\eg \eq{cp0A}, and is thus not transverse on its own. As discussed
in \sec{total}, the culprit, \eq{totalD},
takes precisely the form of the one-loop gluon self-energy in Feynman 
gauge, but through an accident of simplicity. We could have introduced
longitudinal terms \eq{longAdd} into the flow equation \eq{a0},
giving an apparent general gauge,
which by gauge invariance \eq{gaugeS}, finesses longitudinal parts into
the $F$ sector. All the while, the effective action would remain oblivious
to these rearrangements, and it and all stages in the calculation remain
gauge invariant. However in this limited sense, it would appear that the ghost
contributions, required in the standard treatment,
are here taken over by the $F$ sector.
 
Finally, although the only application of our method as yet is the 
calculation of $\beta_1$, we expect the procedure to be quite general and, above all,
best suited for exploring the non-perturbative domain. It should allow
investigations of instantons, renormalons and other controlled
non-perturbative effects in a manifestly gauge invariant way. 

Generalisations to include space-time supersymmetry and/or fermions
seem possible and would open the door to many 
further investigations, from Seiberg-Witten methods~\cite{SW} through
to fully non-perturbative approximations. 

\acknowledgments{ T.R.M. and S.A. acknowledge financial support
from PPARC Rolling grant\hfill\\ PPA/G/O/2000/00464. It is a pleasure to thank
Hugh Osborn and Jean Zinn-Justin for useful discussions.
}


\end{document}